\documentclass[a4paper,11pt]{article}
\pdfoutput=1 %

\usepackage{jheppub} 
\usepackage{amsmath,amssymb,graphicx,float,slashed,xcolor,multicol}
\usepackage{tabularx}
\usepackage{url}
\usepackage{footmisc}
\usepackage{amsfonts}
\usepackage{cancel}
\usepackage{color}
\usepackage{multirow} 
\usepackage{pifont}
\usepackage{epstopdf}
\usepackage{comment}
\usepackage{booktabs} 
\usepackage{natbib}
\usepackage{array}
\usepackage{mathrsfs}
\usepackage[toc,page]{appendix}
\usepackage{mathtools}
\usepackage{romannum}
\usepackage{ulem}
\usepackage{bbold}
\usepackage{enumitem}
\usepackage{multirow}
\usepackage{caption,subcaption}
\usepackage{float}
\usepackage{multirow}
\usepackage{upgreek}
\usepackage[toc,page]{appendix}
\usepackage{romannum}
\allowdisplaybreaks
\usepackage{bbm}                
\usepackage{xspace}	
\usepackage{amssymb}
\usepackage{pifont}
\usepackage{multirow}

\newcommand{\cmark}{\ding{51}}%
\newcommand{\xmark}{\ding{55}}%

\newcommand*{\rom}[1]{\expandafter\@slowromancap\romannumeral #1@}

\def\lag{\mathscr{L}}

\def\d{\partial}
\def\beq{\begin{equation}}
\def\eeq{\end{equation}}
\def\beqa{\begin{eqnarray}}
\def\eeqa{\end{eqnarray}}
\title{Electroweak Phase Transition with an SU(2) Dark Sector}
\author[a,b]{Tathagata Ghosh,}
\author[c]{Huai-Ke Guo,}
\author[b]{Tao Han,}
\author[b]{Hongkai Liu}
\affiliation[a]{Instituto de F\'{i}sica, Universidade de S\~{a}o Paulo, S\~{a}o Paulo, Brazil}
\affiliation[b]{Department of Physics and Astronomy, University of Pittsburgh, Pittsburgh, PA 15260, USA }
\affiliation[c]{Department of Physics and Astronomy, University of Oklahoma, Norman, OK 73019, USA}
\emailAdd{ghoshtatha@usp.br}
\emailAdd{ghk@ou.edu}
\emailAdd{than@pitt.edu}
\emailAdd{hol42@pitt.edu}

\preprint{
\begin{flushright}
PITT-PACC-2011
\end{flushright}
}

\abstract{We consider a non-Abelian dark SU(2)$_{\rm D}$ model where the dark sector 
couples to the Standard Model (SM) through a Higgs portal. We investigate two different scenarios of the dark sector scalars with $Z_2$ symmetry, 
with Higgs portal interactions that can introduce mixing between the SM Higgs boson and the SM singlet scalars in the dark sector. We utilize the existing collider results of the Higgs signal rate, direct heavy Higgs searches, and electroweak precision observables to constrain the model parameters. The $\text{SU(2)}_{\text{D}}$ partially breaks into $\text{U(1)}_{\text{D}}$ gauge group by the scalar sector. The resulting two stable massive dark gauge bosons and pseudo-Goldstone bosons can be viable cold dark matter candidates, while the massless gauge boson from the unbroken $\text{U(1)}_{\text{D}}$ 
subgroup is a dark radiation and can introduce long-range attractive dark matter (DM) self-interaction, which can alleviate the small-scale structure issues.  
We study in detail the pattern of strong first-order phase transition and gravitational wave (GW) production triggered by the dark sector symmetry breaking, 
and further evaluate the signal-to-noise ratio for several proposed space interferometer missions. 
We conclude that the rich physics in the dark sector may be observable with the current and future measurements at colliders, DM experiments, and GW interferometers. 
}

\begin{document}

\titlepage

\maketitle



\flushbottom

\section{Introduction}

The milestone discovery of the Higgs boson predicted in the Standard Model (SM) at the CERN Large Hadron Collider (LHC) has deepened our understanding of nature at the shortest distances, and in the same time sharpened our questions about the Universe. One of the most pressing mysteries in contemporary particle physics and cosmology is the nature of the dark matter (DM). 
There is mounting evidence for the existence of DM through its gravitational effects. However, the null results of the last fifty years of searches challenge the most theoretically attractive candidates, namely, the standard weakly interacting massive particles (WIMPs), that are charged under the SM weak interactions (see Ref.~\cite{Jungman:1995df} for review). On the other hand, it is quite conceivable that the DM particles live in a dark sector that are not charged under the SM gauge group. 
Furthermore, the dark sector may have a rich particle spectrum, leading to other observable consequences~\cite{Alexander:2016aln}. A massless dark gauge field, dubbed as the dark radiation (DR), is one of the quite interesting extensions that could help to alleviate the tension between Planck and HST measurements of the Hubble constant~\cite{Freedman:2017yms}. DM-DR interactions and DM self-interactions can provide solutions to the small-scale structure problems which challenge the cold dark matter (CDM) paradigm~\cite{Spergel:1999mh,Boehm:2001hm, Vogelsberger:2015gpr}. 

In this paper, we would like to explore the potentially observable effects beyond the gravitational interactions from a hypothetical dark sector. We assume that the dark sector interacts with the SM particles only through the Higgs portal~\cite{Patt:2006fw}.
An immediate consequence of this would be the modification of the Higgs boson properties that will be probed in the on-going and future high energy experiments \cite{deFlorian:2016spz, deBlas:2019rxi}. The DM searches from the direct and indirect detection experiments will provide additional tests for the theory~\cite{Jungman:1995df}. Perhaps, an even more significant impact would be on the nature of the electroweak phase transition (EWPT) at the early Universe (see, e.g., ~\cite{Ramsey-Musolf:2019lsf,Mazumdar:2018dfl,Huang:2016cjm} for recent reviews), which could shed light on another profound mystery: the origin of baryon asymmetry in the Universe. Indeed, one of the best-motivated solutions to this mystery is the electroweak baryogenesis (EWBG) \cite{Kuzmin:1985mm,Shaposhnikov:1986jp,Shaposhnikov:1987tw,Morrissey:2012db} (see also~\cite{Cline:2006ts,White:2016nbo} for pedagogical introductions). 
For a successful generation of the baryon asymmetry during the EWPT, all of the three Sakharov conditions~\cite{Sakharov:1967dj} have to be satisfied. One of the three Sakharov conditions is to assure a strong first-order phase transition (FOPT), that is absent within the minimal SM, but could be achieved by the Higgs portal to a sector beyond the SM. It is important to note that many well-motivated extensions of the SM predict gravitational wave (GW) signals through a strong FOPT, that are potentially detectable at LIGO and future 
LISA-like space-based GW detectors.


\begin{table}[tb]
	\resizebox{\columnwidth}{!}{%
		\begin{tabular}{|c |  c|  c| c| c|  }
			\hline
			Models  & Strong 1$^{st}$ order & GW signal & Cold DM & Dark Radiation and \\
			& phase transition &  & &  small scale structure  \\
			\hline	
			\hline
			\bf{SM charged}	& & &  &   \\
			\hline	
			\hline
			Triplet~\cite{Patel:2012pi,Niemi:2018asa,Bell:2020hnr} & \cmark  & \cmark & \cmark & \xmark   \\
			\hline
			complex and real  Triplet~\cite{Zhou:2020idp} & \cmark  & \cmark & \cmark & \xmark   \\
			(Georgi-Machacek model) & & & &  \\
			\hline	
			Multiplet~\cite{Huang:2017rzf} & \cmark & \cmark  & \cmark  &   \\
			\hline
			2HDM~\cite{Basler:2016obg,Dorsch:2017nza,Bernon:2017jgv,Gorda:2018hvi,Wang:2019pet,Su:2020pjw} & \cmark & \cmark &   & \xmark  \\
			\hline
			MLRSM~\cite{Brdar:2019fur} & \cmark & \cmark &  \xmark & \xmark  \\				
			\hline
			NMSSM~\cite{Huang:2014ifa,Bian:2017wfv,Athron:2019teq,Akula:2017yfr,Baum:2020vfl} & \cmark  & \cmark & \cmark  & \xmark  \\
			\hline
			\hline
			\bf{SM uncharged}	& & &  &   \\
			\hline	
			\hline	
			$S_r$ (xSM)~\cite{Barger:2007im,Profumo:2007wc,Profumo:2014opa,Huang:2015tdv,Kotwal:2016tex,Chen:2017qcz,Ellis:2018mja,Gould:2019qek,Alves:2019igs,Alves:2020bpi,Alves:2018jsw,Carena:2019une,Chiang:2020yym} & \cmark  & \cmark & \xmark & \xmark  \\
			\hline
			2 $S_r$'s~\cite{Chao:2017vrq}  & \cmark  & \cmark & \cmark & \xmark  \\
			\hline	
			$S_c$ (cxSM)~\cite{Gonderinger:2012rd,Chiang:2017nmu,Cheng:2018ajh,Chiang:2019oms,Chiang:2020yym}  & \cmark & \cmark & \cmark & \xmark \\
			\hline
			\hline
			$\text{U(1)}_{\text{D}}$ (no interaction with SM)~\cite{Bhoonah:2020oov} & \cmark & \cmark & \cmark & \xmark  \\
			\hline
			$\text{U(1)}_{\text{D}}$ (Higgs Portal)~\cite{Jaeckel:2016jlh} & \cmark & \cmark & \cmark &   \\
			\hline
			$\text{U(1)}_{\text{D}}$ (Kinetic Mixing)~\cite{Addazi:2017gpt} & \cmark & \cmark & \cmark &    \\
			\hline		
			Composite $\text{SU(7)}/\text{SU(6)}$~\cite{Chala:2016ykx} & \cmark & \cmark  & \cmark &    \\
			\hline			
			$\text{U(1)}_{\text{L}}$~\cite{Addazi:2020zcj} & \cmark &\cmark  & \cmark  & \xmark     \\ 
			\hline
			$\text{SU(2)}_{\text{D}} \rightarrow \text{global}\ SO(3)$ &  &  & \cmark & \xmark   \\
			by a doublet~\cite{Hambye:2008bq,Boehm:2014bia,Gross:2015cwa}  &  &  &  &    \\
			\hline
			$\text{SU(2)}_{\text{D}} \rightarrow \text{U(1)}_{\text{D}}$ &  &  & \cmark & \cmark   \\
			by a triplet~\cite{Baek:2013dwa,Khoze:2014woa,Daido:2019tbm}  &  &  &  &    \\
			\hline		
			$\text{SU(2)}_{\text{D}} \rightarrow Z_2$ &  &  & \cmark & \xmark   \\
			by two triplets~\cite{Ko:2020qlt}  &  &  &  &    \\
			\hline		
			$\text{SU(2)}_{\text{D}} \rightarrow Z_3$ &  &  & \cmark & \xmark   \\
			by a quadruplet~\cite{Chen:2015nea,Chen:2015dea}  &  &  &  &    \\
			\hline
			$\text{SU(2)}_{\text{D}} \times \text{U(1)}_{\text{B-L}} \rightarrow Z_2 \times Z_2$ &  &  & \cmark & \xmark   \\
			by a quintuplet and a $S_c$~\cite{Chiang:2013kqa}  &  &  &  &    \\
			\hline	
			$\text{SU(2)}_{\text{D}}$ with two dark Higgs doublets~\cite{Hall:2019ank} & \cmark & \cmark  & \xmark & \xmark   \\
			\hline	
			$\text{SU(3)}_{\text{D}} \rightarrow Z_2 \times Z_2$ by two triplets~\cite{,Gross:2015cwa, Arcadi:2016kmk}  &  &  & \cmark & \xmark \\    
			\hline	
			$\text{SU(3)}_{\text{D}}$ (dark QCD) (Higgs Portal)~\cite{Tsumura:2017knk,Aoki:2017aws} & \cmark & \cmark & \cmark &    \\
			\hline		  
			$G_{\text{SM}} \times G_{\text{D},\text{SM}} \times Z_2$~\cite{Addazi:2016fbj} & \cmark & \cmark & \cmark &   \\
			\hline       
			$G_{\text{SM}} \times G_{\text{D},\text{SM}} \times G_{\text{D},\text{SM}} \cdots$~\cite{Archer-Smith:2019gzq}  & \cmark & \cmark & \cmark &   \\
			\hline 
			\hline
			\bf{Current work}	& & &  &   \\
			\hline	
			\hline	
			$\text{SU(2)}_{\text{D}} \rightarrow \text{U(1)}_{\text{D}}$ (see the text) & \cmark & \cmark  &  \cmark & \cmark   \\
			\hline
			\hline
		\end{tabular}
	}%
	\caption{Theoretical models and their implications of EWPT, detectable GW signals, cold DM candidates, existence of DR and the small-scale structure. Models successful in fulfilling (not fulfilling) the desirable features are marked by checks (crosses).
	}
	\label{Table: Models}
\end{table}

Given the rich physics associated with a dark sector, there have been significant activities in the literature dealing with many different aspects of the theory and phenomenology. We collect a broad class of sample models in Table~\ref{Table: Models}. 
We classify them according to the particle charges under the SM gauge interactions, the dark gauge groups (D), 
and their state representations in the scalar sector, and the gauge symmetry breaking patterns. 
Those extensions with new states charged under the SM gauge group will yield substantial observable effects if kinematically accessible. Prominent examples include the extended Higgs sectors and supersymmetric theories. 
Those extensions uncharged under the SM gauge group will be characterized as dark sector. 
In the dark sector, both Abelian (U(1)$_{\rm D}$) and non-Abelian (SU(2)$_{\rm D}$, SU(3)$_{\rm D}$) gauge sectors have been studied with different symmetry breaking patterns induced by various scalar scenarios, as listed in Table~\ref{Table: Models}.
In particular, we examine their physical implications of FOPT, detectable signals at GW detectors, cold DM candidates, the existence of DR, and the small-scale structure.
Those models that successfully fulfill the desirable features are marked by checks, those that do not are characterized by crosses. 

Building upon the existing literature, in this paper, we will focus on a dark SU(2)$\rm_D$ model un-charged under the SM gauge group. Some early exploration and the phenomenology associated with the model have been examined \cite{Hambye:2008bq, Baek:2013dwa, Chiang:2013kqa,Khoze:2014woa,Boehm:2014bia, Chen:2015nea, Chen:2015dea, Gross:2015cwa,Daido:2019tbm,Ko:2020qlt,Hall:2019ank}. The previous works mainly focused on the DM studies. In this work, we will study the EWPT and GW with this well-motivated DM model.
In this class of models, it remains largely unconstrained on the choice of the dark scalar sector. 
With just one real scalar triplet, we could achieve a FOPT at the early Universe by transitioning from an electroweak symmetric vacuum that breaks the SU(2)$\rm_D$ symmetry to an electroweak broken vacuum that preserves the SU(2)$\rm_D$ symmetry~\cite{Espinosa:2011ax}. 
As such, all the dark sector particles would remain massless, and there would be no cold DM candidate in this simplest scenario. 
Alternatively, we would like to explore the following two cases to facilitate a strong FOPT 
in the early Universe and to have viable cold DM candidates 
\begin{itemize}
	\item[1.] one real scalar triplet and one real scalar singlet; 
	\item[2.] two real scalar triplets.
\end{itemize}
For both cases, at zero temperature, only one scalar triplet gets a nonzero vacuum expectation value (VEV) and partially breaks the SU(2)$\rm _D$ into U(1)$\rm _D$. The massless vector gauge boson associated with the unbroken $\text{U(1)}_{\text{D}}$ symmetry can serve as a dark radiation (DR). The other two massive gauge bosons associated with the symmetry breaking are our vector DM candidates. 
Due to the presence of the non-Abelian gauge boson couplings, the DM-DR and DM-DM interactions can be naturally introduced. 
The other scalar triplet or singlet can develop a non-zero VEV at a finite temperature and can thus trigger a strong FOPT, besides providing the scalar DM candidates. We also list the main features of our model in the last row of Table.~\ref{Table: Models}. 

The rest of the paper is organized as follows. In section~\ref{sec:form}, we introduce our model and particle spectrum, with the 
phenomenological constraints presented in section~\ref{sec:pheno} and DM phenomenology in section~\ref{sec:dm}. In section~\ref{sec:ewpt_GW}, we perform the study of EWPT and the GWs spectrum with two benchmark points (BMs) as shown in Table~\ref{Table: BMPs}. We summarize in section~\ref{sec:conclusion}.

\section{Theoretical framework}
\label{sec:form}
In addition to the SM, we include a non-Abelian SU(2)$\rm_D$ dark sector. We consider two scenarios for the dark scalar sector, a real singlet plus a real triplet (ST), or two real triplets (TT) under the dark gauge group 
SU(2)$\rm _D$: 
\beq
\Phi_1 =
\begin{cases}
	\text{ST} & \frac{1}{\sqrt{2}} (v_1 + \omega)\\
	\text{TT} & \frac{1}{\sqrt{2}}(\omega_1,\,\omega_2,\,v_1+\omega_3)^\text{T} \\
\end{cases},\quad  \Phi_2 = \frac{1}{\sqrt{2}}(\varphi_1,\,v_2 + \varphi_2,\,\varphi_3)^\text{T}.
\eeq
We assume that the dark sector does not carry SM charges but rather interacts with the SM particles 
through the Higgs portal interactions. Therefore, the Lagrangian of the model consists of three parts
\beqa
\lag &=& \lag_{\text{SM}} + \lag_{\text{portal}} + \lag_{\text{DS}},\\
-\lag_{\text{SM}}  &\supset& V_{\text{SM}} = m_H^2 |H|^2 + \frac{\lambda_H}{2} |H|^4,\\
-\lag_{\text{portal}}  &\supset& V_{\text{portal}} =  \lambda_{H11} |H|^2|\Phi_1|^2+\lambda_{H22} |H|^2|\Phi_2|^2,\label{eq:hportal}\\
\lag_{\text{DS}}  &=& -\frac{1}{4}\tilde{W}^a_{\mu\nu} \tilde{W}^{a\mu\nu} +|D_\mu\Phi_1|^2+|D_\mu\Phi_2|^2 - V_{\text{DS}}, \label{eq:ds}
\eeqa
where $\tilde{W}^a_{\mu\nu} = \d_\mu \tilde{W}^a_\nu - \d_\nu \tilde{W}^a_\mu + \tilde{g}f^{abc}\tilde{W}_\mu^b\tilde{W}_\nu^c$ is the 
dark gauge field strength tensor; $D_\mu = \d_\mu  - i\tilde{g}T^a\tilde{W}^a_\mu$ is the covariant derivative in the dark sector with $T^a$ being the SU(2)$\rm_D$ generators, which
is given in the 3-dimensional representation by 
\beq
T_1= \begin{pmatrix}
	0 & 0 & 0\\
	0 & 0 & -i\\
	0 & i & 0\\
\end{pmatrix},\,T_2=\begin{pmatrix}
	0 & 0 & i\\
	0 & 0 & 0\\
	-i & 0 & 0\\
\end{pmatrix},\,T_3= \begin{pmatrix}
	0 & -i & 0\\
	i & 0 & 0\\
	0 & 0 & 0\\
\end{pmatrix} ;
\eeq
and $H^{\text{T}} = (G^+, (v_h+h_0+ i G_0)/\sqrt{2})$, being the SM Higgs doublet.   
The most general renormalizable hidden sector potential with an assumed $Z_2$ symmetry is given by
\beq
V_{\text{DS}} = m_{11}^2|\Phi_1|^2 + m_{22}^2|\Phi_2|^2
+\frac{\lambda_1}{2}|\Phi_1|^4+\frac{\lambda_2}{2}|\Phi_2|^4+\lambda_{3}|\Phi_1|^2|\Phi_2|^2+\lambda_4|\Phi_1^\dagger\Phi_2|^2 ,
\eeq
where $\lambda_{4} = 0$ in the ST model.  
In principle, there can be cubic terms for the singlet scalar, which can change the phase transition dramatically. 
However, we will not consider breaking the $Z_2$ symmetry in this work.\footnote{
	In doing so, there could be the formation of domain walls during the phase transition when the field 
	acquires a non-zero VEV, which serves as another source for GW production when they annihilate (see, e.g.,~\cite{Saikawa:2017hiv}). 
	If they persist and still exist today, that might be problematic. These are interesting questions and needs a dedicated analysis of
	their formation, evolution and annihilation in a specified cosmological context, which however is beyond the scope of the current study and will be left to a future investigation.}

In our phenomenological analyses in the following sections, we choose $v_1 = 0$ at the zero temperature. An important consequence of this choice is to leave the dark U(1)$\rm _D$ unbroken so that there will be a massless dark gauge field, DR, which would have observational implications. 


\subsection{Mass spectrum}

With the choice of $v_1 = 0$, the SM Higgs boson mixes only with the SU(2)$\rm_D$ 
dark scalar $\varphi_2$. In the TT scenario, the mass terms for the scalar bosons are
\beq
-\lag^{\text{mass}}_{\text{scalar}}\supset\frac{1}{2}\mathbf{h}^T\mathbf{M_h} \mathbf{h} + \frac{1}{2} m_{\omega_2}^2\omega_2^2 + m_{\omega^\pm}^2\omega^+\omega^-,
\eeq
where $\mathbf{h}=\{h_0,\varphi_2\}$ are two neutral scalars with the mass matrix
\beq
\mathbf{M_h}=\begin{pmatrix}
	\lambda_{H} v_h^2 &  \lambda_{H22}v_2v_h\\
	\lambda_{H22}v_2v_h & \lambda_{2} v_2^2 
	\label{Eq:Ms} 
\end{pmatrix},
\eeq
and $m_{\omega^\pm}^2 = \frac{1}{2}(\lambda_3 v_2^2+2m_{11}^2+\lambda_{H11}v_h^2)$ is the mass of the SU(2)$\rm _D$ charged scalars.
The mass  for another neutral scalar $\omega_2$ is $m_{\omega_2}^2 = \frac{1}{2}((\lambda_3+\lambda_4) v_2^2+2m_{11}^2+\lambda_{H11}v_h^2)$. 
The scalar fields $\omega^\pm$ are defined as
\beq
\omega^{+} \equiv \frac{\omega_1 - i \omega_3}{\sqrt{2}},\quad \omega^{-} \equiv \frac{\omega_1 + i \omega_3}{\sqrt{2}}.
\eeq
In the ST scenario, there is only one massive scalar with mass 
\beq
m_{\omega}^2 = m_{\omega^{\pm}}^2.
\eeq
Please note that the sign $\pm$ refers to the dark SU(2)$\rm_D$ charge.
The neutral scalars $h_0$ and $\varphi_2$ are mixed. The mass eigenstates $\mathbf{h}^{\prime} =\{h_1,h_2\} $ can be obtained from a rotation on $\mathbf{h}$
\beq
\begin{pmatrix}
	h_1\\
	h_2\\
\end{pmatrix} = \mathcal{R}(\theta) \begin{pmatrix}
	h_0\\
	\varphi_2\\
\end{pmatrix}.
\eeq
The rotation matrix can be parametrized by one mixing angle $\theta$ as 
\beq
\mathcal{R}(\theta)=\begin{pmatrix}
	\cos\theta  & \sin\theta \\
	-\sin\theta  & \cos\theta
\end{pmatrix}.
\label{eq:hmix}
\eeq
The mass eigenvalues are
\beq
\mathcal{R} \mathbf{M_h} \mathcal{R}^T = \begin{pmatrix}
	m_{h_1}^2 & 0\\
	0 & m_{h_2}^2\\
\end{pmatrix}.
\eeq
Here and henceforth, we identify $h_1$ as the SM-like Higgs boson with $m_{h_1} = 125$ GeV, and $h_2$ is a heavier scalar  in the model.

The scalar fields $\varphi_1$ and $\varphi_3$ are the Nambu-Goldstone (NG) bosons absorbed by two of the $\text{SU(2)}_{\text{D}}$ gauge bosons $\tilde{W}_1$ and $\tilde{W}_3$. 
The mass terms of dark gauge bosons are contained in $(D_\mu\Phi_1)^2$ and $(D_\mu\Phi_2)^2$ in Eq.~(\ref{eq:ds})
\beq
-\lag^{\text{mass}}_{\text{vector}} \supset\frac{1}{2} m_{\tilde{W}}^2 \sum_{i=1,3} \tilde{W}_i^2 = m_{\tilde{W}^{\pm}}^2 \tilde{W}^{+}\tilde{W}^{-},
\eeq
where
\beq
\tilde{W}^{+} \equiv \frac{\tilde{W}_1 - i \tilde{W}_3}{\sqrt{2}},\, \tilde{W}^{-} \equiv \frac{\tilde{W}_1 + i \tilde{W}_3}{\sqrt{2}}, \,m_{\tilde{W}^{\pm}}  = \tilde{g} v_2 ,
\eeq
and $\tilde{W}_2$ remains massless.

\subsection{Interactions}
The interactions between the SM and the dark sector are generated through the Higgs portal as in Eq.~(\ref{eq:hportal}), specifically 
\beq
\begin{split}
	\lag^{\text{int}}_{\text{DS-SM}}&\supset  2 \tilde{g}^2v_2 (\sin\theta h_1 + \cos\theta h_2) \tilde{W}^+\tilde{W}^- + \tilde{g}^2 (\sin\theta h_1 + \cos\theta h_2)^2\tilde{W}^+\tilde{W}^-\\
	&-\sum_{i=1,2}(c_{i}h_i \omega^+\omega^- - d_{i}h_i \omega_2^2) - \sum_{\underset{i<j}{i,j=1,2}}(c_{ij}h_ih_j \omega^+\omega^- - d_{ij}h_ih_j \omega_2^2)
\end{split},
\label{eq:DS-SM}
\eeq
where the scalar couplings are given in terms of the mixing angle and the other model parameters
\beqa
c_1 &=&  \lambda_{3}v_2\sin\theta +\lambda_{H11}v_h\cos\theta,\quad
d_1 = \frac{1}{2}((\lambda_{3}+ \lambda_{4})v_2\sin\theta +\lambda_{H11}v_h\cos\theta),\label{Eq: c111}\\
c_{2} &=&  \lambda_{3}v_2\cos\theta-\lambda_{H11}v_h\sin\theta,\quad
d_{2}= \frac{1}{2}((\lambda_{3}+ \lambda_{4})v_2\cos\theta-\lambda_{H11}v_h\sin\theta),\\
c_{11} &=&  \frac{1}{2}(\lambda_{3}\sin^2\theta + \lambda_{H11}\cos^2\theta),\quad
d_{11} = \frac{1}{4}((\lambda_{3}+\lambda_{4})\sin^2\theta + \lambda_{H11}\cos^2\theta),\\ 
c_{12} &=& \frac{1}{2}(\lambda_{3} - \lambda_{H11})\sin2\theta,\quad
d_{12} = \frac{1}{4}(\lambda_{3} +\lambda_{4} - \lambda_{H11})\sin2\theta,\\ 
c_{22} &=& \frac{1}{2}(\lambda_{3}\cos^2\theta + \lambda_{H11}\sin^2\theta),\quad
d_{22} = \frac{1}{4}((\lambda_{3}+\lambda_{4})\cos^2\theta + \lambda_{H11}\sin^2\theta).
\eeqa
In the ST scenario, $c_i$, $c_{ij}$, and $\lambda_4$ are zero. The above interactions govern the phenomenology relevant for the potential experimental observations, such as the Higgs properties, 
the DM relic density and direct detections, and EWPT at the early Universe, as we will explore in the following sections. 

\section{Phenomenological constraints}
\label{sec:pheno}
%
The scalar potential of the model is 
\beq
V_{S}= \frac{m_H^2}{2}h_0^2 + \frac{\lambda_{H}}{8} h_0^4+\frac{m_{11}^2}{2}\omega_3^2 + \frac{\lambda_{1}}{8} \omega_3^4+\frac{m_{22}^2}{2}\varphi_2^2 + \frac{\lambda_{2}}{8} \varphi_2^4+\frac{\lambda_{H11}}{4}h_0^2\omega_3^2+\frac{\lambda_{H22}}{4}h_0^2\varphi_2^2+\frac{\lambda_{3}}{4}\omega_3^2\varphi_2^2.
\label{eq:pot}
\eeq
The two minima conditions $\frac{\d V_{S}}{\d h_0} = 0$ and $\frac{\d V_{S}}{\d \varphi_2} = 0$ evaluated at the VEVs are
\beqa 
v_h(2 m_H^2+ \lambda_{H}v_h^2 + \lambda_{H11}v_1^2+ \lambda_{H22}v_2^2)&=&0\,,
\label{Eq:MC1}
\\
v_2(2 m_{22}^2+\lambda_{H22}v_h^2 + \lambda_{3}v_1^2+ \lambda_{2}v_2^2)&=&0\,.
\label{Eq:MC2}
\eeqa
The mass parameters $m_H$ and $m_{22}$ can be solved by using these two minima conditions
\beqa
m_H^2 &=& -\frac{1}{2} (\lambda_{H}v_h^2+\lambda_{H11}v_1^2+\lambda_{H22}v_2^2),\\
m_{22}^2 &=& -\frac{1}{2} (\lambda_{H22}v_h^2+\lambda_{3}v_1^2+\lambda_{2}v_2^2).
\eeqa
In the TT model as described in the last section, there are fourteen parameters 
$$ \tilde{g},\,v_h,\,v_1,\,v_2,\,m_H^2,\,m_{11}^2,\,m_{22}^2,\,\lambda_{H},\,\lambda_{H11},\,\lambda_{H22},\,\lambda_{1},\,\lambda_{2},\,\lambda_{3},\,\lambda_{4}.$$
By applying the two extrema conditions in Eqs.~(\ref{Eq:MC1}) and~(\ref{Eq:MC2}) for the scalar potential and $v_1=0$, we can get rid of three parameters. Adopting the SM values $m_{h_1} = 125$~GeV, $v_h = 246$~GeV, we are left with nine independent parameters, which can be chosen as 
\beq
\sin\theta,\,\tilde{g},\, m_{\tilde{W}^+},\, m_{h_2},\, m_{\omega^+},\,m_{\omega_2},\,\lambda_1,\,\lambda_{H11},\,\lambda_3.
\label{eq:para}
\eeq
In the ST model, we have one less free parameter as $m_{\omega^+}$ and $m_{\omega_2}$ are replaced by one parameter $m_{\omega}$.

We wish to have observable imprints from the dark sector in the current and future experiments. We thus take the $\text{SU(2)}_{\text{D}}$ symmetry breaking not too far from the electroweak scale in the SM, and 
vary the mass of the second Higgs boson $m_{h_2}$ in the range of 200 GeV$-$1 TeV. We will not consider $m_{h_2} > 1$ TeV, as the perturbative GW calculations are not reliable.
We examine the possible bounds on the other model parameters from the existing experiments in the following sessions. 
For the purpose of illustration, we choose two benchmark points (BMs) for the input parameters as shown in Table~\ref{Table: BMPs}. Both BM1 and BM2 possess the desirable features as listed in the last row of Table~\ref{Table: Models}. Some other calculated physical quantities are also summarized in the table.

\begingroup
\setlength{\tabcolsep}{10pt} 
\renewcommand{\arraystretch}{1.} 
\begin{table}
	\centering
	\begin{tabular}{|c |  c|  c|}
		\toprule
		Parameters & BM1 & BM2 \\
		\midrule
		\midrule
		$\sin\theta$ & $-0.25$ &  $-0.12$  \\
		\midrule
		$\tilde{g}$   & 0.094 & 0.133  \\
		\midrule
		$m_{\tilde{W}^{\pm}}$   & 94 GeV & 133 GeV  \\
		\midrule
		$m_{h_2}$   & 200 GeV & 290 GeV   \\
		\midrule
		$m_{\omega^{\pm}}$ & 1.2 TeV & 1.3 TeV   \\
		\midrule
		$m_{\omega_2}$ & 2.0 TeV & 1.9 TeV  \\
		\midrule
		$\lambda_1$    & $3.5$ & $3.5$  \\
		\midrule
		$\lambda_{H11}$    & $2.0$ & $2.0$  \\
		\midrule
		$\lambda_3$    & $3.0$ &  $3.5$ \\
		\midrule
		\midrule
		$\lambda_H$  & $0.28 $ & $0.27$  \\
		\midrule
		$\lambda_2$    & $3.8 \times 10^{-2}$ & $8.3 \times 10^{-2}$ \\
		\midrule
		$\lambda_{H22}$    & $2.4 \times 10^{-2}$ & $3.2 \times 10^{-2}$ \\
		\midrule
		$\lambda_4$    & $5.0$ & $4.0$  \\
		\midrule
		$v_2$ & 1 TeV & 1 TeV \\
		\midrule
		\midrule
		$\Omega_{\tilde{W}^{\pm}} h^2$ & 0.096 &0.12 \\
		\midrule
		$\sigma_{\text{SI}}~(\text{cm}^2)$  & $7.8\times 10^{-47}$ & $8.0\times 10^{-47}$ \\
		\midrule
		\midrule
		$T_c~ (\rm{GeV})$   & 177 & 252  \\
		\midrule
		$T_n~ (\rm{GeV})$   & 147 & 234  \\
		\midrule
		$\beta/H_n$   & 297 & 760  \\
		\midrule
		$\alpha$   & 0.32 & $5.1 \times 10^{-2}$ \\
		\midrule
		phase transition pattern   & 2-step~(\ref{eq:2step}) & 3-step~(\ref{eq:3step})  \\		
		\toprule
	\end{tabular}
	\caption{Model parameters and calculated physical quantities with two benchmark points, BM1 and BM2. The independent model parameters in Eq.~(\ref{eq:para}) are listed in the upper part of the table. }
	\label{Table: BMPs}
\end{table}
\endgroup
\subsection{Vacuum stability}
A stable physical vacuum has to be bounded from below keeping the scalar fields from running away. The behavior of the scalar potential is dominant by the quartic part when the field strength approaches infinity. The conditions of vacuum stability are given in Ref.~\cite{Arhrib:2011uy,Poulin:2018kap}. Following their procedure, we find the following conditions
\beqa
& \lambda_{H}>0,\quad \lambda_{1}>0,\quad \lambda_{2}>0,\\
&\lambda_{3} + \lambda_{4}  + \sqrt{\lambda_{1}\lambda_{2}}>0,\\
&\lambda_{H11} + \sqrt{\lambda_{H}\lambda_{1}}>0,\quad
\lambda_{H22} + \sqrt{\lambda_{H}\lambda_{2}}>0.
\eeqa

\subsection{Partial wave unitarity}
The scattering amplitudes for spin-less $2\to 2$ processes can be decomposed into a sum over the partial waves $a_j$ as
\beq
{\cal A}(\alpha) = 16\pi \sum_{j=0}^{\infty}a_j(2j+1)P_j(\cos\alpha),
\eeq
where $P_j(\cos\alpha)$ are the Legendre polynomials in terms of the scattering angle $\alpha$. The perturbative unitarity requires $\text{Im}(a_j) = |a_j|^2$, which implies 
\beq
|a_j|\leq 1,\quad |\text{Re}(a_j)|\leq \frac{1}{2}.
\eeq  
We will adopt the second condition as it turns out to be more constraint. The $s$-wave amplitude can be computed by
\beq
a_0 = \frac{1}{32\pi}\int_{-1}^{1}{\cal A}(\alpha)d\cos\alpha, \quad a_j = 0 \,(j>0).
\eeq
For a spin-less $2\to 2$ elastic scattering process, the unitarity bound can be rephrased as  
\beq
|{\cal A}| < 8\pi.
\eeq
Owing to the Goldstone-boson equivalence theorem, the scattering of the longitudinal gauge bosons can be approximated by the pseudo-Goldstone boson scattering in the high-energy limit. Given the fact that the high energy scattering is dominated by the four-scalar contact interactions, we only need to evaluate the quartic or bi-quadratic terms. There are ten scalar fields in the TT scenario, namely, $\omega_{i}\,(i=\text{1 to 3}), \varphi_{j}\,(j=\text{1 to 3})$, $G_{k}\,(k=\text{0 to 2})$, and $h_0$. So there are 55 pair combinations and 1540 scattering channels. An additional symmetric factor $1/\sqrt{2}$ needs to be included for each pair of identical particles in the initial or final states. 
The unitarity bounds from scattering amplitude matrix $\mathbf{A_{55 \times 55}}$ are
\beqa
|\lambda_{H}| &<& 8 \pi \label{Eq:Un1},\quad
|\lambda_{H11}| < 8 \pi,\quad
|\lambda_{H22}| < 8 \pi,\nonumber\\
|\lambda_{3}-\frac{1}{2}\lambda_{4}| &<& 8 \pi,\quad
|\lambda_{3}+\frac{1}{2}\lambda_{4}| < 8 \pi,\quad
|\lambda_{3}+2\lambda_{4}| < 8 \pi,\\
|\lambda_{1}+\lambda_{2}-\sqrt{(\lambda_{1}-\lambda_{2})^2+\lambda_{4}^2}| &<& 16 \pi,\quad
|\lambda_{1}+\lambda_{2}+\sqrt{(\lambda_{1}-\lambda_{2})^2+\lambda_{4}^2}| < 16 \pi,\nonumber\\
& &|\text{Eigenvalues}[\mathcal{P}] |< 8 \pi \nonumber,
\eeqa
where 
\beq
\mathcal{P} =\frac{1}{2} \begin{pmatrix}
	5\lambda_{1} & 3\lambda_{3}+\lambda_{4} & 2\sqrt{3}\lambda_{H11}\\
	3\lambda_{3}+\lambda_{4}& 5\lambda_{2} & 2\sqrt{3}\lambda_{H22}\\
	2\sqrt{3}\lambda_{H11}& 2\sqrt{3}\lambda_{H22} &6\lambda_{H}\\
\end{pmatrix}.
\eeq
Similarly, for the ST case, there are a total of eight scalar fields and therefore 36 pair combinations. The unitarity bounds from scattering amplitude matrix $\mathbf{A_{36 \times 36}}$ are
\beqa
|\lambda_{H}| &<& 8 \pi,\quad  |\lambda_{H11}| < 8 \pi,\quad
|\lambda_{H22}| < 8 \pi,\\
|\lambda_{2}| &<& 8 \pi,\quad  |\lambda_{3}| < 8 \pi,\quad |\text{Eigenvalues}[\mathcal{P}^{\prime}] |< 8 \pi,
\eeqa
where
\beq
\mathcal{P}^{\prime} =\frac{1}{2} \begin{pmatrix}
	3\lambda_{1} & 3\lambda_{3} & 2\lambda_{H11}\\
	\lambda_{3}& 5\lambda_{2} & 2\sqrt{3}\lambda_{H22}\\
	2\lambda_{H11}& 2\sqrt{3}\lambda_{H22} & 6\lambda_{H}\\
\end{pmatrix}.
\eeq

\subsection{Electroweak precision observables}
\begin{figure}
	\centering
	\includegraphics[width=0.7\columnwidth]{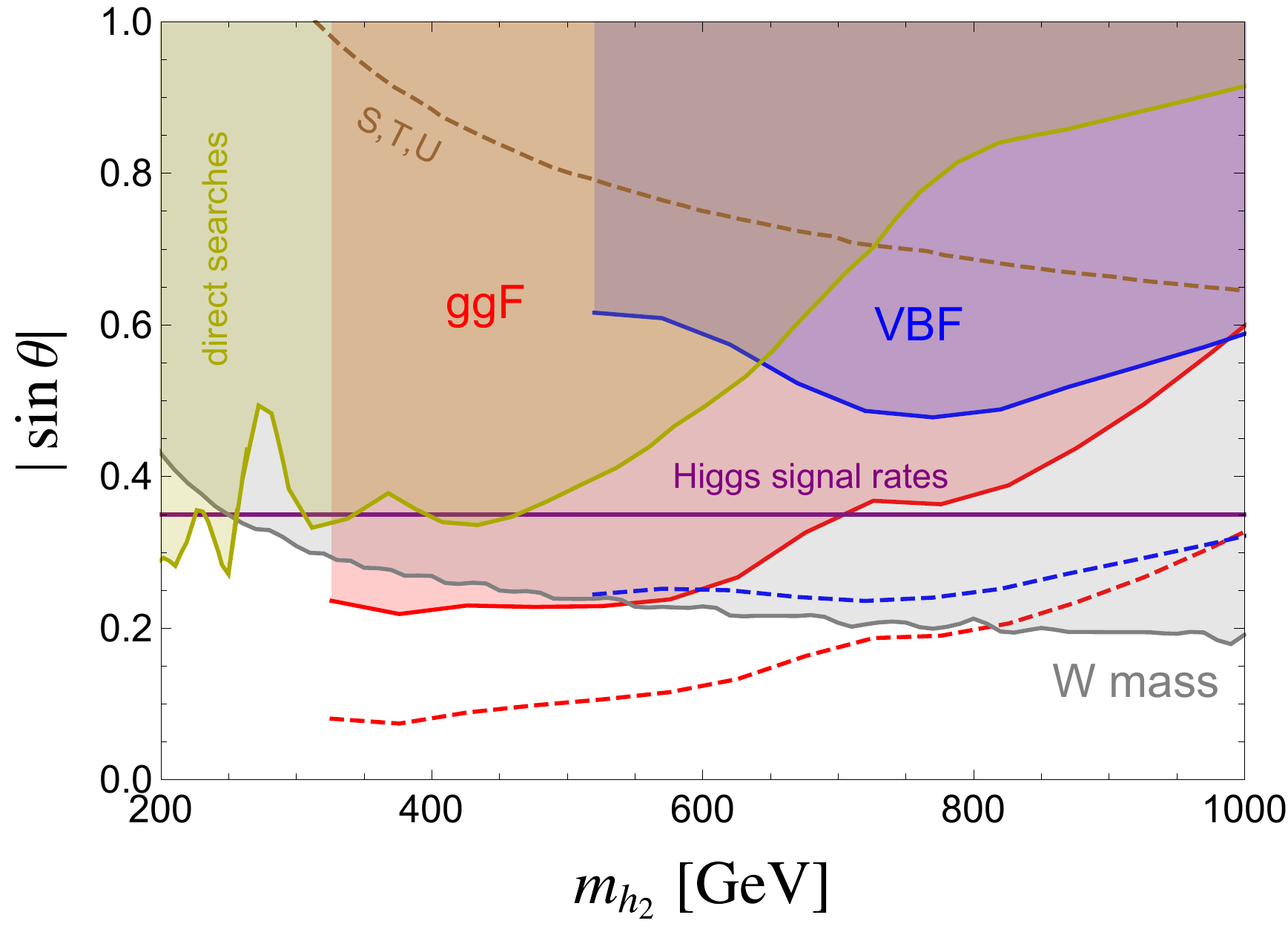}
	\caption{Upper bounds on the mixing angle $|\sin\theta|$ versus the heavy Higgs mass $m_{h_2}$. The horizontal purple line is from the Higgs signal rate measurement~\cite{Aad:2015pla}. The yellow shaded region shows the upper bound from the direct searches for the heavy Higgs at LEP and LHC ($\sqrt{s} = 7$ TeV)~\cite{Falkowski:2015iwa}. The blue (red) shaded regions are excluded by the LHC di-boson searches with VBF (ggF) channels. The blue and red dashed lines correspond to the HL-LHC projection for these two channels, respectively \cite{Aaboud:2018bun}. The grey shaded area labelled by $W$ mass, and the area above the brown dashed line labelled by $S,T,U$ are excluded by the electroweak precision observables \cite{Lopez-Val:2014jva}.}
	\label{fig:HVV}
\end{figure}

Quantum corrections to the $W$ boson mass~\cite{Lopez-Val:2014jva} and the electroweak oblique parameters~\cite{Bian:2014cja}, from the mixing between SM Higgs and the dark massive eigenstates, can put constraints on the model parameters $\sin\theta$ and $m_{h_2}$. The bound from $W$ boson mass constraint, which is shown by the gray shaded region in Fig.~\ref{fig:HVV}, turns out to be more stringent than that from the oblique parameters~\cite{Lopez-Val:2014jva, Robens:2015gla}. The bound from oblique parameters are shown by the dashed brown line in Fig.~\ref{fig:HVV} for comparison. 

\subsection{Higgs phenomenology}
\label{sec:higgs}
The scalar state $h_0$ mixes with $\varphi_2$ after the electroweak symmetry breaking. We identify that the lighter  mass eigenstate $h_1$ is the observed SM-like Higgs boson with a mass of 125 GeV. The couplings of the physical scalars $h_1$ and $h_2$ to the SM particles are
\beq
\lag \supset \frac{h_1\cos\theta-h_2\sin\theta}{v_h} (2 m_W^2 W^+_\mu W^{\mu-}+m_Z^2 Z_\mu Z^\mu - \sum_{f} m_f \bar{f}f ).
\label{eq:h2}
\eeq
The SM-like Higgs boson coupling to the SM particles are modified by a universal factor $\cos\theta$. 
The relevant Higgs self-interactions in the scalar sector are
\beqa
\label{eq:h3}
&& \lag \supset -\kappa_{111} h_1^3- \kappa_{112} h_1^2h_2 - \kappa_{122} h_1h_2^2- \kappa_{222} h_2^3,\\
\nonumber
&&\kappa_{111} =\frac{m^2_{h_1} (v_2 \cos^3\theta + v_h \sin^3\theta) }{2 v_2 v_h},\ \ 
\kappa_{112} = -\frac{\sin 2\theta  (2 m^2_{h_1} + m^2_{h_2}) (v_2 \cos\theta - v_h \sin\theta) }{4 v_2 v_h} , \\ 
\nonumber
&&\kappa_{122} = \frac{\sin 2\theta  (m^2_{h_1} + 2 m^2_{h_2}) (v_2 \sin\theta + v_h \cos\theta) }{4 v_2 v_h} ,\ \
\kappa_{222} = \frac{m^2_{h_2} (v_h \cos^3\theta - v_2 \sin^3\theta) }{2 v_2 v_h},
\eeqa
where $v_2 = m_{\tilde{W}^+}/\tilde{g}$. 
These couplings are important for the DM annihilation at the early Universe through the Higgs portal. 
The Higgs phenomenology at colliders is similar to that of one real singlet scalar extension of the SM, which has been extensively studied (see~\cite{Kotwal:2016tex,Huang:2017jws,Alves:2019igs,Adhikari:2020vqo} and references therein). The most relevant parameters are the mixing angle $\theta$ and the mass of the second Higgs $m_{h_2}$ as shown in Eq.~(\ref{eq:h2}). The current bounds on $\sin\theta$ and $m_{h_2}$ from the Higgs phenomenology are shown in Fig.~\ref{fig:HVV}. We will discuss the details of each bound in the following subsections.

\subsubsection{Higgs invisible decay}
\begin{figure}[tb]
	\centering
	\begin{subfigure}{.24\textwidth}
		\centering
		\includegraphics[width=\textwidth]{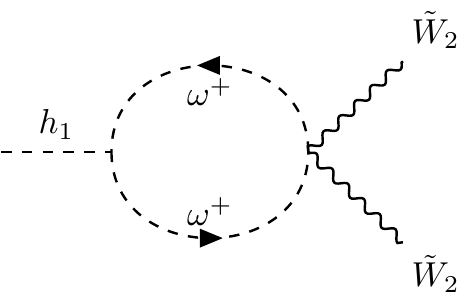}
	\end{subfigure}
	\begin{subfigure}{.24\textwidth}
		\centering
		\includegraphics[width=\textwidth]{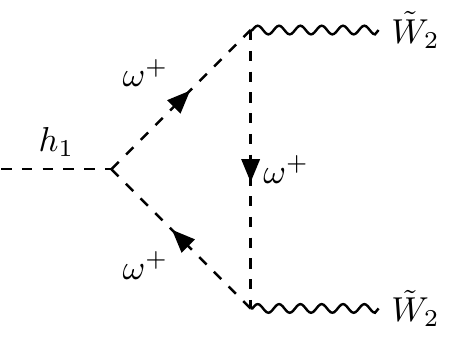}
	\end{subfigure}
	\begin{subfigure}{.24\textwidth}
		\centering
		\includegraphics[width=\textwidth]{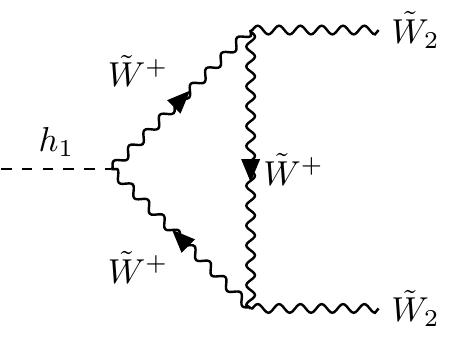}
	\end{subfigure}
	\begin{subfigure}{.24\textwidth}
		\centering
		\includegraphics[width=\textwidth]{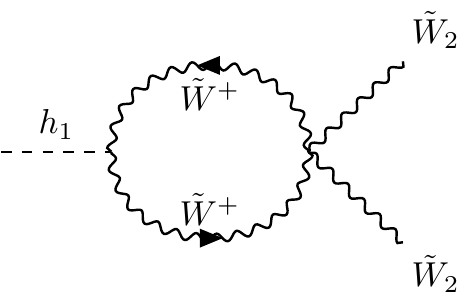}
	\end{subfigure}
	\caption{Feynman diagrams for the Higgs invisible decay to the dark radiation.}
	\label{fig:Hinv}
\end{figure}

In the case that DM masses are larger than the half of the Higgs boson mass, the invisible decay of the Higgs boson is to the DR $\tilde{W}_2$ through the $\text{SU(2)}_{\text{D}}$ charged scalar and gauge bosons loops as shown in Fig.~\ref{fig:Hinv}. The decay width through dark gauge bosons can be calculated as
\beq
\Gamma_{\tilde{W}}(h_1\rightarrow \tilde{W}_2\tilde{W}_2) = \frac{\tilde{\alpha}^3\sin^2\theta m_{h_1}^3}{64\pi^2m_{\tilde{W}^+}^2} (2+3\tau ^{-1}+ 3\tau^{-1} (2 -\tau^{-1}) f(\tau) )^2,
\eeq
where
\begin{equation}
\tilde{\alpha}=\frac{\tilde{g}^2}{4\pi},\quad\tau =  \frac{m_{h_1}^2}{4 m_{\tilde{W}^+}^2},\quad 
{\rm and} \quad
f(\tau)=
\begin{cases}
\arcsin^{-1} (\sqrt{\tau})& \text{for $\tau \leq 1$,}\\
-\frac{1}{4}[\text{ln}\frac{1+\sqrt{1-\tau^{-1}}}{1-\sqrt{1-\tau^{-1}}}-i\pi]^2 & \text{for $\tau > 1$.}\\
\end{cases}       
\end{equation}
In the limit $m_{h_1} \ll m_{\omega^+}$, the decay width through dark scalars can be calculated as
\beq
\Gamma_\omega(h_1\rightarrow \tilde{W}_2\tilde{W}_2) = \frac{5 \tilde{\alpha}^2 c_{1}^2}{\pi^3 m_{h_1}}(\frac{m_{h_1}}{8\sqrt{3}m_{\omega^+}})^4,
\eeq
where $c_{1}$ is the coupling of vertex $h_1\omega^+\omega^-$ given in Eq.~(\ref{Eq: c111}). The Higgs invisible decay width for our benchmark points shown in Table~\ref{Table: BMPs} are
\beqa
&\text{BM1: }& \Gamma(h_1\rightarrow \tilde{W}_2\tilde{W}_2) = 3.1\times10^{-7} ~ \text{MeV},\\
&\text{BM2: }& \Gamma(h_1\rightarrow \tilde{W}_2\tilde{W}_2) = 2.1\times10^{-5} ~ \text{MeV},
\eeqa
which are dominated by the last two diagrams in Fig.~\ref{fig:Hinv}.
The Higgs invisible decay is highly suppressed by the small mixing angle,  dark-sector gauge coupling, and the one-loop suppression. The branching fractions of the invisible Higgs decay are far beyond the reach of current and future experiments. 

\subsubsection{Higgs coupling measurements}
Higgs couplings with SM particles have been measured with good precisions at the LHC. The Higgs signal strength is defined as~\cite{Baek:2011aa}
\beq
\mu_{h_1} \equiv \frac{\sigma_{h_1}~\text{BR}(h_1\rightarrow \text{SM})}{\sigma_{h_1}^{\text{SM}}~\text{BR}^{\text{SM}}(h_1\rightarrow \text{SM})},
\eeq
where $\sigma_{h_1} = \cos^2\theta \sigma_{h_1}^{\text{SM}}$, $\text{BR}(h_1\rightarrow \text{SM}) = \frac{ \Gamma_{h_1}^{\text{SM}}\cos^2\theta}{ \Gamma_{h_1}^{\text{SM}}\cos^2\theta +\Gamma_{h_1}^{\text{DS}} }$, and by definition $\text{BR}^{\text{SM}}(h_1\rightarrow \text{SM}) \equiv 1$. Therefore, the signal strength can be written as
\beq
\mu_{h_1} = \frac{\Gamma_{h_1}^{\text{SM}}\cos^4\theta}{\Gamma_{h_1}^{\text{SM}}\cos^2\theta+\Gamma_{h_1}^{\text{DS}}}.
\eeq
As we learned from the previous section, $\Gamma_{h_1}^{\text{DS}}$ are highly suppressed, as the SM-like Higgs $h_1$ can only decay to DR through one-loop diagrams in Fig.~\ref{fig:Hinv}.
The signal strength simply scales as $\cos^2\theta$. The bound on the mixing angle, from the Higgs couplings measurement by ATLAS~\cite{Aad:2015pla}, is $|\sin\theta |\lesssim 0.35$, which is shown by the purple line in Fig.~\ref{fig:HVV}.

\begin{figure}[tb]
	\centering
	\includegraphics[width=.6\textwidth]{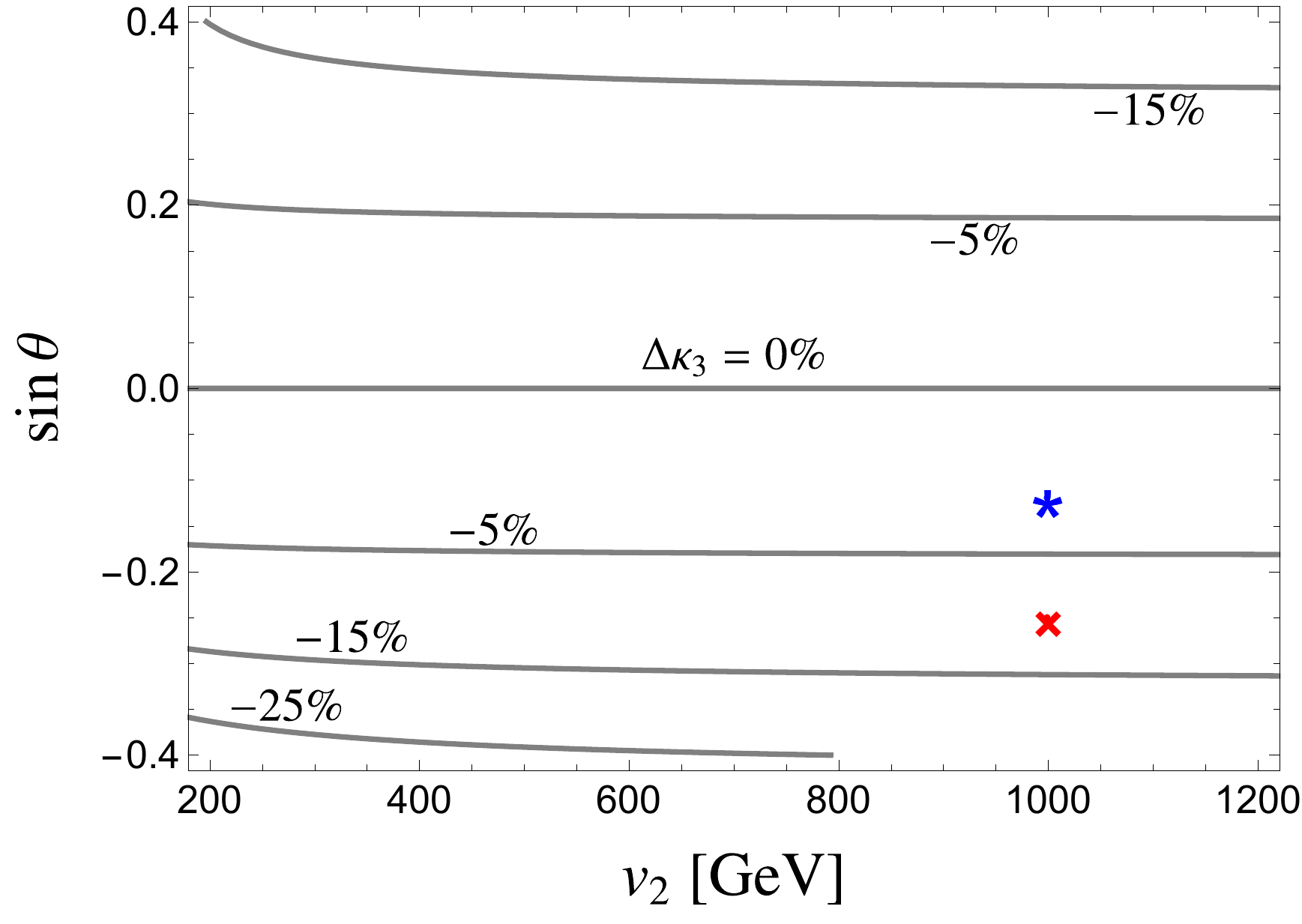}
	\caption{Predicted deviation of $\Delta \kappa_{3}$ in the $v_2$-$\sin\theta$ plane as defined in Eq.~(\ref{eq:deltaK}). The red-cross and blue-star indicate the predictions for our BM1 and BM2 points, respectively.}
	\label{fig:k3}
\end{figure}

Of special interest is the SM-like Higgs triple coupling $\kappa_{111}$ as in Eq.~(\ref{eq:h3}) because of its sensitivity to the BSM new physics and its crucial role in EWPT. We write the derivation from the SM prediction as
\beq
\Delta \kappa_{3} = \frac{\kappa_{111}- \kappa^{\text{SM}}_{111}}{\kappa^{\text{SM}}_{111}}
= -1 + \cos^3\theta + {v_h\over v_2} \sin^3\theta .
\label{eq:deltaK}
\eeq
We depict the resultant deviation of $\Delta \kappa_{3}$ in the $v_2$-$\sin\theta$ plane in  Fig.~\ref{fig:k3} by the gray solid lines. For most of the viable parameter space, the magnitude of $\Delta\kappa_{3}$ is less than 
$25\%$. We also mark the predictions of our benchmark points BM1 for about $-10\%$ by the red-cross and 
MB2 for about $-2\%$ by the blue-star, respectively. 
The achievable sensitivity to probe $\Delta\kappa_{3}$ in the future collider experiments has been extensively studied. While the HL-LHC will only have a moderate sensitivity to $\kappa_3$ \cite{ATLAS:2017muo, CMS:2017cwx}, future improvements are highly anticipated, reaching a $1\sigma$ sensitivity of
$13\%$ at a 1-TeV ILC \cite{Tian:2013yda} and $10\%$ at CLIC \cite{Roloff:2019crr}, 
and $2\sigma$ sensitivity of 
$5\%$ at FCC$_{hh}$/SPPC \cite{Benedikt:2018csr}, 
$2\%$ at a multi-TeV muon collider \cite{Han:2020pif}.
The precision measurement for $\kappa_3$ would provide important indirect test of the model as well as BSM theories in general.  

\subsubsection{Direct searches for the heavy Higgs boson} 

\begin{figure}
	\centering
	\includegraphics[width=0.7\columnwidth]{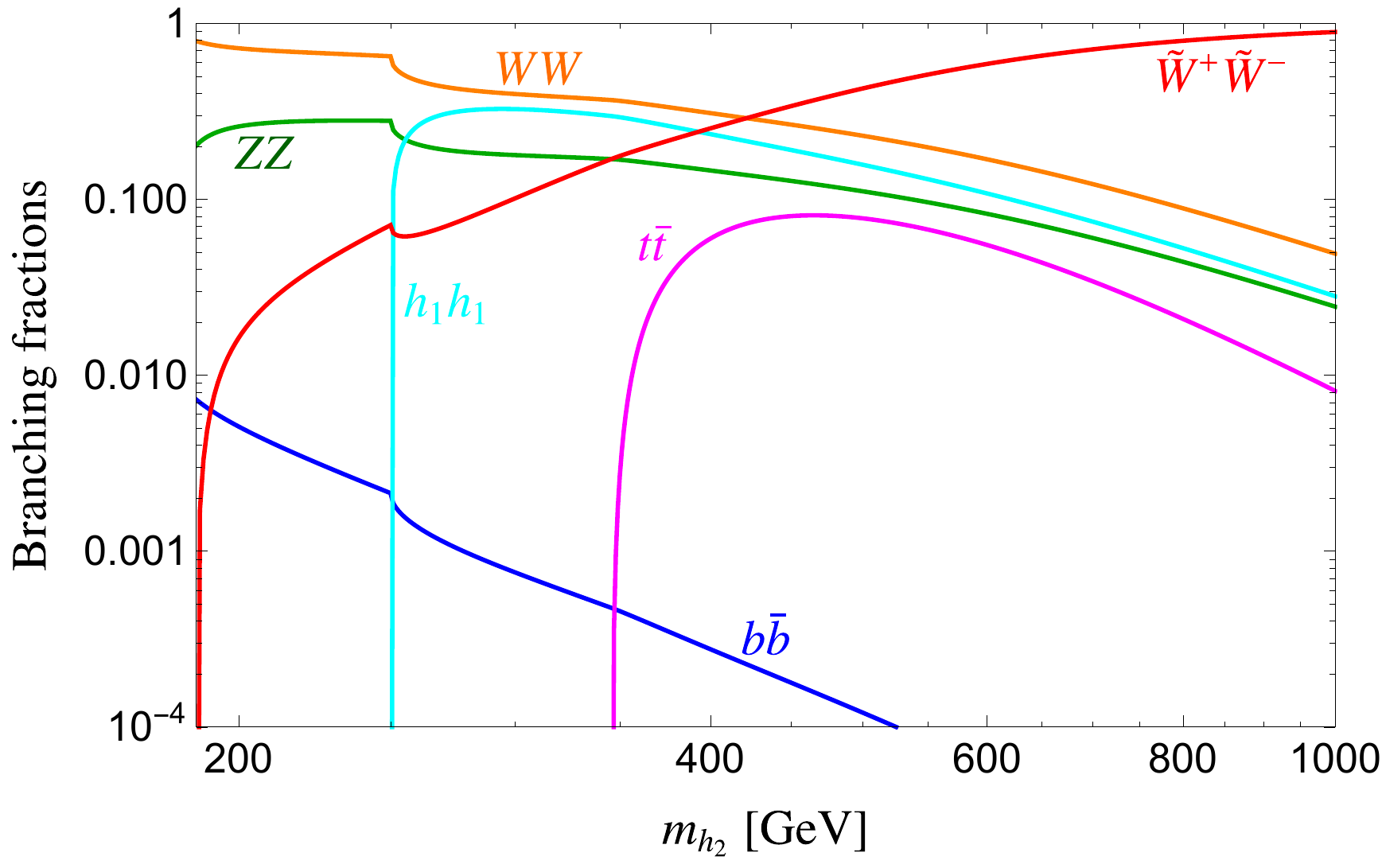}
	\caption{Branching fractions of heavy Higgs $h_2$ decay versus $m_{h_2}$. The other  parameters are fixed as BM1 in Table~\ref{Table: BMPs}.}
	\label{fig:Hdecay}
\end{figure}

The heavy Higgs boson in the model, $h_2$, can interact with the SM particles via the mixing
as shown in Eq.~(\ref{eq:hmix}). The coupling strength is proportional to $\sin\theta$. The heavy Higgs searches at the high-energy colliders can put strong constraints in this scenario.
Heavy Higgs $h_2$ mainly decay to heavy particles when they are kinematically allowed, such as $b\bar b$,  
top quarks, massive gauge bosons, and the dark gauge bosons. The branching fractions of the heavy Higgs decay versus $m_{h_2}$ are shown in Fig.~\ref{fig:Hdecay}, where the other parameters are fixed as BM1 in Table~\ref{Table: BMPs} for illustration. 
The heavy Higgs decay channels are to di-bosons $WW + ZZ$ until the threshold for $\tilde{W}^+\tilde{W}^-$ is open, as shown in Fig.~\ref{fig:Hdecay}. 

The LHC di-boson resonance search in gluon-gluon fusion (ggF) and vector boson fusion (VBF)~\cite{Aaboud:2018bun} can put strong bounds on the mixing angle $\theta$ and the heavy Higgs mass $m_{h_2}$. We evaluate the resonance production rate as
\beq
\sigma(pp\rightarrow VV) = \sigma(pp\rightarrow h_2)~\text{BR}(h_2\rightarrow VV).
\eeq
The bounds on the plane in $m_{h_2}$-$\sin\theta$ with $v_2 = 1000$~GeV are shown by the red (ggF) and blue (VBF) shaded regions in Fig.~\ref{fig:HVV}. 
The dashed lines with the same color scheme are the projected limit from HL-LHC with 3 $\text{ab}^{-1}$ integrated luminosity, obtained by rescaling the current bounds by the square root of luminosity ratio $\sqrt{3000/36.1}$. 
For the mass below 350 GeV, we adopted the bounds provided in Ref.~\cite{Falkowski:2015iwa} from a combination of various decay channels at LEP and LHC with $\sqrt{s} = 7$~TeV. The bounds are shown by the yellow shaded region.

\section{Dark radiation and dark matter phenomenology}
\label{sec:dm}

The dark sector in our model possesses rich phenomenology. 
There are two self-interacting vector dark matter candidates $\tilde{W}^{\pm}$. The massless state $\tilde{W}_2$ is the DR. In addition, there is one scalar DM $\omega$ in the ST scenario, or there are three self-interacting scalar DM candidates $\omega^\pm$ and $\omega_2$ in the TT scenario. 
The DM interactions with SM particles are through the mixing between $\varphi_2$ and $h_0$. The relevant Lagrangian of DM-SM interactions are shown in Eq.~(\ref{eq:DS-SM}). Due to the non-Abelian nature of the dark sector, there exist nontrivial self-interactions inside the dark sector among scalar DM, vector DM, and DR, which are from the dark gauge couplings and the scalar potential.
For simplicity we decoupled the scalar DM by assuming the mass hierarchy to be 
$m_{\tilde{W}^+} \ll m_\omega$ in ST, or $m_{\tilde{W}^+} \ll m_{\omega^+} \lesssim m_{\omega_2}$ in TT.

\subsection{Dark radiation}
The massless DR $\tilde{W}_2$ associated with the unbroken $\text{U(1)}_{\text{D}}$ can contribute to the energy density of the Universe, regulating the Universe expansion rate.
In the radiation-dominated era, the expansion rate of the Universe depends on the relativistic energy density
\beq
\rho = g_*(T) \frac{\pi^2}{30} T^4,
\eeq   
where $g_*$ is the total relativistic degrees of freedom defined as
\beq
g_*(T) \equiv \sum_{m_i<T}C_i g_i (\frac{T_i}{T})^4,
\label{eq:dofe}
\eeq
where the coefficients are $C_{i}$=1 (7/8) for bosons (fermions), and $g_i$ is the internal degrees of freedom for particle $i$.  
$\tilde{W}_2$ 
can contributes to $g_*$ and it is conventional to define this extra energy density by
\beq
\Delta N_{\text{eff}}\equiv\frac{\rho_{\tilde{W}_2}}{\rho_\nu}= \frac{8}{7} (\frac{\tilde{T}}{T_\nu})^4,
\label{eq:dneff}
\eeq
where $\tilde{T}$ is the dark sector's temperature, $T_\nu$ is the SM neutrinos' temperature. After neutrinos  decouple from the thermal bath, the ratio $\tilde{T}/T_\nu$ is fixed as they evolve in the same way. We thus evaluate this temperature ratio at the epoch of neutrino decoupling. Before the DM decouples, the dark sector and visible sector are in thermal equilibrium, $\tilde{T}_{\text{dec},\chi}$ =  $T_{\text{dec},\chi}$. After decoupling of DM, the dark sector and visible sector lost thermal contact, the entropy is conserved in each sector separately. So we have~\cite{Feng:2008mu}
\beq
\frac{g^{\text{DS}}_{*s} (\tilde{T}_{\text{dec},\nu}) \tilde{T}^3_{\text{dec},\nu}}{g^{\text{DS}}_{*s} (\tilde{T}_{\text{dec},\chi}) \tilde{T}^3_{\text{dec},\chi}} = \frac{g^{\text{SM}}_{*s} (T_{\text{dec},\nu}) T^3_{\text{dec},\nu}}{g^{\text{SM}}_{*s} (T_{\text{dec},\chi}) T^3_{\text{dec},\chi}},
\label{eq:entropy}
\eeq
where $g_{*s}$ is the relativistic degrees of freedom for entropy
\beq
g_{*s}(T) \equiv \sum_{m_i<T}C_i g_i (\frac{T_i}{T})^3.
\label{eq:dofs}
\eeq
At the DM decoupling, $T_{\text{dec},\chi} \ll m_{\chi}$. The only relativistic particle is the DR. So that $g^{\text{DS}}_{*s}(\tilde{T}_{\text{dec},\chi}) = g^{\text{DS}}_{*s}(\tilde{T}_{\text{dec},\nu}) = 2$. In the visible sector, $g^{\text{SM}}_{*s}(T_{\text{dec},\chi}) = 106.75$, $g^{\text{SM}}_{*s}(T_{\text{dec},\nu}) = 10.75$. Combining Eqs.~(\ref{eq:dneff}) and (\ref{eq:entropy}), $\Delta N_{\text{eff}}$ can be evaluated as
\beq
\Delta N_{\text{eff}} = \frac{8}{7}(\frac{10.75}{106.75})^{4/3} \approx 0.054.
\eeq
Currently, the strongest bounds on $N_{\text{eff}}$ come from the Planck satellite~\cite{Akrami:2018vks, Aghanim:2018eyx} which measured $N_{\text{eff}} = 2.99 \pm 0.17$ including baryon acoustic oscillation data. The projected limit of CMB Stage IV experiments is $\Delta N_{\text{eff}} = 0.03$~\cite{Abazajian:2019eic}, which has sufficient sensitivity to explore this scenario.

\begin{figure}[tb]
	\centering
	\begin{subfigure}{.24\textwidth}
		\centering
		\includegraphics[width=\textwidth]{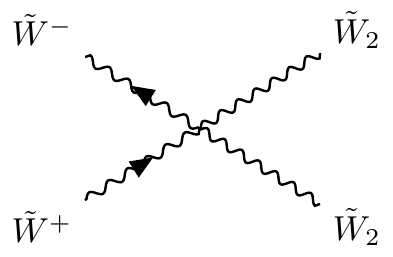}
	\end{subfigure}
	\begin{subfigure}{.24\textwidth}
		\centering
		\includegraphics[width=\textwidth]{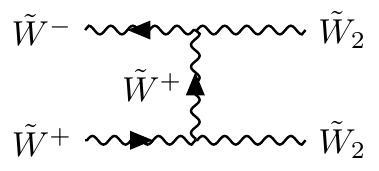}
	\end{subfigure}
	\begin{subfigure}{.24\textwidth}
		\centering
		\includegraphics[width=\textwidth]{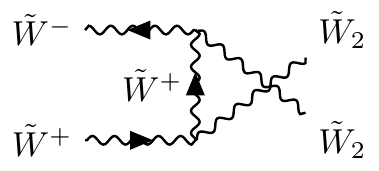}
	\end{subfigure}
	\begin{subfigure}{.24\textwidth}
		\centering
		\includegraphics[width=\textwidth]{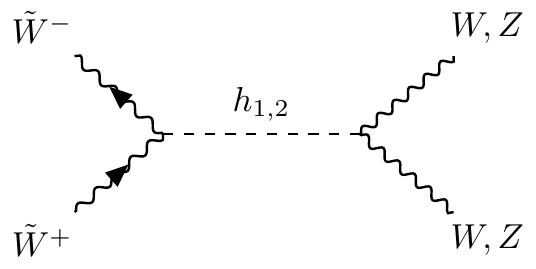}
	\end{subfigure}
	\begin{subfigure}{.24\textwidth}
		\centering
		\includegraphics[width=\textwidth]{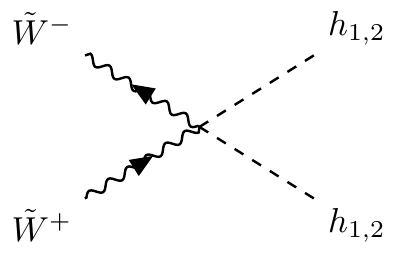}
	\end{subfigure}
	\begin{subfigure}{.24\textwidth}
		\centering
		\includegraphics[width=\textwidth]{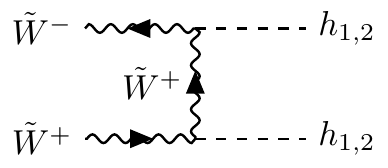}
	\end{subfigure}
	\begin{subfigure}{.24\textwidth}
		\centering
		\includegraphics[width=\textwidth]{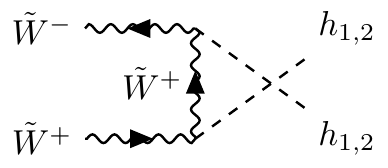}
	\end{subfigure}
	\begin{subfigure}{.24\textwidth}
		\centering
		\includegraphics[width=\textwidth]{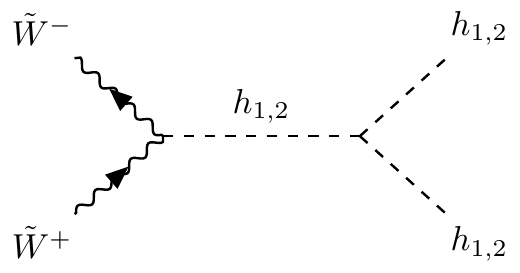}
	\end{subfigure}
	\begin{subfigure}{.24\textwidth}
		\centering
		\includegraphics[width=\textwidth]{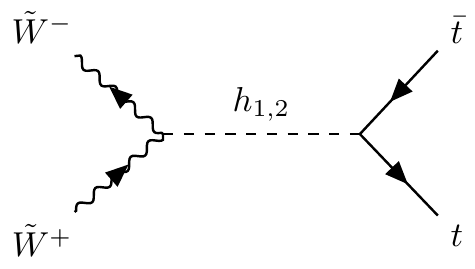}
	\end{subfigure}
	\caption{Representative Feynman diagrams for vector DM $\tilde{W}^+\tilde{W}^-$ pair annihilation.}
	\label{fig:chi1_ann}
\end{figure} 

\subsection{Relic density}
The observed value of the DM relic density  $\Omega_{\text{obs}}h^2 \simeq 0.12$ inferred by the  Planck collaboration from the analysis of Cosmic Microwave Background (CMB)~\cite{Ade:2015xua}. The vector DM candidates $\tilde{W}^{\pm}$ and scalar DM candidates  $\omega$ ($\omega^{\pm}$ and $\omega_2$) in the ST (TT) model can account for the DM relic density we observed today.\footnote{An SU(2)$_{\rm D}$ theory broken down to U(1)$_{\rm D}$ by an adjoint scalar gives rise to dark magnetic monopoles, which may also contribute to the relic density calculation~\cite{Khoze:2014woa}. However, for our choices of triplet VEVs and $\tilde{g}$, it is unlikely that monopoles will contribute significantly to the observed relic density (see Fig.~3 of Ref.~\cite{Khoze:2014woa}). Hence, 
	for simplicity we do not include monopoles in our consideration.} By solving the Boltzmann equation in the standard freeze-out scenario, the relic density of our DM candidates can be estimated by~\cite{Ackerman:mha}
\beq
\Omega_{\text{DM}}h^2 = 1.07\times 10^9 \frac{x_f~\text{GeV}^{-1}}{(g_{*S}/\sqrt{g_*}) M_{pl} \langle\sigma v_{\text{rel}}\rangle } ,
\eeq
where $x_f \equiv m_\chi/T_f$, which can be estimated by 
\beq
x_f = \text{ln}~[0.038\frac{g}{\sqrt{g_*}}M_{pl} m_\chi \langle\sigma v_{\text{rel}}\rangle] - \frac{1}{2}\text{ln ln}~[0.038\frac{g}{\sqrt{g_*}}M_{pl} m_\chi \langle\sigma v_{\text{rel}}\rangle].
\eeq
Here
$g_*\ (g_{*S})$ is the effective degree of freedom in energy density (entropy) at freeze-out defined in Eq.~(\ref{eq:dofe}) ((\ref{eq:dofs})). We evaluate the s-wave annihilation cross section at the leading order~\cite{Srednicki:1988ce}
\beq
\langle \sigma v_{\text{rel}}\rangle = \frac{1}{32\pi}
\frac{\sqrt{1- 4 M_W^2 /s}}{m_\chi^2}
|M_{\text{annihilation}}(s)|^2.
\eeq
The attractive longe-range force between the vector DM $\tilde{W}^{\pm}$ introduced by the exchange of massless DR $\tilde{W}_2$ can increase the annihilation cross section, which is the so-called Sommerfeld enhancement. The Sommerfeld factor is given by~\cite{Cirelli:2007xd} 
\beq
\hat{S} = \frac{\tilde{\alpha}\pi}{v}\frac{1}{1-\text{exp}[-\tilde{\alpha}\pi / v]}.
\eeq
When the DM freezes out, $x_f \approx 25$, $v = 1/\sqrt{x_f} \approx 0.2$. With $g \sim 0.3$, $\hat{S} - 1 \sim 6\times 10^{-2}$. So, we can safely ignore the effects of the Sommerfeld enhancement in this work for the relic density calculation. 
We calculated the annihilation cross section of the process
\begin{enumerate}[label={}]
	\item $\tilde{W}^+\tilde{W}^- \rightarrow W^+W^-,\,Z Z,\, \bar{t}t,\,h_1 h_1,\,h_1 h_2,\,h_2 h_2,\, \tilde{W}_2 \tilde{W}_2,$
	\item $\omega^+\omega^- \rightarrow W^+W^-,\,Z Z,\, \bar{t}t,\,h_1 h_1,\,h_1 h_2,\,h_2 h_2,\, \tilde{W}^+\tilde{W}^-,\tilde{W}_2\tilde{W}_2$,
	\item $\omega_2\omega_2  \rightarrow W^+W^-,\,Z Z,\, \bar{t}t,\,h_1 h_1,\,h_1 h_2,\,h_2 h_2,\, \tilde{W}^+\tilde{W}^-,\,\omega^+\omega^-$.
\end{enumerate}
The representative Feynman diagrams for the vector DM $\tilde{W}^\pm$ pair annihilation are shown in Fig.~\ref{fig:chi1_ann}. Scalar DM pair annihilations have similar diagrams. 

\begin{figure}[tb]
	\begin{subfigure}{.5\textwidth}
		\centering
		\includegraphics[width=\textwidth]{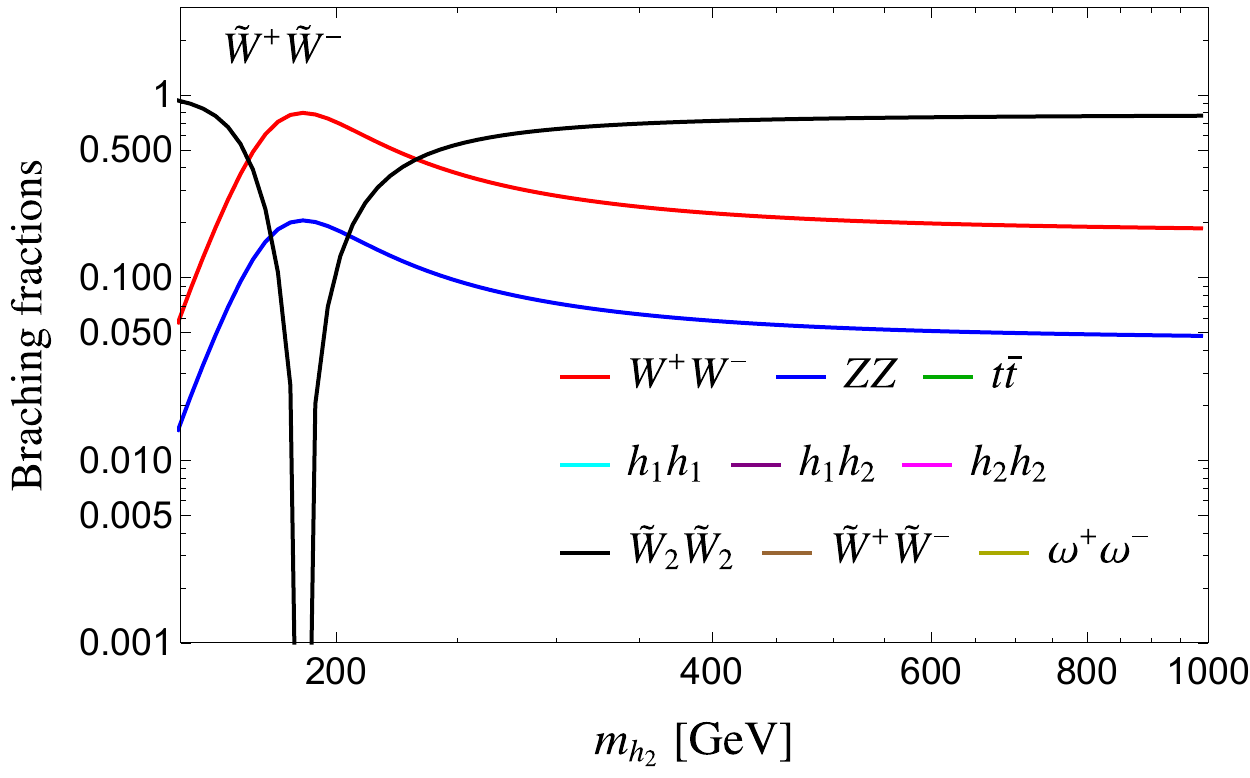}
	\end{subfigure}
	\begin{subfigure}{.5\textwidth}
		\centering
		\includegraphics[width=\textwidth]{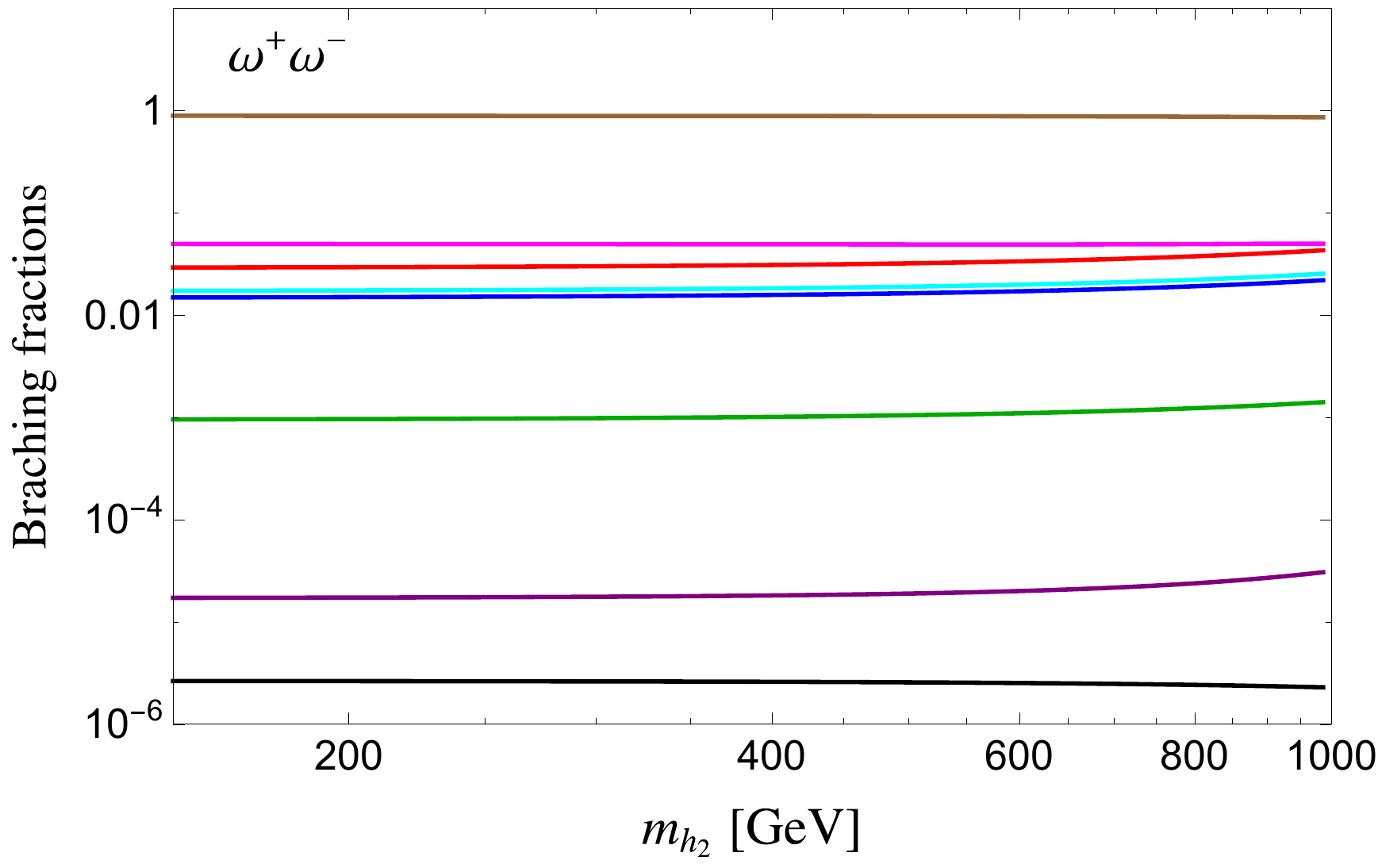}
	\end{subfigure}
	\begin{subfigure}{.5\textwidth}
		\centering
		\includegraphics[width=\textwidth]{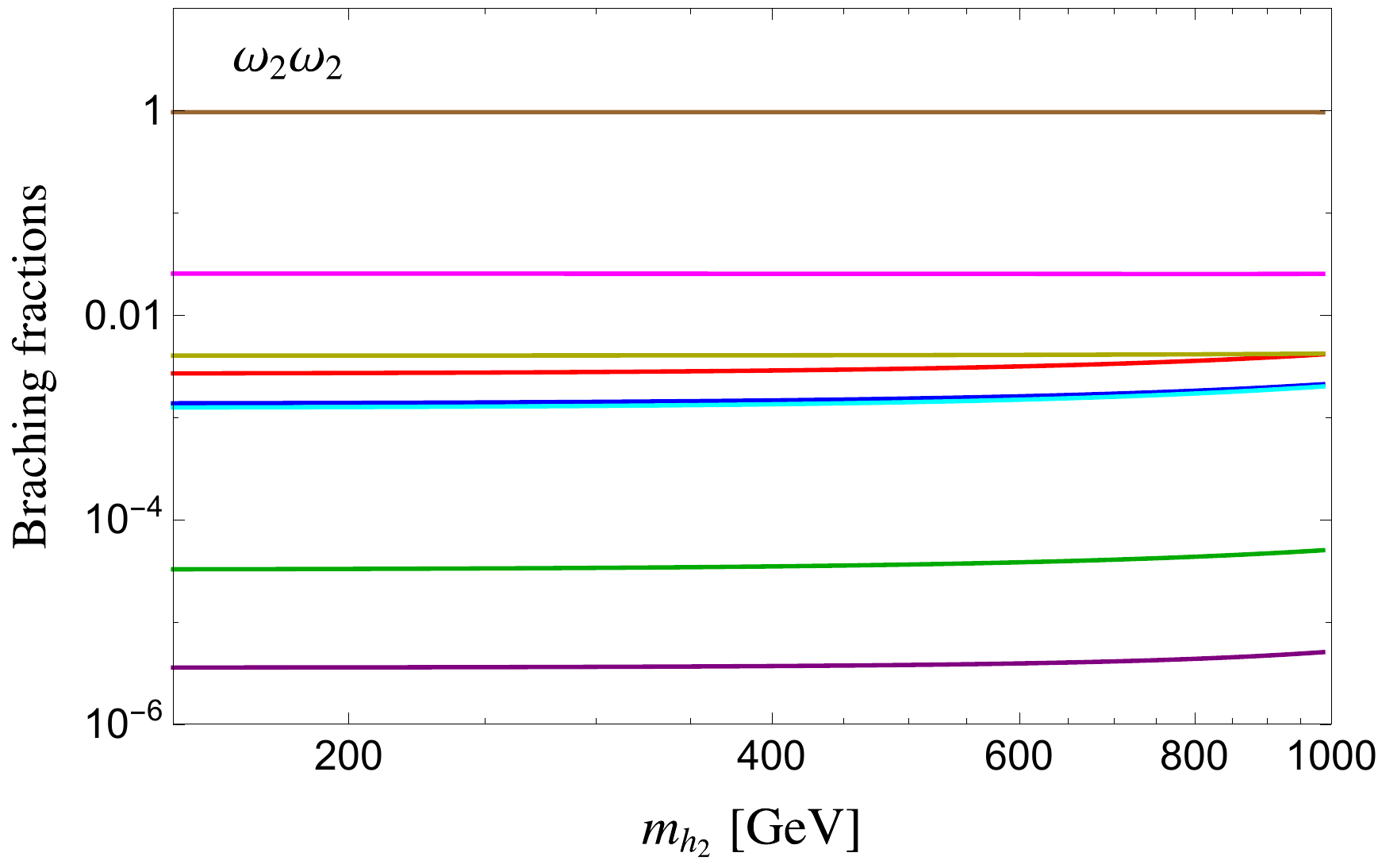}
	\end{subfigure}
	\begin{subfigure}{.5\textwidth}
		\centering
		\includegraphics[width=\textwidth]{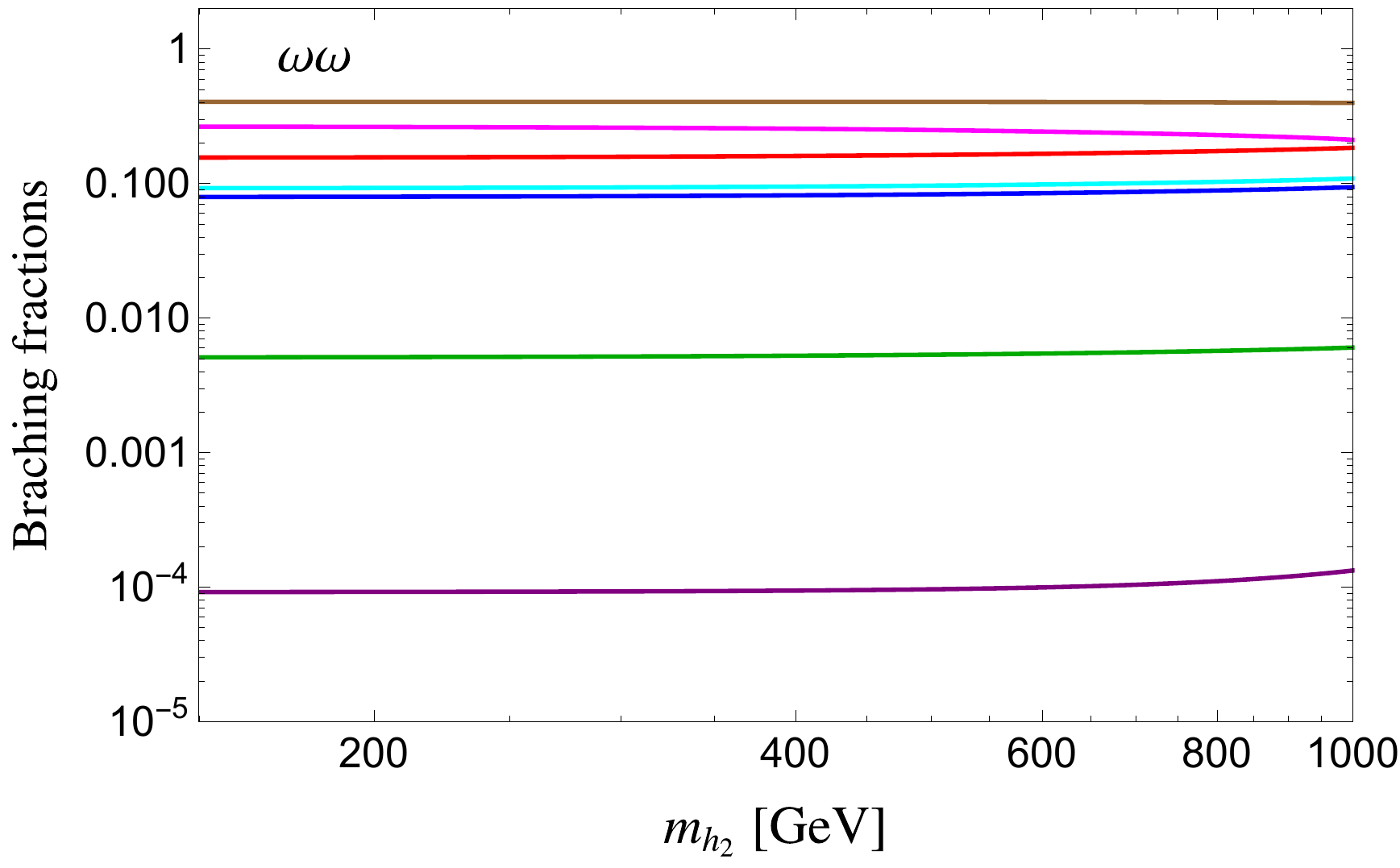}
	\end{subfigure}
	\caption{Annihilation branching fractions of vector DM pair $\tilde{W}^+ \tilde{W}^-$ (upper left), scalar DM pair $\omega^+ \omega^-$ (upper right),  $\omega_2 \omega_2$ (lower left), and  $\omega \omega$ (lower right).  The other  parameters are fixed as BM1 in Table~\ref{Table: BMPs}. }
	\label{fig:chi1_ann_r}
\end{figure} 
Since we choose $m_{\omega^\pm}, m_{\omega_2} \gg m_{\tilde{W}^\pm}$, scalar DM candidates $\omega^\pm$ and $\omega_2$ will be decoupled much earlier than vector DM $\tilde{W}^\pm$. 
The scalar DM states in the TT model annihilate dominantly into the vector DM $\tilde{W}^{\pm}$. While, in the ST model, the scalar DM annihilation channel is dominated by $\omega\omega\rightarrow h_2 h_2$ as it does not carry any charge.
At the decoupling of $\omega^\pm$ and $\omega_2$, $n_{\tilde{W}^\pm} = n^{\text{eq}}_{\tilde{W}^\pm}$. Therefore, including the DM self-interacting processes can further reduce the relic density of $\omega^\pm$ and $\omega_2$.
The number densities of $\omega^\pm$ and $\omega_2$ are much less than $\tilde{W}^\pm$ at the decoupling of $\tilde{W}^\pm$. Therefore, we ignore the processes $\omega^+ \omega^- \rightarrow \tilde{W}^+\tilde{W}^-$ and $\omega_2 \omega_2 \rightarrow \tilde{W}^+\tilde{W}^-$ when we evaluate the number density of $\tilde{W}^\pm$. The vector DM mainly annihilates into the DR $\tilde{W}_2$ except in the resonance region $m_{\tilde{W}^{\pm}} \approx m_{h_2}/2$. 
The branching fractions to a specific final state from an initial state annihilation of both vector and scalar DM pairs are shown in Fig.~\ref{fig:chi1_ann_r}. 

\begin{figure}
	\begin{subfigure}{.5\textwidth}
		\centering
		\includegraphics[width=\textwidth]{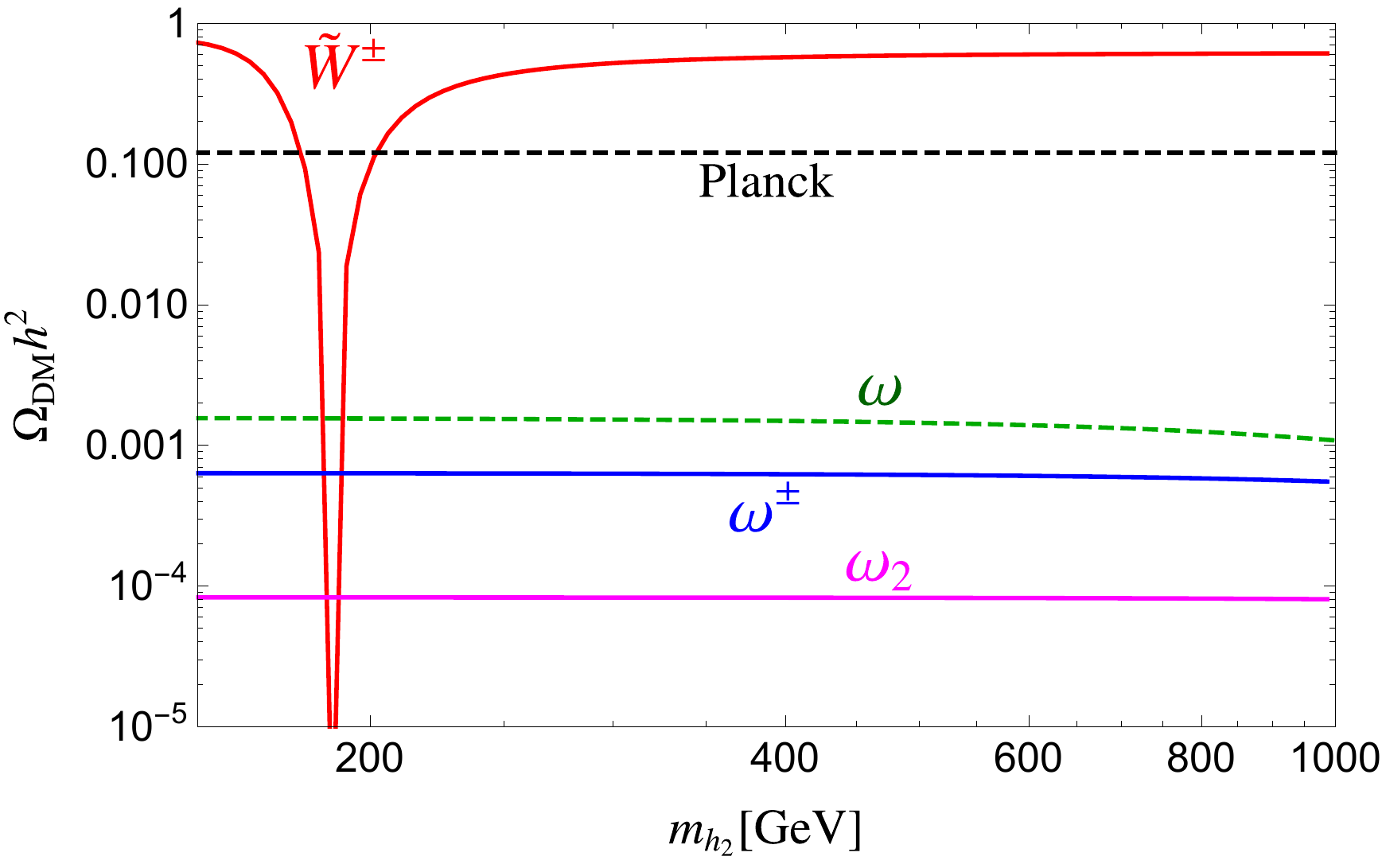}
	\end{subfigure}
	\begin{subfigure}{.5\textwidth}
		\centering
		\includegraphics[width=\textwidth]{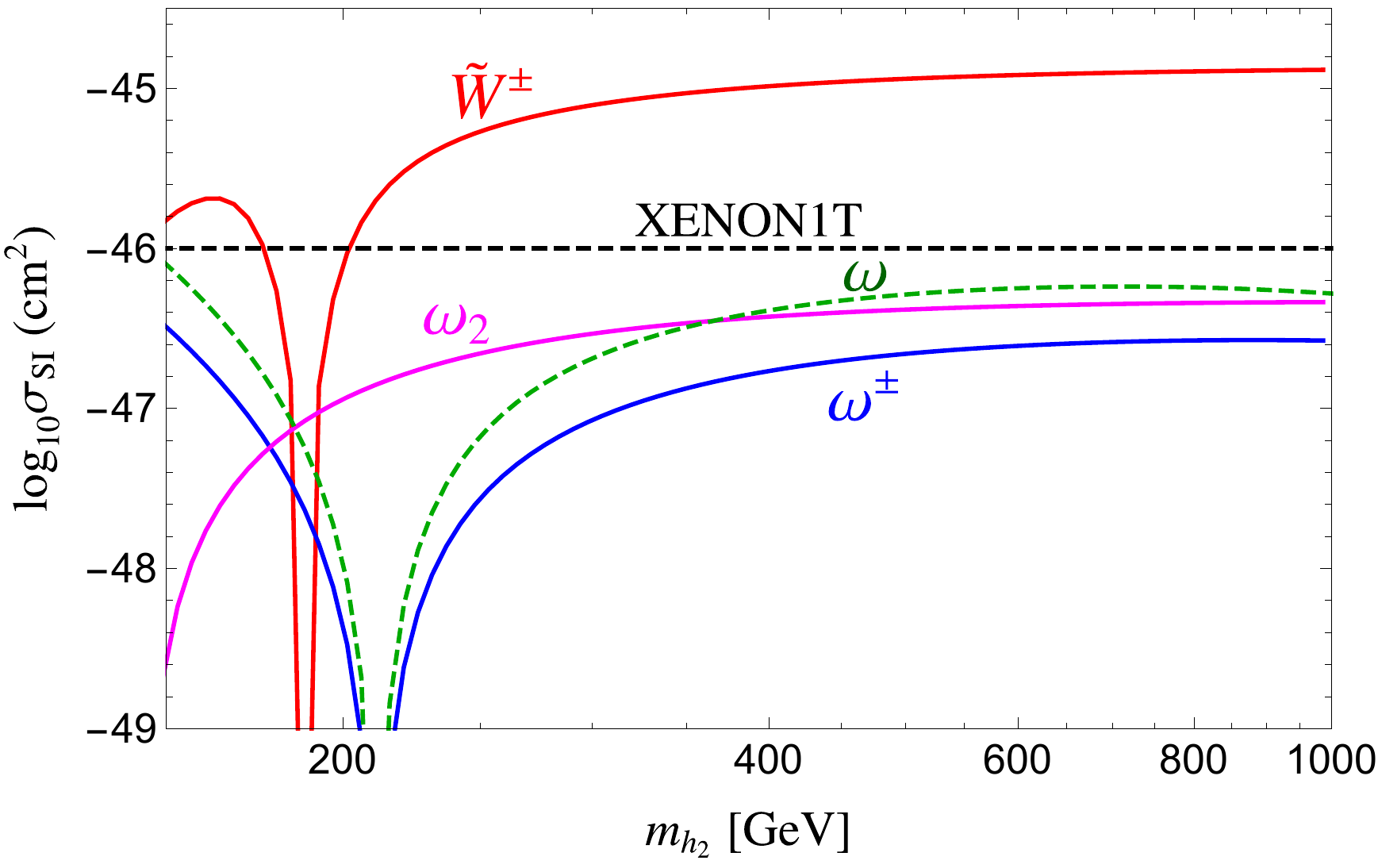}
	\end{subfigure}
	\caption{DM relic densities $\Omega_{\text{DM}} h^2$ (left) and the SI cross section $\sigma_{\text{SI}}$ (right) for the vector and scalar DM candidates versus $m_{h_2}$.
		The dashed green lines are the scalar DM $\omega$ from the ST model. The solid blue and magenta lines are the scalar DM $\omega^{\pm}, \omega_2$ from the TT model, respectively. The solid red lines are from  the vector DM $\tilde{W}^{\pm}$. The dashed horizontal lines indicate the current bounds from PLANCK (left) and XERNON1T (right), respectively. The other parameters are fixed as BM1 in Table~\ref{Table: BMPs}. }
	\label{fig:relicdensity}
\end{figure}

The relic densities for some benchmark points are shown in the left panel of Fig.~\ref{fig:relicdensity} as functions of heavy Higgs mass $m_{h_2}$. The vector DM relic density is highly suppressed at the resonance region. The scalar DM contributions to the total relic density are negligible. The dashed green lines are the scalar DM from the ST scenario, which mostly overlaps with $\omega^{\pm}$ as they have the same masses and similar annihilation channel as shown in Fig.~\ref{fig:chi1_ann_r}.  
We require the DM not to be overly produced $\Omega_{\text{DM}} h^2 \lesssim 0.12$.
The dashed horizontal line in the left panel of Fig.~\ref{fig:relicdensity} indicates the current relic density bound from PLANCK. In the resonance region $m_{\tilde{W}^{\pm}} \approx m_{h_2}/2$, the annihilation cross sections via an $s$-channel $h_2$ are enhanced, and the relic density is much less than the observed value. Away from the resonant region, $\tilde{W}^{\pm}$ could be adequate as a CDM candidate.


\subsection{Direct detection}
The null results of direct detection experiments can set strong bounds on our dark sector parameter space. 
In this model, the DM candidates $\chi$ couple to the dark scalar $\varphi_2$. $\varphi_2$ couples to the SM particles through the Higgs portal. 
The dominant contributions to the spin-independent (SI) scattering cross section come from the exchange of the SM-like Higgs bosons $h_1$ and the heavy Higgs bosons $h_2$.  
The effective interactions of DM ($\chi=\tilde{W}^\pm, \omega^{\pm}, \omega_2, \omega$) with light quarks and gluons are given as~\cite{Jungman:1995df}
\beq
\lag^{\text{eff}}_{q,g} = \sum_{q=u,d,s}f^{\chi}_q m_q \chi\chi \bar{q} q + f^{\chi}_G \chi \chi \frac{\alpha_s}{\pi} G^{a\mu\nu}G^a_{\mu\nu},
\eeq 
where $G^a_{\mu\nu}$ is the field strength tensor of gluon and $\alpha_s$ is the strong coupling constant. $f^{\chi}_q$ is the effective couplings between DM $\chi$ and light quarks, which, in our model, are
\beqa
f^{\tilde{W}^\pm}_q  &=& \tilde{g}^2\frac{v_2}{v_h} \sin\theta\cos\theta (\frac{1}{m^2_{h_2}}-\frac{1}{m^2_{h_1}}),\\ 
f^{\omega^\pm}_q  &=& \frac{1}{v_h}(\frac{c_{2}\cos\theta}{m^2_{h_2}}-\frac{c_{1}\sin\theta}{m^2_{h_1}}),\\
f^{\omega_2}_q  &=& \frac{1}{v_h}(\frac{d_{2}\cos\theta}{m^2_{h_2}}-\frac{d_{1}\sin\theta}{m^2_{h_1}}).
\eeqa
The coupling between DM and gluon comes from the effective coupling after integrating-out of heavy quarks
\beq
f^{\chi}_G=-\frac{1}{12}\sum_{Q=c,b,t}f^{\chi}_Q = -\frac{1}{4} f^{\chi}_q.
\eeq
The interactions between DM and nucleon can be evaluated by using the nucleon matrix elements 
\beq
\langle N | m_q \bar{q}q | N \rangle \equiv f_{Tq}^N m_N, \, \langle N|\frac{\alpha_s}{\pi}GG|N\rangle =-\frac{8}{9} m_N f^N_{TG},
\eeq
where $f_{Tq}^N$ and $f_{TG}^N$ are the mass-fraction parameters of the quarks and the gluon in the nucleon $N$,respectively. In our numerical calculations, we adopt $f_{Td}^p = 0.0191$, $f_{Tu}^p = 0.0153$, $f_{Ts}^p = 0.0447$, and $f^p_{TG}\equiv 1-\sum_{q=u,d,s}f^p_{Tq} = 0.925$~\cite{Belanger:2013oya}. The effective interactions of DM and nucleon can be expressed as
\beq
\lag^{\text{eff}}_{N} = f^{\chi}_N\chi\chi \bar{N} N,
\eeq 
where the effective coupling $f_N$ can be calculated by
\beq
f^{\chi}_N = m_N (\sum_{q=u,d,s} f^N_{Tq} f^{\chi}_q - \frac{8}{9} f^N_{TG} f^{\chi}_G).
\eeq
The SI cross section of DM with nucleon can be calculated with~\cite{Hisano:2010yh}
\beq
\hat{\sigma}^{\chi}_{\text{SI}} = \frac{1}{\pi} \left(\frac{m_N}{m_{\chi} + m_N}\right)^2(f^{\chi}_N)^2,
\label{Eq: sigmaSI}
\eeq 
where $m_N$ is the mass of nucleon and $m_{\chi}$ is the mass of DM candidate. To derive the experimental upper bound, we scale the SI cross sections with the density fractions
\beq
\sigma_{\text{SI}} = \left(\frac{\Omega_{\chi}h^2}{\Omega_{\text{obs}}h^2}\right)\hat{\sigma}^{\chi}_{\text{SI}}.
\label{Eq: sigmaSI_exp}
\eeq 
The XENON1T~\cite{Aprile:2018dbl} and the SI cross sections are shown in the right panel of Fig.~\ref{fig:relicdensity}. In the resonance region $m_{\tilde{W}^{\pm}} \approx m_{h_2}/2$, the relic density is much less than the observed value, hence the direct detection bound can be easily evaded. Away from the resonant region however, $\tilde{W}^{\pm}$ could lead to a detectable cross section.

\subsection{Dark matter self-interactions}
The collision-less and cold DM can successfully describe the large scale structure of the Universe~\cite{Springel:2005nw}. There are, however, some challenges for the cold and collision-less DM model at the small-scale (see Ref.~\cite{Bullock:2017xww} for a review). Rather than going to the warm DM scenario, there are generally two mechanisms which can alleviate the CDM challenges: (i) DM-DR interactions~\cite{Boehm:2001hm}; (ii) DM self-interactions~\cite{Spergel:1999mh}.

In our model, the leading DM self-interaction is mediated by the massless DR. This scenario has been studied carefully in Ref.~\cite{Feng:2009mn, Agrawal:2016quu}.
The most relevant DM self-interactions are through $t/u$-channel mediated by the massless DR. The differential cross section of $t$- and $u$-channel in the center-of-mass (CM) frame is 
\beq
\frac{d\sigma}{d\Omega} \propto \frac{\tilde{\alpha}^2}{16 m_{\tilde{W}^{\pm}}^2v^4_r\sin^4\frac{\theta_{\text{cm}}}{2}},
\eeq
leading to $\sigma\sim \pi\tilde{\alpha}^2 / (m_{\tilde{W}^{\pm}}^2 v^4_r)$, where $v_r$ is the relative velocity of the two colliding DM particles in the CM frame. The cross sections of the DM self-interactions quickly drop at higher velocities to evade impacts on the large scale structure, hence, maintain the effective collision-less descriptions. 
From the observed ellipticity of galactic DM halos~\cite{Feng:2009mn, Agrawal:2016quu}, a bound  on the dark gauge couplings can be estimated as 
\beq
(\frac{\tilde{g}}{0.1})^4(\frac{200~\text{GeV}}{m_{\tilde{W}^{\pm}}})^3\lesssim 50. 
\eeq
This constraint can potentially be overly strong and depends on the assumptions of DM relic density~\cite{Agrawal:2016quu}. The constraints from the Bullet Cluster are much weaker~\cite{Feng:2009mn, Agrawal:2016quu}. To solve the small-scale structure problems, we need $\sigma/m_{\tilde{W}} \sim 0.1-10~\text{cm}^2/g$ at dwarf galaxies~\cite{Zavala:2012us,Khoze:2014woa}, which gives
\beq
(\frac{\tilde{g}}{0.1})^4(\frac{200~\text{GeV}}{m_{\tilde{W}^{\pm}}})^3\sim 0.01 - 1.
\eeq

DM can also interact with themselves through four-gauge-boson contact and $s$-channel interactions. The cross sections of contact interactions are $\sigma\sim \pi\tilde{\alpha}^2/m_{\tilde{W}^{\pm}}^2$; the $s$-channel cross sections are $\sigma\sim \pi\tilde{\alpha}^2 v_r^4/ m_{\tilde{W}^{\pm}}^2$. Therefore they are irrelevant compared to the contributions of $u/t$-channel for the DM self-interactions for low-velocity systems such as dwarf galaxies. {It is evident from the discussion above that the DM-DR interaction cross sections are suppressed by the DM mass. So, for the parameter space of our interest in this work, DM and DR are decoupled very early and cannot significantly change the small-scale structures of the Universe.}
Before closing the DM section, we would like to mention that we will not study the indirect detection aspects of this model due to the complication with the Sommerfeld enhancement in low-velocity systems.

\section{Electroweak phase transition and gravitational waves}
\label{sec:ewpt_GW}

\subsection{Electroweak phase transition}
\label{sec:ewpt}

The dynamics of the phase transition is determined by the effective potential at the finite temperature (see, {\it e.g.}, Ref.~\cite{Hindmarsh:2020hop} for a recent review),  
which can be calculated perturbatively or non-perturbatively on the lattice with dimensional reduction~\cite{Farakos:1994xh,Kajantie:1995kf,Moore:2000jw}. While the latter
approach provides a gauge independent result and is free of the infrared problem~\cite{Linde:1980ts}, it is computationally
expensive and so far has been adopted for only a few models with a simple extended Higgs sector~\cite{Brauner:2016fla,Andersen:2017ika,Niemi:2018asa,Gorda:2018hvi,Gould:2019qek,Niemi:2020hto}. Therefore the perturbative method was predominant in the literature on the analysis of a thermal phase transition. 
In the standard perturbative approach, the effective potential receives contributions from the tree-level potential, the one-loop Coleman-Weinberg correction and its finite-temperature counterpart, as well as Daisy resummations, which
together leads to a gauge dependent result (see, {\it e.g.}, Refs.~\cite{Croon:2020cgk,Papaefstathiou:2020iag} for a study of the uncertainties with this approach). 
A gauge independent result nevertheless can still be obtained if only the leading order
thermal correction at the high temperature is kept~\cite{Patel:2011th}. This also makes an analytical understanding of the otherwise
complicated effective potential possible and can better guide the exploration of the phase history. Thus we follow this gauge independent perturbative approach. 
The finite temperature effective potential can thus be written in the following simplified form
\beq
V^{(1)}(T) = V_{\text{tree}} + \Delta V^{(1)}(T) ,
\eeq
where $V_{\text{tree}}$ is given in Eq~(\ref{eq:pot}) and $\Delta V^{(1)}(T)$ is the leading thermal correction given by~\cite{Quiros:1999jp}
\beq
\Delta V^{(1)}(T) = \frac{T^4}{2 \pi^2} \left\{\sum_{b} n_b J_B [\frac{m_b^2(\phi_i)}{T^2}] - \sum_{f} n_f J_F [\frac{m_f^2(\phi_i)}{T^2}]\right\},
\eeq
where $\phi_i (i=1,2,3)$ indicates any of the three fields.
Here the functions $J_B$ and $J_F$ have the following high-temperature limit, {\it i.e.,} for $y\equiv m/T \ll 1$,
\beq
J_B (y^2) \simeq \frac{-\pi^4}{45} + \frac{\pi^2}{12}y^2 -\frac{\pi}{6} y^3+ O(y^4),\quad
J_F (y^2) \simeq \frac{7\pi^4}{360} - \frac{\pi^2}{24}y^2 + O(y^4).
\eeq
Therefore at order $y^2$, the thermal corrections reduce to a simpler polynomial form
\beq
\Delta V^{(1)}(T) =  \frac{T^2}{24}[n_s \text{Tr}(\mathbf{M_S}^2) + n_{\tilde{W}}\text{Tr}(\mathbf{M_V}^2)+n_Wm_W^2 + n_Z m_Z^2 + \frac{n_t}{2}m_t^2],
\eeq
where $\mathbf{M_S} \text{ and } \mathbf{M_V}$ are the field-dependent masses for scalar and dark gauge bosons, which are given in Appendix~\ref{sec:fdm}.
From the finite temperature effective potential, the details of the phase transition can be studied. 
In particular, one can determine the thermal mass terms. For the TT model, they are given by
\beqa
m^2_{H} (T) &=& m^2_{H} + \frac{T^2}{16}(g_1^2 + 3g_2^2 +2(2\lambda_{H}+\lambda_{H11}+\lambda_{H22}+2 y_t^2)),\label{eq:m1}\\
m^2_{11} (T) &=& m^2_{11} + \frac{T^2}{24}(12 \tilde{g}^2+5 \lambda_{1} + 3\lambda_{3}+\lambda_{4}+4\lambda_{H11}),\label{eq:m2}\\
m^2_{22} (T) &=& m^2_{22} +  \frac{T^2}{24}(12 \tilde{g}^2+5 \lambda_{1} + 3\lambda_{3}+\lambda_{4}+4\lambda_{H22})\label{eq:m3}.
\eeqa
In the ST model, the thermal mass terms are
\beqa
m^2_{H} (T) &=& m^2_{H} + \frac{T^2}{16}(g_1^2 + 3g_2^2 +2(2\lambda_{H}+\frac{1}{3}\lambda_{H11}+\lambda_{H22}+2 y_t^2)),\label{eq:m4}\\
m^2_{11} (T) &=& m^2_{11} + \frac{T^2}{24}(3 \lambda_{1} + 3\lambda_{3}+4\lambda_{H11}),\label{eq:m5}\\
m^2_{22} (T) &=& m^2_{22} +  \frac{T^2}{24}(12 \tilde{g}^2+5 \lambda_{1} + 3\lambda_{3}+4\lambda_{H22})\label{eq:m6}.
\eeqa
\begingroup
\setlength{\tabcolsep}{10pt} 
\renewcommand{\arraystretch}{1.} 
\begin{table}
	\centering
	\begin{tabular}{|c |  c|  c|  c|c|c|}
		\toprule
		Extrema Type  &  $h_0$ &  $\omega_3$ or $\omega$  & $\varphi_2$ & potential value $V_{\text{min}}$ & stableness \\
		\midrule
		Type-1   & 0 & 0 & 0 & 0 &  condition (\ref{Eq:cdt1}) \\
		\midrule
		Type-2   & $v_h$ & 0 & 0 &$-\frac{m_H^4}{2\lambda_H}$ &  condition (\ref{Eq:cdt2})  \\
		\midrule
		Type-3   & 0 & $v_1$ & 0 & $-\frac{m_{11}^4}{2\lambda_{1}}$ &  condition (\ref{Eq:cdt3}) \\
		\midrule
		Type-4   & 0 & 0 & $v_2$ & $-\frac{m_{22}^4}{2\lambda_{2}}$ &  condition (\ref{Eq:cdt4})  \\
		\midrule
		Type-5   & $v_h$ & $v_1$ & 0 & $-\frac{\lambda_{H}m_{11}^4-2\lambda_{H11}m_{11}^2m_H^2+\lambda_{1}m_H^4}{2\lambda_{1}\lambda_{H}-2\lambda_{H11}^2}$ &  condition (\ref{Eq:cdt5}) \\
		\midrule
		Type-6   & 0 & $v_1$ & $v_2$ & $-\frac{\lambda_{1}m_{22}^4-2\lambda_{3}m_{11}^2m_{22}^2+\lambda_{2}m_{11}^4}{2\lambda_{1}\lambda_{2}-2\lambda_{3}^2}$ &   condition (\ref{Eq:cdt6}) \\
		\midrule
		Type-7   & $v_h$ & 0 & $v_2$ & $-\frac{\lambda_{H}m_{22}^4-2\lambda_{H22}m_{22}^2m_H^2+\lambda_{2}m_H^4}{2\lambda_{2}\lambda_{H}-2\lambda_{H22}^2}$ &  condition (\ref{Eq:cdt7}) \\
		\midrule
		Type-8   & $v_h$ & $v_1$ & $v_2$ & see details in Ref.~\cite{Vieu:2018nfq} & Ref.~\cite{Vieu:2018nfq} \\
		\toprule
	\end{tabular}
	\caption{Eight possible types of stable vacuum extrema in the three VEVs scenario. }
	\label{Table: extrema}
\end{table}
\endgroup

\begin{figure}
	\begin{subfigure}{.5\textwidth}
		\centering
		\includegraphics[width=\textwidth]{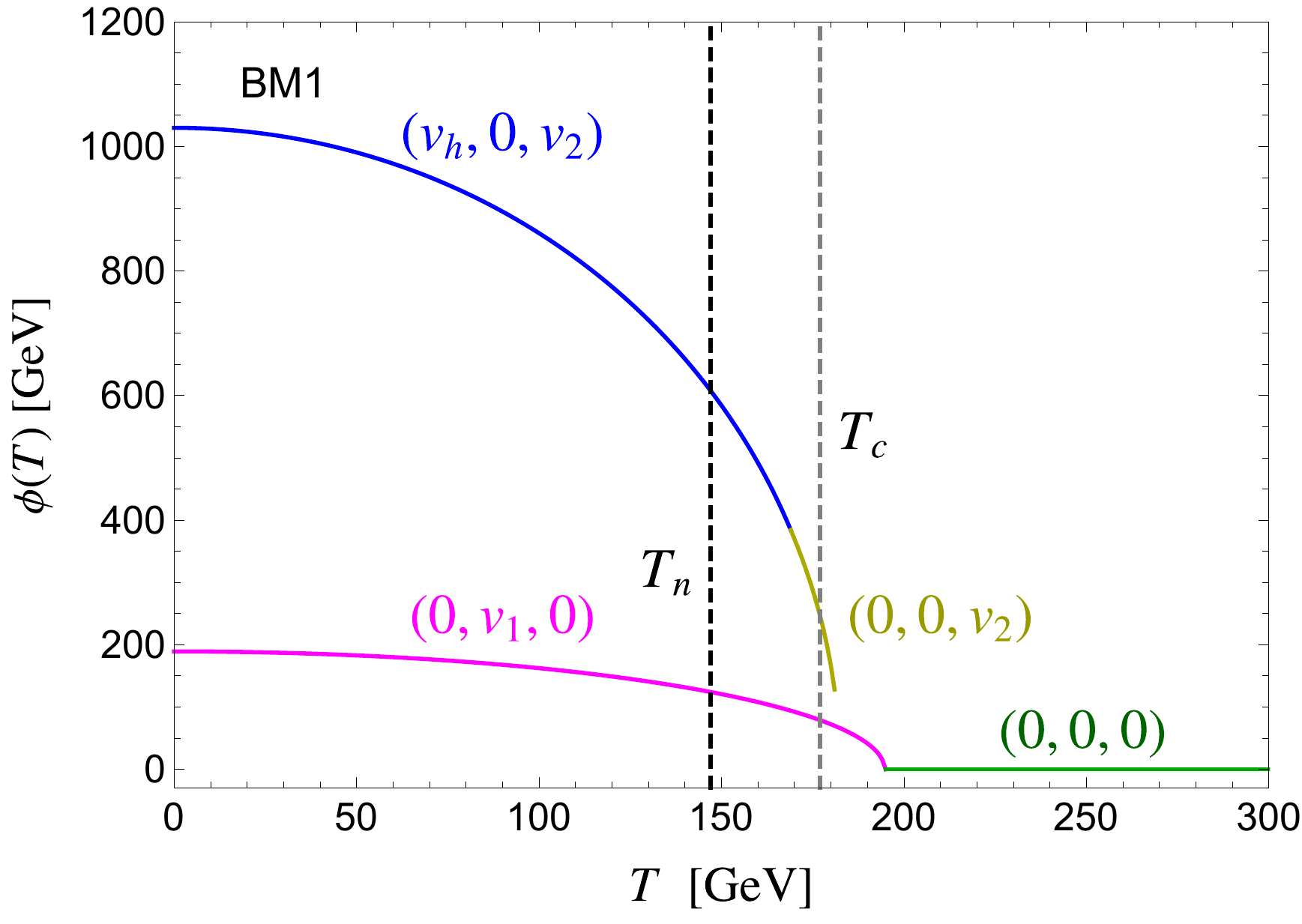}
	\end{subfigure}
	\begin{subfigure}{.48\textwidth}
		\centering
		\includegraphics[width=\textwidth]{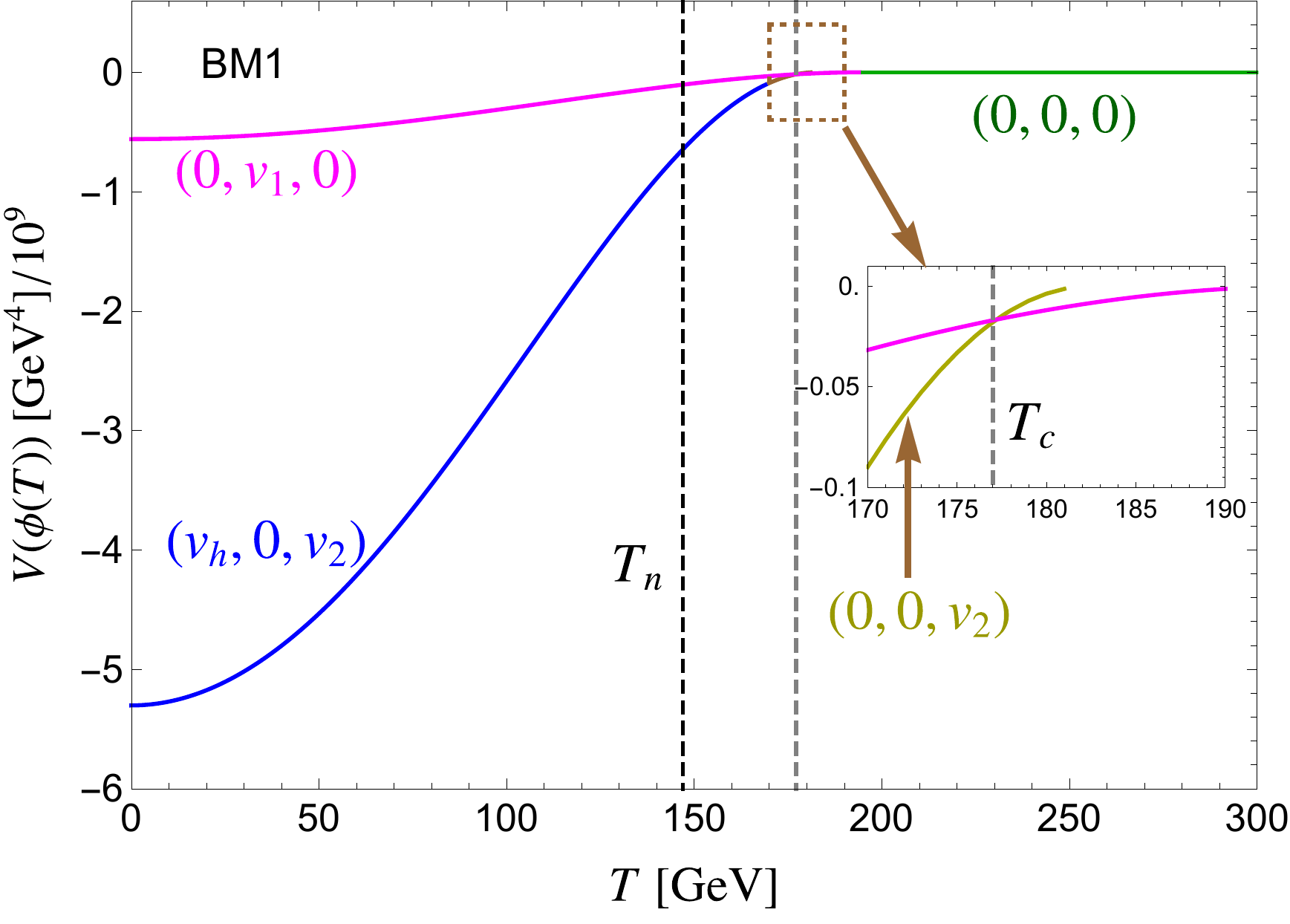}
	\end{subfigure}
	\begin{subfigure}{.5\textwidth}
		\centering
		\includegraphics[width=\textwidth]{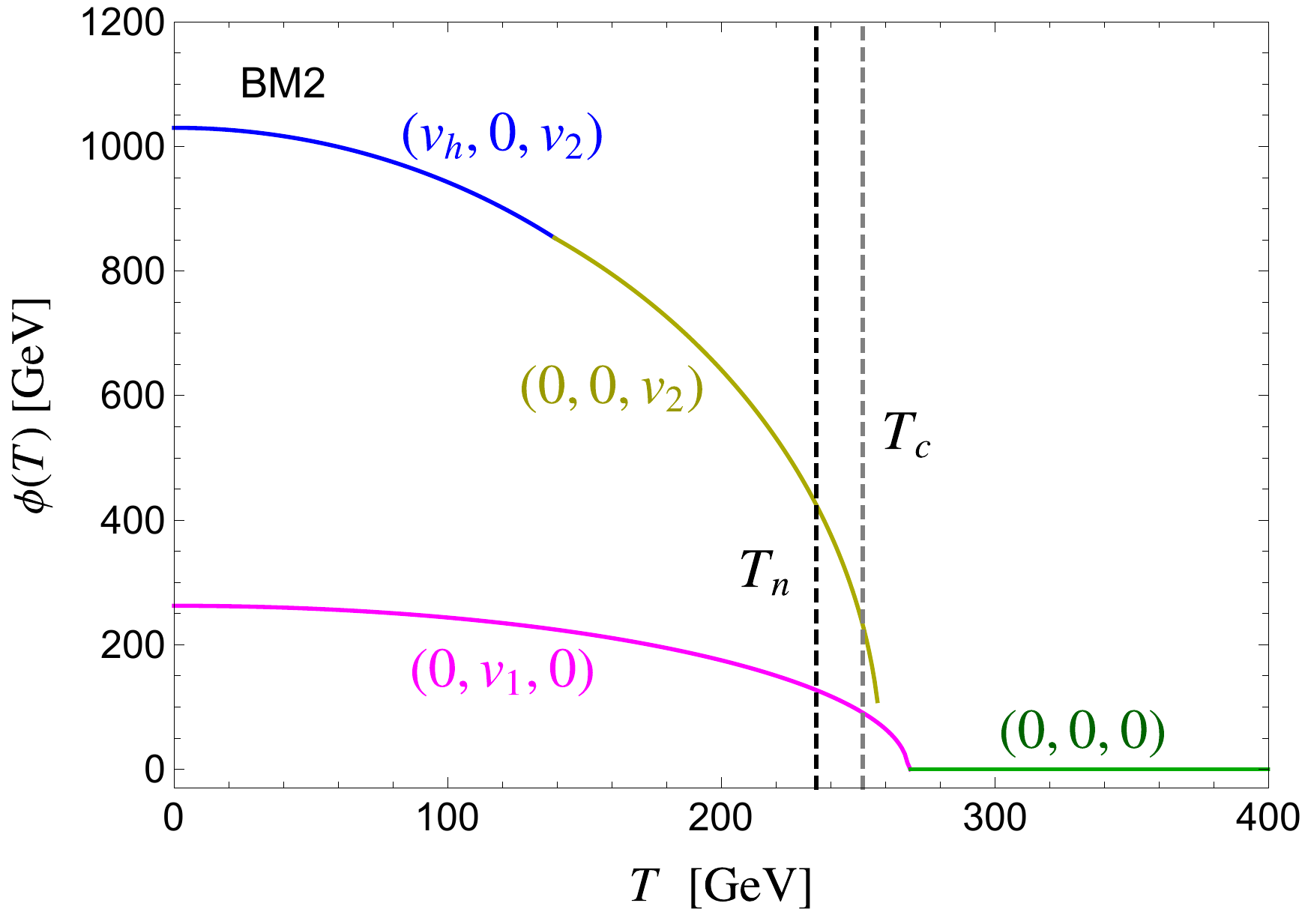}
	\end{subfigure}
	\begin{subfigure}{.48\textwidth}
		\centering
		\includegraphics[width=\textwidth]{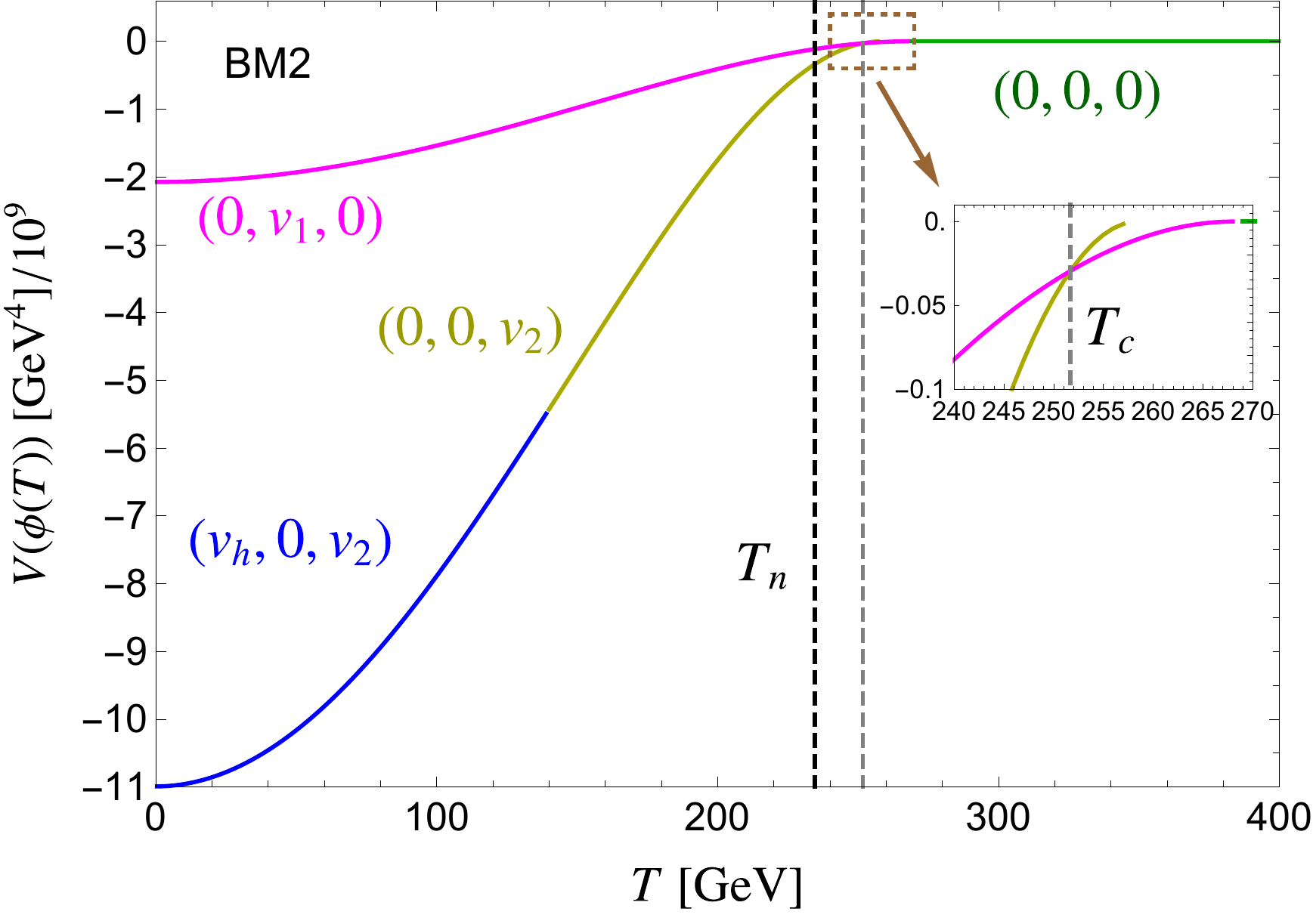}
	\end{subfigure}
	\caption{The evolution of the vacuum ($\phi \equiv \sqrt{v_h^2 + v_1^2 + v_2^2}$, left) 
		as a function of the temperature $T$, and their corresponding potential values (right) 
		are shown for BM1 (upper panels) and BM2 (lower panels). 
		Here the critical and nucleation temperatures are denoted by the dashed vertical lines, respectively.}
	\label{fig:evo}
\end{figure}
Even though those two scenarios have the same zero-temperature potential in Eq.~(\ref{eq:pot}), the mass parameters evolve differently with temperature as shown in Eqs.~(\ref{eq:m1}) to (\ref{eq:m6}). The parameter space for FOPT in those two scenarios is not the same, though the phase transition pattern should not be qualitatively different. For the rest of this paper, we will focus on the two BMs in Table~\ref{Table: BMPs} in the TT scenario as an illustration for the phase transition and GW generation. Given the three possible non-zero VEVs $(v_h,v_1,v_2)$, there are eight combinations of possible extrema. Those and their stable conditions are listed in Appendix B and summarized in 
Table~\ref{Table: extrema}. With the desirable features from the extra DR, we require that at $T=0$, the stable vacuum be  in Type-7: $(v_h,0,v_2)$.
From the scanning, we found mainly two possible paths of the phase transitions to achieve this pattern
\beqa
\text{two-step: }(v_h,v_1,v_2): &(0,0,0)&\rightarrow (0,v_1,0) \Rightarrow (v_h,0,v_2),\label{eq:2step}\\
\text{three-step: }(v_h,v_1,v_2): &(0,0,0)&\rightarrow (0,v_1,0) \Rightarrow (0,0,v_2)\rightarrow(v_h,0,v_2),\label{eq:3step}
\eeqa
where ``$\Rightarrow$'' indicates a first-order phase transition and ``$\rightarrow$'' for a continuous transition.\footnote{See a remark on this in Appendix \ref{app:PT}.}
The two-step transition as in Eq.~(\ref{eq:2step}) can yield an electroweak FOPT \cite{Patel:2012pi}, while the second path in Eq.~(\ref{eq:3step}) would not lead to an electroweak FOPT and can wash out any previously existing baryon asymmetry. 
%
To give a clearer picture of the above transitions, we illustrate the vacuum evolution in detail for the case of BM1 as defined in Table~\ref{Table: BMPs}.
In this case, the phase transition is a two-step process as shown in Eq.~(\ref{eq:2step}).
The evolution of the vacuum and the corresponding potential values in BM1 are shown in the upper panel of Fig.~\ref{fig:evo}. 
We see that at high temperatures, the stable vacuum is in a symmetric phase of Type-1: $(0, 0, 0)$. At $T \approx 200$~GeV, the field $\Phi_1$ develops a VEV and the stable phase becomes of Type-3: $(0, v_1, 0)$ through a continuous transition, where the order parameter, the VEV $v_1$, undergoes a continuous change. As the temperature further decreases, another minimum  appears via $\Phi_2$ at $(0,0,v_2)$, which eventually evolves into a minimum of Type-7 $(v_h, 0, v_2)$ continuously. 
At $T=T_c$, corresponding to the right of the vertical dashed line in the left panel of Fig.~\ref{fig:evo}, these two 
types of vacua  $(0,v_1,0)$ and $(0, 0, v_2)$ are degenerate, and are separated by a barrier, characteristic for a FOPT. 
At $T<T_c$, the initially stable vacuum at $(0,v_1,0)$ now becomes metastable while the phase corresponding to $(0, 0, v_2)$ becomes 
energetically preferable, and the Universe becomes supercooled as $T$ decreases.
During the coexistence of these two phases, while the probability for the Universe to make a transition from the former to the latter 
becomes increasingly higher, it remains significantly small during this period.
The temperature at which the phase transition happens can be quantified by the temperature when there is about one bubble per Hubble volume, 
and is called the nucleation temperature $T_n$, corresponding to the left of the vertical dashed line in Fig.~\ref{fig:evo}. 
As $T$ decreases towards $T_n$, the minimum at $(0, 0, v_2)$ evolves into $(v_h, 0, v_2)$.
At $T\approx T_n$, the transition then proceeds through the formation of bubbles, with the vacuum inside
being the more stable one $(v_h,0,v_2)$, and that outside the metastable one $(0,v_1,0)$. Thus the VEV changes non-continuously.
The BM2 has a three-step phase transition shown in Eq.~(\ref{eq:3step}) and the lower panels of Fig.~\ref{fig:evo}. 
It is similar to BM1 but different in that it has a prolonged phase at $(0,0,v_2)$ coexisting with the metastable $(0,v_1,0)$.
The tunneling probability is thus high enough for a FOPT from $(0, v_1, 0)$ to $(0,0,v_2)$ before 
the latter evolves into $(v_h, 0, v_2)$. After this step, the vacuum at $(0,0,v_2)$ makes a further continuous electroweak transition to $(v_h, 0, v_2)$. Further description of the process is provided in an Appendix \ref{app:PT}.


\subsection{Gravitational waves}

From studies of the above phase transition and its evolution at different temperatures, one can determine a set of portal parameters that determine the resulting GW signals \cite{Caprini:2015zlo}
\begin{eqnarray}
T_n, \quad \alpha_e, \quad \beta/H_n, \quad v_w ,
\end{eqnarray}
where $T_n$, as introduced previously, is the nucleation temperature denoting 
roughly the time for the onset of phase transition when there is one bubble per Hubble volume;
$\alpha_e$ is a dimensionless quantity characterizing the energy fraction released from the phase transition in the unit of the total radiation energy
density at $T_n$; $\beta$ is roughly the inverse time duration of the phase transition determining the peak frequency of the GWs and $H_n$ is the Hubble rate $H$ at $T_n$; $v_w$ is the wall velocity. 

The calculations start with the determination of the tunneling probability per unit time per unit volume given by~\cite{Turner:1992tz}
\beq
\Gamma(T)\simeq T^4 \left( \frac{S_3}{2\pi T} \right)^{3/2}e^{-S_3/T},
\eeq
where $S_3$ is the three-dimensional Euclidean action corresponding to the critical bubble:
\beq
S_3 = \int_{0}^{\infty} dr\ r^2 \left[\frac{1}{2}(\frac{d\phi(r)}{dr})^2 + V(\phi,T)\right],
\eeq
with the scalar field minimizing the action and corresponding to the solution of the following equation of motion:
\beq
\frac{d^2\phi}{dr^2}+\frac{2}{r}\frac{d\phi}{dr}=\frac{dV(\phi,T)}{dr} ,
\eeq
subjected to the bounce boundary conditions
\beq
\lim\limits_{r\rightarrow \infty} \phi(r)=0,\quad \frac{d\phi}{dr}\Bigr|_{r=0}=0.
\eeq
In this work, we employ the $\mathbf{CosmoTransitions}$~\cite{Wainwright:2011kj}
to solve the above bounce equation and thus 
compute the Euclidean action $S_3$. From the nucleation rate, the nucleation temperature is usually determined by solving
the following equation,\footnote{It can be more precisely determined by directly calculating the number of bubbles in a generic expanding Universe as shown in Ref.~\cite{Guo:2020grp}.}
\beq
\int_{T_n}^{\infty}\frac{dT}{T}\frac{\Gamma(T)}{H(T)^4} = 1 , 
\eeq
which says that there is about one bubble in a Hubble volume. 
A rough estimation of nucleation temperature $T_n$ is usually obtained using the condition $S_3(T_n)/T_n = 140$~\cite{Apreda:2001us}. 
One can further calculate the parameter $\beta$ where
\beq
\beta = H_*T_* \frac{d(S_3/T)}{dT}\Bigr|_{T_*}, 
\eeq
where $T_*$ is the GW generation temperature and is approximately equal to the nucleation temperature $T_n$. Similar to the definition of $H_n$, $H_*$ is the Hubble rate $H$ at $T_*$. 
The dimension of $\beta$ is hertz and it is related to the mean bubble separation at the phase transition(see, e.g., ~\cite{Hindmarsh:2019phv,Guo:2020grp} 
for the derivation in Minkowski and FLRW spacetimes), which in turn
gives the typical scale for GW production and thus its peak frequency.
Moreover, $\alpha_e$ is the vacuum energy released from the EWPT normalized by the total radiation energy density
\beq
\alpha_e = \frac{\rho_{\text{vac}}}{\rho^*_{\text{rad}}} = \frac{1}{\rho^*_{\text{rad}}}\left[T\frac{\d \Delta V(T)}{\d T} - \Delta V(T)\right]\Bigr|_{T_*},
\eeq
where $\Delta V(T) = V_{\text{low}}(T)-V_{\text{high}}(T)$ is the difference between lower and higher phases, and $\rho^*_{\text{rad}} = g_* \pi^2 T^4/30$, $g_*$ is the relativistic degrees of freedom at $T=T_*$. 
For a phase transition in a thermal plasma, as is considered here, the energy released goes in part into the kinetic energy of the plasma, with 
energy fraction $\kappa_v$, which sources
gravitational waves, and into the heat of the plasma. The flow can also go turbulent, with energy fraction $\kappa_{\text{turb}}$, 
which becomes another source for GW production. 
A fraction of released energy can also go into the gradient of the scalar fields, which however is believed to be of negligible fraction~\cite{Bodeker:2017cim} and 
we will not consider it here.


With these portal parameters, we are ready to calculate the GW energy density spectrum. 
The GW from a FOPT, as in most cosmic processes, is a stochastic
background and can be searched for using the cross correlation method $-$ see recent reviews
on theories~\cite{Caprini:2015zlo,Cai:2017cbj,Caprini:2018mtu} and on detection methods~\cite{Romano:2016dpx,Christensen:2018iqi}.
It is now generally accepted that there are mainly three sources for GW production during 
a cosmological FOPT: bubble wall collisions, sound waves, and 
magnetohydrodynamic (MHD) turbulence. For bubble collisions, the GW is sourced by the stress
energy located at the wall and can be understood very well both analytically~\cite{Jinno:2016vai} and numerically~\cite{Huber:2008hg} under
the envelope approximation~\cite{Kosowsky:1992rz,Kosowsky:1991ua,Kosowsky:1992vn}, where the wall is assumed to be thin and 
contribution from the overlapped regions is neglected. 
There has also been recent progress for simulations going beyond the envelope approximation~\cite{Jinno:2017fby,Child:2012qg,Cutting:2018tjt}. However,
for a phase transition proceeding in a thermal plasma, it is believed to be of negligible contribution~\cite{Bodeker:2017cim}.
A significant fraction of the energy released from the phase transition goes to the kinetic energy of the plasma, while the rest heats
up the plasma. The kinetic energy of the plasma corresponds to the velocity perturbations of the plasma, which are sound
waves in a medium consisting of relativistic particles. This relatively long-living acoustic production of GW is generally accepted to be the dominant one. 
GW spectrum from this source typically relies on large
scale lattice simulations~\cite{Hindmarsh:2013xza,Hindmarsh:2015qta,Hindmarsh:2017gnf,Cutting:2019zws}. However,
an analytical modeling reproduces the spectra from simulations reasonably well based on the sound shell model~\cite{Hindmarsh:2016lnk,Hindmarsh:2019phv} 
(see Ref.~\cite{Guo:2020grp} for the generalization to an expanding Universe), which assumes the plasma velocity field is a linear
superposition of the sound shells from all bubbles.
The fully ionized fluid can go turbulent for a sufficiently large Reynolds number and corresponds to the third 
source~\cite{Hindmarsh:2013xza,Hindmarsh:2015qta}. We will thus include only the contributions from the sound waves and the MHD turbulence, with the present 
dimensionless GW energy fraction spectrum given by
\beq
\Omega_{\text{GW}}h^2 \simeq \Omega_{\text{sw}}h^2+\Omega_{\text{turb}}h^2 ,
\eeq
where $h \approx 0.673$, the Hubble rate today $H_0$ in unit of 100 $\text{km} \text{s}^{-1} \text{Mpc}^{-1}$.
The sound wave's contribution is~\cite{Weir:2017wfa,Caprini:2015zlo}
\beq
\Omega_{\text{sw}}h^2 = 2.65\times 10^{-6} \left(\frac{H_*}{\beta}\right)
\left(\frac{\kappa_v \alpha_e}{1+\alpha_e}\right)^2 \left(\frac{100}{g_s}\right)^{\frac{1}{3}}v_w\left(\frac{f}{f_\text{sw}}\right)^3
\left[\frac{7}{4+3(f/f_\text{sw})^2}\right]^\frac{7}{2} \times \Upsilon(\tau_{\text{sw}}).
\label{Eq:GWsw}
\eeq
Here $g_s$ is the relativistic degrees of freedom for entropy; $T_{\ast}$ is the temperature right after GW production stops;
$f_{\text{sw}}$ is the present peak frequency:
\beq
f_{\text{sw}}= 1.9 \times 10^{-2}~\text{mHz}\frac{1}{v_w}\left(\frac{\beta}{H_*}\right)\left(\frac{T_*}{100~\text{GeV}}\right)\left(\frac{g_s}{100}\right)^{\frac{1}{6}}.
\eeq
Here, $\kappa_v$ can be calculated from a semi-analytical hydrodynamic analysis of the velocity profile of a single bubble for given $v_w$ and $\alpha_e$~\cite{Espinosa:2010hh}.
This determination gives a good estimate of $\kappa_v$ for relatively weak transitions, i.e., $\alpha_e \ll 1$. However, for strong transitions
and for small $v_w$, a recent simulation found that $\kappa_v$ as determined this way gives an overestimation~\cite{Cutting:2019zws}. Therefore care should be taken
when calculating $\kappa_v$ from the hydrodynamic analysis. Moreover, the multiplication factor $\Upsilon$ was only discovered in a recent 
study~\cite{Guo:2020grp} (which was also adopted in Ref.~\cite{Hindmarsh:2020hop}), and originates from the finite lifetime of the source.
\beq
\Upsilon = 1 - \frac{1}{\sqrt{1 + 2 \tau_{\text{sw}} H_{\ast}}} ,
\label{eq:upsilon}
\eeq
and the usually adopted spectrum corresponds to $\tau_{\text{sw}} \rightarrow \infty$ for which $\Upsilon$ takes the asymptotic value $1$.
However, the lifetime of the sound waves is certainly finite which leads to a suppression of the spectrum. 
We note that before the discovery of $\Upsilon$, a similar suppression factor  $\text{min}(1,\tau_{\text{sw}} H_{\ast})$ was 
adopted~\cite{Ellis:2019oqb,Ellis:2020awk,Caprini:2019egz} based on a Minkowski derivation of the spectrum~\cite{Hindmarsh:2015qta}, which
corresponds to the limit of $\Upsilon$ when $\tau_{\text{sw}} H_{\ast} \ll 1$. 
The lifetime $\tau_{\text{sw}}$ can be taken as the time scale when the turbulence develops, roughly given by~\cite{Pen:2015qta,Hindmarsh:2017gnf}:
\begin{eqnarray}
\tau_{\text{sw}} \sim \frac{R_{\ast}}{\bar{U}_f} ,
\end{eqnarray}
where $R_{\ast}$ is the mean bubble separation and is related to $\beta$ through the relation
$R_{\ast} = (8\pi)^{1/3} v_w /\beta$ for an exponential nucleation of the bubbles (see, {\it e.g.}, Ref.~\cite{Hindmarsh:2019phv} for a derivation in Minkowski spacetime and see Ref.~\cite{Guo:2020grp} for an analysis in the expanding Universe); $\bar{U}_f$ is
the root-mean-squared fluid velocity and can be determined from the hydrodynamic analysis, with the result $\bar{U}_f = \sqrt{3 \kappa_v \alpha/4}$~\cite{Hindmarsh:2019phv,Weir:2017wfa}.

\begin{figure}
	\centering
	\includegraphics[width=0.7\columnwidth]{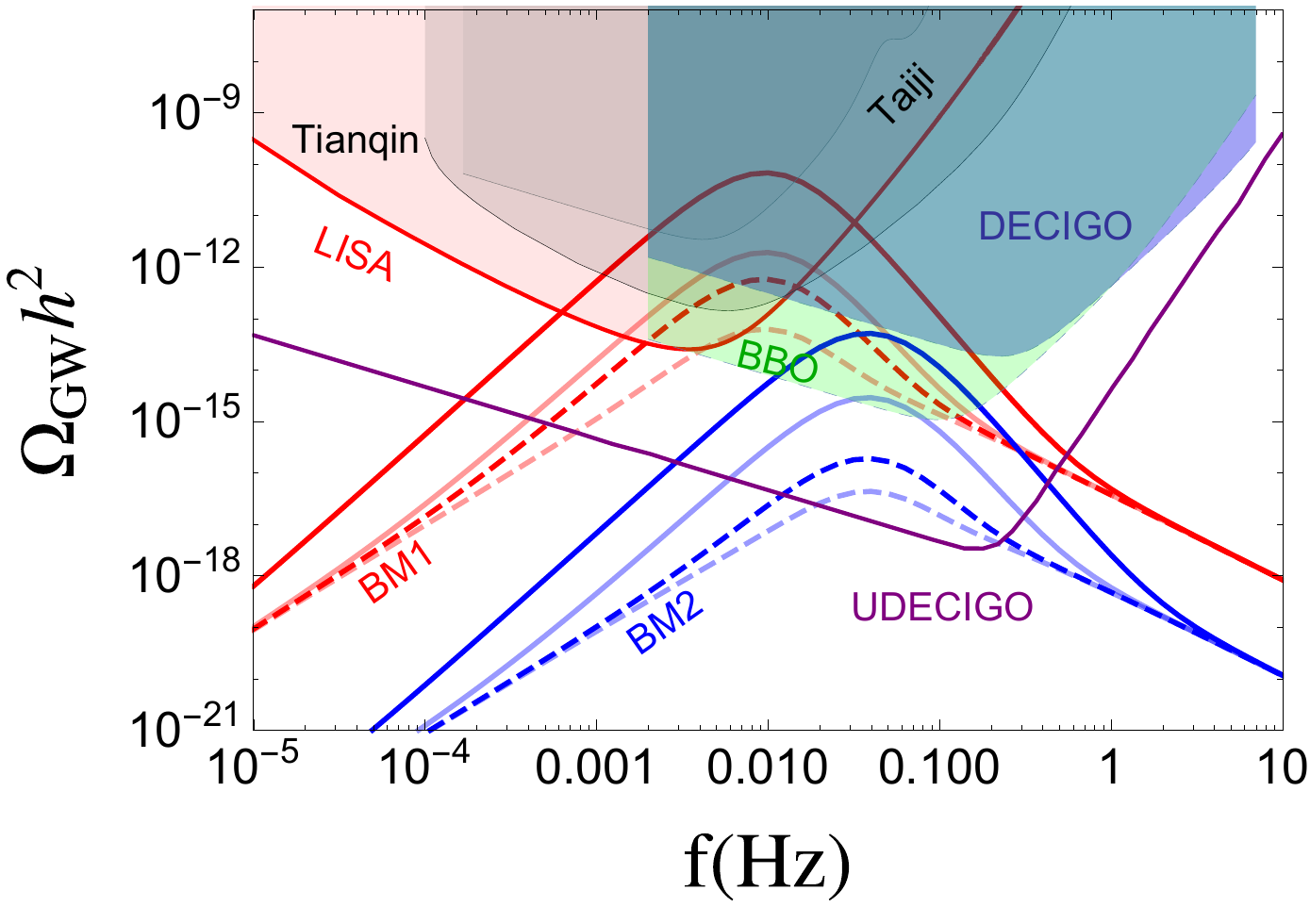}
	\caption{Gravitational wave energy spectrum versus the frequency for our two benchmark points and experimental sensitivities of some GW detectors. The lighter lines show the corresponding spectrum with the suppression factor $\Upsilon$. The dashed lines show the spectrum beyond the bag model.}
	\label{fig:GWBM}
\end{figure}

The approach above is based on the bag model. As pointed out in a recent work~\cite{Giese:2020rtr}, a more robust way to quantify the strenght of phase transition is going beyond the bag model and using the pseudotrace $\bar\theta$ to define the strength parameter $\alpha_{\bar\theta}$
\beq
\bar\theta \equiv e - p / c^2_{s,b},\quad \alpha_{\bar\theta} \equiv \frac{D \bar\theta}{3 \omega_s(T_s)},
\eeq
where the subscript $s$ ($b$) means evaluating the corresponding variables at the symmetric (broken) phase. The difference in $\bar\theta$ in two phases is defined as $D\bar\theta \equiv \bar\theta_s(T_s) - \bar\theta_b(T_s)$.
The speed of sound in the broken phase $c_{s,b}$ is related to the pressure $p$ and energy density $e$ in the broken phase
\beq
c^2_{s,b} \equiv \left.\frac{dp_b/dT}{de_b/dT}\right|_{T_s},
\eeq
and $\omega = e + p$ is the enthalpy density. Beyond the bag model, the GWs power spectrum produced by sound wave can still be obtained from Eq.~(\ref{Eq:GWsw}) by
the replacement
\beq
\frac{\alpha_e \kappa_\nu}{\alpha_e+1} \rightarrow \left(\frac{D\bar\theta}{4 e_s}\right)\kappa_{\bar\theta}.
\eeq
In Fig.~\ref{fig:GWBM}, we show the GWs spectrum within and beyond the bag model and with and without the suppression factor $\Upsilon$ defined in Eq.~(\ref{eq:upsilon}).  

The contributions from MHD turbulence can be modeled as~\cite{Caprini:2015zlo}
\beq
\Omega_{\text{turb}}h^2 =3.35\times 10^{-4} \left(\frac{H_*}{\beta}\right)\left(\frac{\kappa_\text{turb} \alpha}{1+\alpha}\right)^{\frac{3}{2}}
\left(\frac{100}{g_s}\right)^{\frac{1}{3}}v_w\frac{(f/f_\text{turb})^3}{[1+(f/f_\text{turb})]^\frac{11}{3}(1+8\pi f/H_0)},
\eeq
where $\kappa_\text{turb}$ is the energy going to turbulence and $f_{\text{turb}}$ is the present day peak frequency:
\beq
f_{\text{turb}}= 2.7 \times 10^{-2}~\text{mHz}\frac{1}{v_w}\left(\frac{\beta}{H_*}\right)\left(\frac{T_*}{100~\text{GeV}}\right)\left(\frac{g_s}{100}\right)^{\frac{1}{6}}.
\eeq
We note that the contribution from MHD is currently the least understood and might witness significant changes in the future. Indeed recent direct numerical
simulations show significantly different result~\cite{Pol:2019yex}. Also the value of $\kappa_\text{turb}$ is unknown and we take 
tentatively $\kappa_\text{turb} \approx (5\sim 10) \% \kappa_v$~\cite{Hindmarsh:2015qta} .
For both contributions, while in principle the wall velocity $v_w$ can be calculated from micro-dynamics of particle interactions with the Higgs condensate, its precise value remains
undetermined due to the theoretical uncertainties in the calculations. 
On the other hand, if baryon asymmetry were to be generated during the phase transition, then a subsonic 
value is needed. However a supersonic value of $v_w$ might still be compatible with EWBG due to the outflowing fluid around the wall~\cite{no:2011fi},
as adopted in~\cite{Alves:2018oct,Alves:2018jsw,Alves:2019igs,Alves:2020bpi}, 
though a definitive justification of this argument is still missing, which would require a thorough scrutiny of the particle transport near the wall. 
So we choose tentatively a value $v_w=0.9$. For the benchmark points in Table~\ref{Table: BMPs}, the GW spectrum are shown in Fig.~\ref{fig:GWBM}. To illustrate the suppression effect of $\Upsilon$, We present the results without considering it by the dashed lines. Some space-based interferometers sensitivities: LISA~\cite{Audley:2017drz}, Taiji~\cite{Gong:2014mca}, TianQin~\cite{Luo:2015ght}, Big Bang Observer (BBO), DECi-hertz Interferometer GW Observatory (DECIGO) and  Ultimate-DECIGO~\cite{Kudoh:2005as} are overlaid in Fig.~\ref{fig:GWBM}. To quantify the detectability of the signals, we define the signal-to-noise ratio (SNR)~\cite{Caprini:2015zlo}:
\beq
\text{SNR} = \sqrt{\delta\times \mathcal{T}\int_{f_{min}}^{f_{max}}df[\frac{h^2\Omega_{\text{GW}}(f)}{h^2\Omega_{\text{exp}}(f)}]^2},
\eeq
where $\mathcal{T}$ is the duration of the mission in years. Here we adopt $\mathcal{T} = 5$. $h^2\Omega_{\text{exp}}(f)$ denotes the experimental sensitivities as shown in Fig.~\ref{fig:GWBM}. $\delta = 2$ for BBO and UDECIGO, and $\delta = 1$ for the rest, indicating the number of independent channels for the GWs detector. The values of SNR with LISA and BBO configuration based on the bag model are
\beqa
&\text{BM1: }& \text{SNR}= 5.4\ (\text{LISA}),\,\text{SNR}= 95\ (\text{BBO}),\\
&\text{BM2: }& \text{SNR}= 6.3 \times 10^{-4}\ (\text{LISA}),\,\text{SNR}= 1.1\ (\text{BBO}).
\eeqa
The threshold value of SNR for detection is 10 or 50~\cite{Caprini:2015zlo}, and thus the BM1 can produce strong GW signal detectable at BBO, though
the corresponding values of SNR obtained using the beyond bag model are significantly reduced.

\section{Summary and conclusions}
\label{sec:conclusion}
\begin{figure}[tb]
	\centering
	\begin{subfigure}{.48\textwidth}
		\centering
		\includegraphics[width=\textwidth]{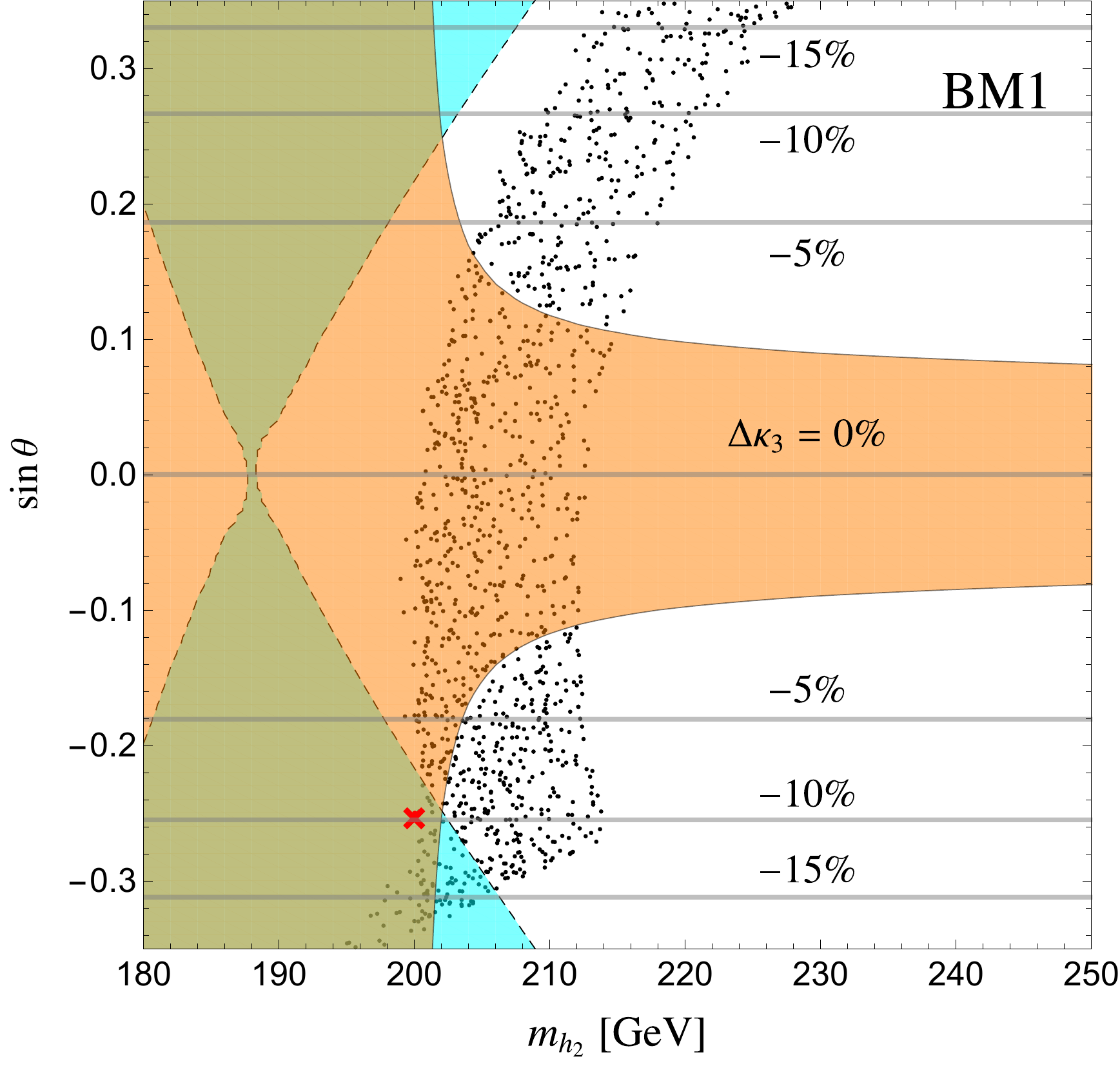}
	\end{subfigure}
	\begin{subfigure}{.48\textwidth}
		\centering
		\includegraphics[width=\textwidth]{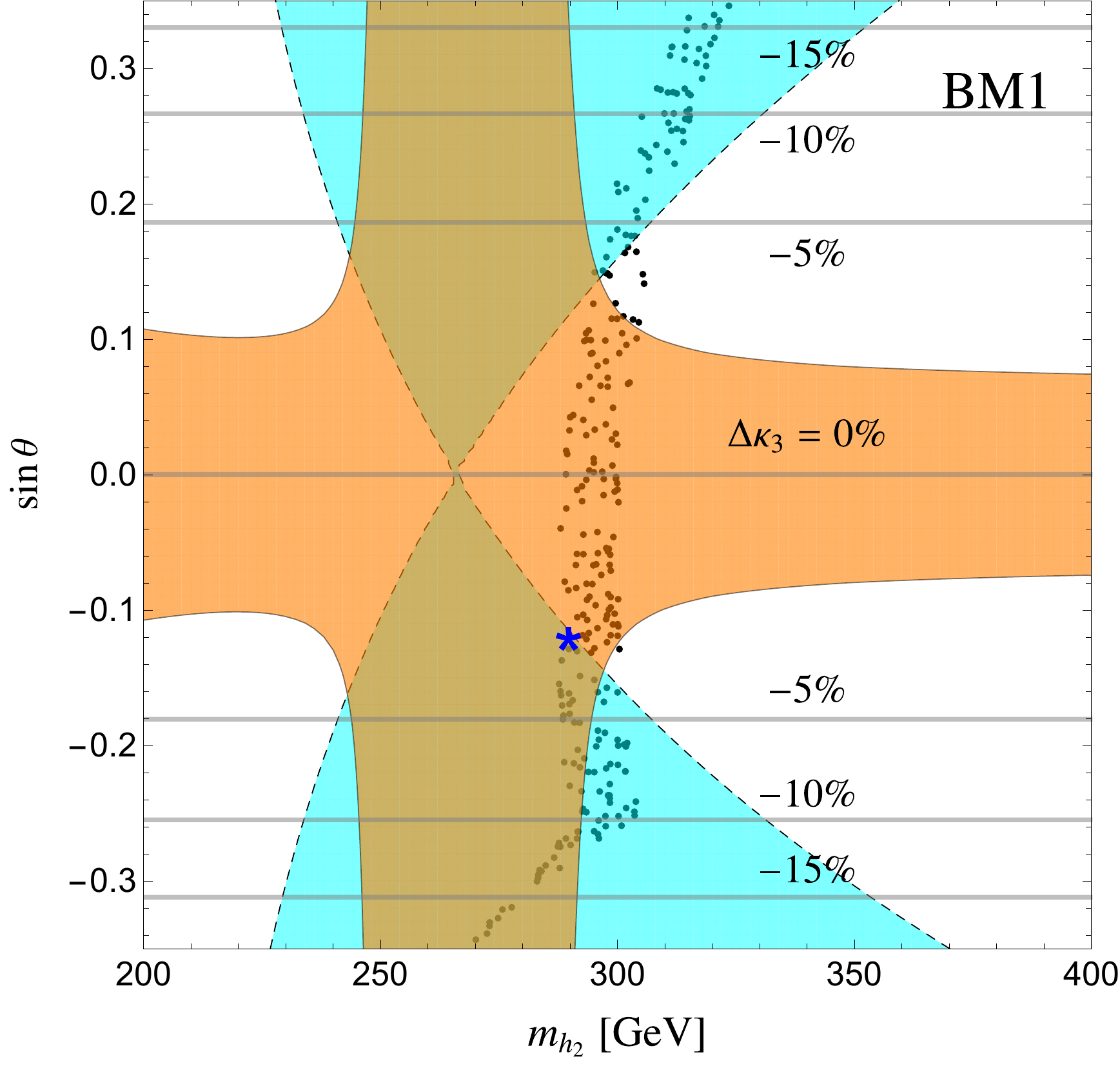}
	\end{subfigure}
	\caption{Contour plot on $\sin\theta$ - $m_{h_2}$ plane. The orange (cyan) shaded regions are allowed by DM direct detection (relic density). The dashed lines indicate the value of $\Delta \kappa_{3}$ defined in Eq.~(\ref{eq:deltaK}). The black points give strong FOPT. The red cross at left panel and blue star at right panel are our BM1 and BM2 points, respectively. }
	\label{fig:thetam}
\end{figure}

In this work, we extended the Standard Model with a dark $\text{SU(2)}_{\text{D}}$ gauge sector and a dark scalar sector, 
beyond the existing literature shown in Table \ref{Table: Models}. We imposed $Z_2$ symmetry in the dark scalar sector. We considered two different scalar scenarios under the dark $\text{SU(2)}_{\text{D}}$ gauge charge, namely, a model with two scalar triplets (TT) and one with a scalar singlet plus a scalar triplet (ST). 
The dark sector couples to the SM through the Higgs portal $-$ the mixing between the SM Higgs boson and the dark scalars. We worked out the existing constraints on the dark sector model-parameters from the vacuum stability, perturbative unitarity, Higgs physics at the LHC, and the cosmological bounds from CMB measurements and the DM relic abundance and its direct detections. For illustration, we chose two  representative benchmark points as shown in Table~\ref{Table: BMPs}, which satisfy all the constraints, possess the desirable features, and could lead to observable effects. 
We summarize our novel results as follows:
\begin{itemize}
	\item[$\bullet$] Higgs boson physics: 
	Via the Higgs portal,  the properties of the SM Higgs boson would be modified, including the couplings and an invisible decay. It is particularly interesting to test the potentially large deviation of the Higgs boson triple-self coupling from the SM prediction.
	Direct searches for the heavy Higgs boson decaying to the SM heavy particles may also be fruitful. We showed those in Figs.~\ref{fig:HVV}, ~\ref{fig:k3}, and \ref{fig:Hdecay}.
	\item[$\bullet$] DR: 
	Because of the existence of a massless DR associated with the unbroken subgroup $\text{U(1)}_{\text{D}}$, it  can introduce the velocity-dependent DM self-interaction, which would be desirable to resolve the small-scale structure problems. 
	\item[$\bullet$] DM: 
	The two stable massive gauge bosons associated with the broken dark gauge group and the pseudo-Goldstone boson can serve as cold DM candidates. The acceptable relic densities were shown in the left panel of Fig.~\ref{fig:relicdensity}. We explored the prospects of their detection in the direct DM searches as shown in the right panel of Fig.~\ref{fig:relicdensity}. 
	\item[$\bullet$] EWPT: 
	The nontrivial scalar potential has eight types of vacuum pattern for the vacuum structure as shown in Table~\ref{Table: extrema}. We have found both the two-step and three-step phase transitions with the cooling of the Universe. Due to the rich vacuum pattern, the scalar sectors can introduce a strong FOPT, as illustrated in Fig.~\ref{fig:evo} for the benchmark points BM1 with a successful EW FOPT, and BM2 with a FOPT in the dark sector. 
	\item[$\bullet$] GW: 
	Our benchmark GW spectra are shown in Fig.~\ref{fig:GWBM}. We found that the two-step EWPT in our BM1 can produce strong GW signals and can be detectable using the future space-based interferometers  BBO, while the GW signal for BM2 may be difficult to observe at BBO due to the rather low signal-to-noise ratio.   
\end{itemize}
We summarize our results on the $m_{h_2}$-$\sin\theta$ plane in Fig.~\ref{fig:thetam},
fixing the other parameters according to our BM1 (left panel) and BM2 (right panel). 
The orange shaded regions are allowed by the DM direct detections. Outside the cyan shaded regions, DM would over-close our Universe. The black points are the viable FOPT points which can enable GW production. The gray solid lines show the predicted deviation of the SM triple Higgs coupling. Our BM1 and BM2 points sit in the red-cross and blue-star symbols, respectively.

In conclusion, given the outstanding puzzles we are facing now such as the identity of the DM and the nature of the EWPT, it is prudent to consider the possibility of a dark sector uncharged under the SM interactions. We demonstrated with a well-motivated example of a dark SU(2)$\rm _D$ sector, that rich physics may exist that is potentially observable with the current and future measurements at colliders, DM experiments, and GW interferometers. 

\acknowledgments
We would like to thank Michael Ramsey-Musolf and Daniel Vagie for helpful discussions. 
The work of TH and HL was supported by the U.S.~Department of Energy under grant No.~DE-FG02- 95ER40896 and by the PITT PACC. 
TG was supported by the U.S.~Department of Energy under grant No.~DE-FG02- 95ER40896, the PITT PACC, and by the FAPESP process no.~2019/17182-0.
HG was supported by the U.S.~Department of Energy under grant No.~DE-SC0009956.

\begin{appendices}
	
	\section{Field-dependent mass}
	\label{sec:fdm}
	The field-dependent masses for the scalar degrees of freedom with $n_S = 1$ are
	\beqa
	m_{h}^2&=&m_H^2 + \frac{3\lambda_{H}}{2} h_0^2+ \frac{\lambda_{H11}}{2}\omega_3^2+ \frac{\lambda_{H22}}{2}\varphi_2^2,\\
	m_{G_0}^2&=&m_{G_1}^2=m_{G_2}^2=m_H^2 + \frac{\lambda_{H}}{2} h_0^2+ \frac{\lambda_{H11}}{2}\omega_3^2+ \frac{\lambda_{H22}}{2}\varphi_2^2,\\
	m_{\varphi_1}^2&=&m_{22}^2 + \frac{\lambda_{H22}}{2}h_0^2+ \frac{\lambda_{2}}{2}\varphi_2^2+ \frac{\lambda_{3}}{2}\omega_3^2,\\
	m_{\varphi_2}^2&=&m_{22}^2 + \frac{\lambda_{H22}}{2}h_0^2+ \frac{3\lambda_{2}}{2}\varphi_2^2+ \frac{\lambda_{3}}{2}\omega_3^2,\\
	m_{\varphi_3}^2&=&m_{22}^2 + \frac{\lambda_{H22}}{2}h_0^2+ \frac{\lambda_{2}}{2}\varphi_2^2+ \frac{\lambda_{3}+\lambda_{4}}{2}\omega_3^2,\\
	m_{\omega_1}^2&=&m_{11}^2 + \frac{\lambda_{H11}}{2}h_0^2+ \frac{\lambda_{3}}{2}\varphi_2^2+ \frac{\lambda_{1}}{2}\omega_3^2,\\
	m_{\omega_2}^2&=&m_{11}^2 + \frac{\lambda_{H11}}{2}h_0^2+ \frac{\lambda_{3}+\lambda_{4}}{2}\varphi_2^2+ \frac{\lambda_{1}}{2}\omega_3^2,\\
	m_{\omega_3}^2&=&m_{11}^2 + \frac{\lambda_{H11}}{2}h_0^2+ \frac{\lambda_{3}}{2}\varphi_2^2+ \frac{3\lambda_{1}}{2}\omega_3^2.
	\eeqa
	Similarly the field-dependent masses for the vector degrees of freedom with $n_{\tilde{W}} = n_Z =3$ and $n_W = 6$ are
	\beq
	\begin{split}
		&m_{\tilde{W}_1}^2 = \tilde{g}^2 (\varphi_1^2+\varphi_2^2),\, m_{\tilde{W}_2}^2 = \tilde{g}^2 \varphi_1^2,\, m_{\tilde{W}_3}^2 = \tilde{g}^2 \varphi_2^2,\\
		& m^2_W =\frac{g_2^2}{4}h^2,\,m^2_Z =\frac{g_2^2+g_1^2}{4}h^2.\\
	\end{split}
	\eeq
	Finally, the field-dependent masses for the fermion degrees of freedom with $n_t = 12$ is
	\beq
	m^2_t = \frac{y_t^2}{2}h^2.
	\eeq
	
	\section{Stable conditions for all the minima}
	\label{sec:sc}
	The stable conditions for the extrema in Table~\ref{Table: extrema} are
	\beqa
	&\text{Type-1 scenario: }& m_{H}^2,\,m_{11}^2,\,m_{22}^2 >0, \label{Eq:cdt1}\\
	&\text{Type-2 scenario: }& m_{H}^2<0,\,\lambda_{H11} m_{H}^2 - \lambda_{H}m^2_{11}<0, \, \lambda_{H22} m_{H}^2 - \lambda_{H}m^2_{22}<0\label{Eq:cdt2}\\
	&\text{Type-3 scenario: }& m_{11}^2<0,\,\lambda_{H11} m_{11}^2- \lambda_{1}m^2_{H}<0, \,
	\lambda_{3} m_{11}^2 - \lambda_{1}m^2_{22}<0\label{Eq:cdt3}\\
	&\text{Type-4 scenario: }& m_{22}^2<0,\,\lambda_{H22} m_{22}^2 - \lambda_{2}m^2_{H}<0, \,
	\lambda_{3} m_{22}^2 - \lambda_{2}m^2_{11}<0\label{Eq:cdt4}\\
	&\text{Type-5 scenario: }& \lambda_H \lambda_1 - \lambda_{H11}^2>0, \lambda_{H11} m_H^2- \lambda_{H} m_{11}^2>0,\,
	\lambda_{H11} m^2_{11} - \lambda_{1} m_H^2>0,\,\nonumber\\
	&&m_{22}^2 + \frac{\lambda_{3}(\lambda_{H11}m_H^2-\lambda_Hm_{11}^2)}{\lambda_H \lambda_1 - \lambda_{H11}^2}+ \frac{\lambda_{H22}(\lambda_{H11}m_{11}^2-\lambda_1m_{H}^2)}{\lambda_H \lambda_1 - \lambda_{H11}^2} > 0\label{Eq:cdt5}\\
	&\text{Type-6 scenario: }& \lambda_1 \lambda_2 - \lambda_{3}^2>0, \lambda_{3} m_2^2- \lambda_{2} m_{11}^2>0,\,
	\lambda_{3} m^2_{22} - \lambda_{2} m_{11}^2>0,\,\nonumber\\
	&&m_{H}^2 + \frac{\lambda_{H11}(\lambda_{3}m_{22}^2-\lambda_2m_{11}^2)}{\lambda_1 \lambda_2 - \lambda_{3}^2}+ \frac{\lambda_{H22}(\lambda_{3}m_{11}^2-\lambda_1m_{22}^2)}{\lambda_1 \lambda_2 - \lambda_{3}^2} > 0,\label{Eq:cdt6}\\
	&\text{Type-7 scenario: }& \lambda_H \lambda_2 - \lambda_{H22}^2>0, \lambda_{H22} m_H^2- \lambda_{H} m_{22}^2>0,\,
	\lambda_{H22} m^2_{22} - \lambda_{2} m_H^2>0,\,\nonumber\\
	&&m_{11}^2 + \frac{\lambda_{3}(\lambda_{H22}m_H^2-\lambda_Hm_{22}^2)}{\lambda_H \lambda_2 - \lambda_{H22}^2}+ \frac{\lambda_{H11}(\lambda_{H22}m_{22}^2-\lambda_2m_{H}^2)}{\lambda_H \lambda_2 - \lambda_{H22}^2} > 0.\label{Eq:cdt7}
	\eeqa
	Those cases are summarized in Table \ref{Table: extrema}. For Type-8 scenario, we refer to Ref.~\cite{Vieu:2018nfq} due to the complicity and irrelevance.

	\section{Further description for the phase transition process}
	\label{app:PT}
	The effective potential is a polynomial of the fields $(h_0, \omega_3, \varphi_2)$ 
	up to 
	quartic terms by renormalizibility. The coefficients of the quartic terms are required to be positive as the potential is bounded from below. For the quadratic terms, they can be generically put into the form 
	\beq
	V \sim D_i (T^2-T_i^2) \phi_i^2 
	\eeq
	with $\phi_i$ denoting one of the three fields. For $D_i >0$ and 
	$T > T_i$, this term remains positive and enforces a minimum at $\phi_i=0$, which is in a symmetric phase. 
	As $T$ decreases below $T_i$, the minimum at $\phi_i=0$ will roll away from the origin and takes a non-zero value, corresponding to a continuous phase transition. This is indeed what happens for the continuous transitions in Eq.~(\ref{eq:2step}) and Eq.~(\ref{eq:3step}), where the potential minimum corresponds to a non-zero field value at some temperature. The same story can happen to any of the three fields. If the parameters are such that two minima coexist across a time duration, then a first order phase transition can happen when the universe tunnels from one minimum to another, characteristic for 
	a first-order phase transition (FOPT), as shown in these two benchmarks in the text. This analytical understanding can provide a way of identifying the parameter space giving a first order phase transition, as demonstrated in~\cite{Chao:2017vrq,Espinosa:2011ax}.
	In practice, however, there is a challenge in this procedure. Whether or not a transition takes place  between two coexisting minima depends on the tunneling probability and it is sensitive to the potential shape such as the height of the barrier separating them and the potential difference
	at the two minima, which however is difficult to understand analytically (see~\cite{Chala:2019rfk,Alves:2020bpi} for relevant analyses and discussions). This presents an uncertainty for the presence of a FOPT even if we perceive the coexistence of two minima at the same time. As such, some numerical techniques, such as scanning over a large parameter space, may be unavoidable, as we did in our analyses.
	
	There are also subtleties in classifying second-order/higher-order phase transitions and a smooth cross-over. A proper classification could be specified by a dimensionless susceptibility, see, {\it e.g.} Ref.~\cite{Niemi:2020hto}. 
	
\end{appendices}

\bibliographystyle{JHEP}
\bibliography{ref,mybib}

\providecommand{\href}[2]{#2}\begingroup\raggedright\begin{thebibliography}{100}

\bibitem{Jungman:1995df}
G.~Jungman, M.~Kamionkowski, and K.~Griest, {\it {Supersymmetric dark matter}},
   {\em Phys. Rept.} {\bf 267} (1996) 195--373,
  [\href{http://arxiv.org/abs/hep-ph/9506380}{{\tt hep-ph/9506380}}].

\bibitem{Alexander:2016aln}
J.~Alexander et~al., {\it {Dark Sectors 2016 Workshop: Community Report}},  8,
  2016.
\newblock \href{http://arxiv.org/abs/1608.08632}{{\tt arXiv:1608.08632}}.

\bibitem{Freedman:2017yms}
W.~L. Freedman, {\it {Cosmology at a Crossroads}},  {\em Nature Astron.} {\bf
  1} (2017) 0121, [\href{http://arxiv.org/abs/1706.02739}{{\tt
  arXiv:1706.02739}}].

\bibitem{Spergel:1999mh}
D.~N. Spergel and P.~J. Steinhardt, {\it {Observational evidence for
  self-interacting cold dark matter}},  {\em Phys. Rev. Lett.} {\bf 84} (2000)
  3760--3763, [\href{http://arxiv.org/abs/astro-ph/9909386}{{\tt
  astro-ph/9909386}}].

\bibitem{Boehm:2001hm}
C.~Boehm, A.~Riazuelo, S.~H. Hansen, and R.~Schaeffer, {\it {Interacting dark
  matter disguised as warm dark matter}},  {\em Phys. Rev. D} {\bf 66} (2002)
  083505, [\href{http://arxiv.org/abs/astro-ph/0112522}{{\tt
  astro-ph/0112522}}].

\bibitem{Vogelsberger:2015gpr}
M.~Vogelsberger, J.~Zavala, F.-Y. Cyr-Racine, C.~Pfrommer, T.~Bringmann, and
  K.~Sigurdson, {\it {ETHOS \textendash an effective theory of structure
  formation: dark matter physics as a possible explanation of the small-scale
  CDM problems}},  {\em Mon. Not. Roy. Astron. Soc.} {\bf 460} (2016), no.~2
  1399--1416, [\href{http://arxiv.org/abs/1512.05349}{{\tt arXiv:1512.05349}}].

\bibitem{Patt:2006fw}
B.~Patt and F.~Wilczek, {\it {Higgs-field portal into hidden sectors}},
  \href{http://arxiv.org/abs/hep-ph/0605188}{{\tt hep-ph/0605188}}.

\bibitem{deFlorian:2016spz}
{\bf LHC Higgs Cross Section Working Group} Collaboration, D.~de~Florian
  et~al., {\it {Handbook of LHC Higgs Cross Sections: 4. Deciphering the Nature
  of the Higgs Sector}},  \href{http://arxiv.org/abs/1610.07922}{{\tt
  arXiv:1610.07922}}.

\bibitem{deBlas:2019rxi}
J.~de~Blas et~al., {\it {Higgs Boson Studies at Future Particle Colliders}},
  {\em JHEP} {\bf 01} (2020) 139, [\href{http://arxiv.org/abs/1905.03764}{{\tt
  arXiv:1905.03764}}].

\bibitem{Ramsey-Musolf:2019lsf}
M.~J. Ramsey-Musolf, {\it {The electroweak phase transition: a collider
  target}},  {\em JHEP} {\bf 09} (2020) 179,
  [\href{http://arxiv.org/abs/1912.07189}{{\tt arXiv:1912.07189}}].

\bibitem{Mazumdar:2018dfl}
A.~Mazumdar and G.~White, {\it {Cosmic phase transitions: their applications
  and experimental signatures}},  \href{http://arxiv.org/abs/1811.01948}{{\tt
  arXiv:1811.01948}}.

\bibitem{Huang:2016cjm}
P.~Huang, A.~J. Long, and L.-T. Wang, {\it {Probing the Electroweak Phase
  Transition with Higgs Factories and Gravitational Waves}},  {\em Phys. Rev.}
  {\bf D94} (2016), no.~7 075008, [\href{http://arxiv.org/abs/1608.06619}{{\tt
  arXiv:1608.06619}}].

\bibitem{Kuzmin:1985mm}
V.~Kuzmin, V.~Rubakov, and M.~Shaposhnikov, {\it {On the Anomalous Electroweak
  Baryon Number Nonconservation in the Early Universe}},  {\em Phys. Lett. B}
  {\bf 155} (1985) 36.

\bibitem{Shaposhnikov:1986jp}
M.~Shaposhnikov, {\it {Possible Appearance of the Baryon Asymmetry of the
  Universe in an Electroweak Theory}},  {\em JETP Lett.} {\bf 44} (1986)
  465--468.

\bibitem{Shaposhnikov:1987tw}
M.~Shaposhnikov, {\it {Baryon Asymmetry of the Universe in Standard Electroweak
  Theory}},  {\em Nucl. Phys. B} {\bf 287} (1987) 757--775.

\bibitem{Morrissey:2012db}
D.~E. Morrissey and M.~J. Ramsey-Musolf, {\it {Electroweak baryogenesis}},
  {\em New J. Phys.} {\bf 14} (2012) 125003,
  [\href{http://arxiv.org/abs/1206.2942}{{\tt arXiv:1206.2942}}].

\bibitem{Cline:2006ts}
J.~M. Cline, {\it {Baryogenesis}},  in {\em {Les Houches Summer School -
  Session 86: Particle Physics and Cosmology: The Fabric of Spacetime Les
  Houches, France, July 31-August 25, 2006}}, 2006.
\newblock \href{http://arxiv.org/abs/hep-ph/0609145}{{\tt hep-ph/0609145}}.

\bibitem{White:2016nbo}
G.~A. White, {\it {A Pedagogical Introduction to Electroweak Baryogenesis}}, .

\bibitem{Sakharov:1967dj}
A.~D. Sakharov, {\it {Violation of CP Invariance, C asymmetry, and baryon
  asymmetry of the universe}},  {\em Pisma Zh. Eksp. Teor. Fiz.} {\bf 5} (1967)
  32--35. [Usp. Fiz. Nauk161,no.5,61(1991)].

\bibitem{Patel:2012pi}
H.~H. Patel and M.~J. Ramsey-Musolf, {\it {Stepping Into Electroweak Symmetry
  Breaking: Phase Transitions and Higgs Phenomenology}},  {\em Phys. Rev.} {\bf
  D88} (2013) 035013, [\href{http://arxiv.org/abs/1212.5652}{{\tt
  arXiv:1212.5652}}].

\bibitem{Niemi:2018asa}
L.~Niemi, H.~H. Patel, M.~J. Ramsey-Musolf, T.~V. Tenkanen, and D.~J. Weir,
  {\it {Electroweak phase transition in the real triplet extension of the SM:
  Dimensional reduction}},  {\em Phys. Rev. D} {\bf 100} (2019), no.~3 035002,
  [\href{http://arxiv.org/abs/1802.10500}{{\tt arXiv:1802.10500}}].

\bibitem{Bell:2020hnr}
N.~F. Bell, M.~J. Dolan, L.~S. Friedrich, M.~J. Ramsey-Musolf, and R.~R.
  Volkas, {\it {A Real Triplet-Singlet Extended Standard Model: Dark Matter and
  Collider Phenomenology}},  \href{http://arxiv.org/abs/2010.13376}{{\tt
  arXiv:2010.13376}}.

\bibitem{Zhou:2020idp}
L.~Bian, H.-K. Guo, Y.~Wu, and R.~Zhou, {\it {Gravitational wave and collider
  searches for electroweak symmetry breaking patterns}},  {\em Phys. Rev. D}
  {\bf 101} (2020), no.~3 035011, [\href{http://arxiv.org/abs/1906.11664}{{\tt
  arXiv:1906.11664}}].

\bibitem{Huang:2017rzf}
F.~P. Huang and J.-H. Yu, {\it {Exploring inert dark matter blind spots with
  gravitational wave signatures}},  {\em Phys. Rev. D} {\bf 98} (2018), no.~9
  095022, [\href{http://arxiv.org/abs/1704.04201}{{\tt arXiv:1704.04201}}].

\bibitem{Basler:2016obg}
P.~Basler, M.~Krause, M.~Muhlleitner, J.~Wittbrodt, and A.~Wlotzka, {\it
  {Strong First Order Electroweak Phase Transition in the CP-Conserving 2HDM
  Revisited}},  {\em JHEP} {\bf 02} (2017) 121,
  [\href{http://arxiv.org/abs/1612.04086}{{\tt arXiv:1612.04086}}].

\bibitem{Dorsch:2017nza}
G.~Dorsch, S.~Huber, K.~Mimasu, and J.~No, {\it {The Higgs Vacuum Uplifted:
  Revisiting the Electroweak Phase Transition with a Second Higgs Doublet}},
  {\em JHEP} {\bf 12} (2017) 086, [\href{http://arxiv.org/abs/1705.09186}{{\tt
  arXiv:1705.09186}}].

\bibitem{Bernon:2017jgv}
J.~Bernon, L.~Bian, and Y.~Jiang, {\it {A new insight into the phase transition
  in the early Universe with two Higgs doublets}},  {\em JHEP} {\bf 05} (2018)
  151, [\href{http://arxiv.org/abs/1712.08430}{{\tt arXiv:1712.08430}}].

\bibitem{Gorda:2018hvi}
T.~Gorda, A.~Helset, L.~Niemi, T.~V. Tenkanen, and D.~J. Weir, {\it
  {Three-dimensional effective theories for the two Higgs doublet model at high
  temperature}},  {\em JHEP} {\bf 02} (2019) 081,
  [\href{http://arxiv.org/abs/1802.05056}{{\tt arXiv:1802.05056}}].

\bibitem{Wang:2019pet}
X.~Wang, F.~P. Huang, and X.~Zhang, {\it {Gravitational wave and collider
  signals in complex two-Higgs doublet model with dynamical CP-violation at
  finite temperature}},  {\em Phys. Rev. D} {\bf 101} (2020), no.~1 015015,
  [\href{http://arxiv.org/abs/1909.02978}{{\tt arXiv:1909.02978}}].

\bibitem{Su:2020pjw}
W.~Su, A.~G. Williams, and M.~Zhang, {\it {Strong first order electroweak phase
  transition in 2HDM confronting future Z \& Higgs factories}},
  \href{http://arxiv.org/abs/2011.04540}{{\tt arXiv:2011.04540}}.

\bibitem{Brdar:2019fur}
V.~Brdar, L.~Graf, A.~J. Helmboldt, and X.-J. Xu, {\it {Gravitational Waves as
  a Probe of Left-Right Symmetry Breaking}},  {\em JCAP} {\bf 12} (2019) 027,
  [\href{http://arxiv.org/abs/1909.02018}{{\tt arXiv:1909.02018}}].

\bibitem{Huang:2014ifa}
W.~Huang, Z.~Kang, J.~Shu, P.~Wu, and J.~M. Yang, {\it {New insights in the
  electroweak phase transition in the NMSSM}},  {\em Phys. Rev.} {\bf D91}
  (2015), no.~2 025006, [\href{http://arxiv.org/abs/1405.1152}{{\tt
  arXiv:1405.1152}}].

\bibitem{Bian:2017wfv}
L.~Bian, H.-K. Guo, and J.~Shu, {\it {Gravitational Waves, baryon asymmetry of
  the universe and electric dipole moment in the CP-violating NMSSM}},  {\em
  Chin. Phys.} {\bf C42} (2018), no.~9 093106,
  [\href{http://arxiv.org/abs/1704.02488}{{\tt arXiv:1704.02488}}].

\bibitem{Athron:2019teq}
P.~Athron, C.~Balazs, A.~Fowlie, G.~Pozzo, G.~White, and Y.~Zhang, {\it {Strong
  first-order phase transitions in the NMSSM \textemdash{} a comprehensive
  survey}},  {\em JHEP} {\bf 11} (2019) 151,
  [\href{http://arxiv.org/abs/1908.11847}{{\tt arXiv:1908.11847}}].

\bibitem{Akula:2017yfr}
S.~Akula, C.~Bal\'azs, L.~Dunn, and G.~White, {\it {Electroweak baryogenesis in
  the $ {\mathbb{Z}}_3 $ -invariant NMSSM}},  {\em JHEP} {\bf 11} (2017) 051,
  [\href{http://arxiv.org/abs/1706.09898}{{\tt arXiv:1706.09898}}].

\bibitem{Baum:2020vfl}
S.~Baum, M.~Carena, N.~R. Shah, C.~E. Wagner, and Y.~Wang, {\it {Nucleation is
  More than Critical}: {A Case Study of the Electroweak Phase Transition in the
  NMSSM}},  \href{http://arxiv.org/abs/2009.10743}{{\tt arXiv:2009.10743}}.

\bibitem{Barger:2007im}
V.~Barger, P.~Langacker, M.~McCaskey, M.~J. Ramsey-Musolf, and G.~Shaughnessy,
  {\it {LHC Phenomenology of an Extended Standard Model with a Real Scalar
  Singlet}},  {\em Phys. Rev. D} {\bf 77} (2008) 035005,
  [\href{http://arxiv.org/abs/0706.4311}{{\tt arXiv:0706.4311}}].

\bibitem{Profumo:2007wc}
S.~Profumo, M.~J. Ramsey-Musolf, and G.~Shaughnessy, {\it {Singlet Higgs
  phenomenology and the electroweak phase transition}},  {\em JHEP} {\bf 08}
  (2007) 010, [\href{http://arxiv.org/abs/0705.2425}{{\tt arXiv:0705.2425}}].

\bibitem{Profumo:2014opa}
S.~Profumo, M.~J. Ramsey-Musolf, C.~L. Wainwright, and P.~Winslow, {\it
  {Singlet-catalyzed electroweak phase transitions and precision Higgs boson
  studies}},  {\em Phys. Rev.} {\bf D91} (2015), no.~3 035018,
  [\href{http://arxiv.org/abs/1407.5342}{{\tt arXiv:1407.5342}}].

\bibitem{Huang:2015tdv}
P.~Huang, A.~Joglekar, B.~Li, and C.~E.~M. Wagner, {\it {Probing the
  Electroweak Phase Transition at the LHC}},  {\em Phys. Rev.} {\bf D93}
  (2016), no.~5 055049, [\href{http://arxiv.org/abs/1512.00068}{{\tt
  arXiv:1512.00068}}].

\bibitem{Kotwal:2016tex}
A.~V. Kotwal, M.~J. Ramsey-Musolf, J.~M. No, and P.~Winslow, {\it
  {Singlet-catalyzed electroweak phase transitions in the 100 TeV frontier}},
  {\em Phys. Rev.} {\bf D94} (2016), no.~3 035022,
  [\href{http://arxiv.org/abs/1605.06123}{{\tt arXiv:1605.06123}}].

\bibitem{Chen:2017qcz}
C.-Y. Chen, J.~Kozaczuk, and I.~M. Lewis, {\it {Non-resonant Collider
  Signatures of a Singlet-Driven Electroweak Phase Transition}},  {\em JHEP}
  {\bf 08} (2017) 096, [\href{http://arxiv.org/abs/1704.05844}{{\tt
  arXiv:1704.05844}}].

\bibitem{Ellis:2018mja}
J.~Ellis, M.~Lewicki, and J.~M. No, {\it {On the Maximal Strength of a
  First-Order Electroweak Phase Transition and its Gravitational Wave Signal}},
   {\em Submitted to: JCAP} (2018) [\href{http://arxiv.org/abs/1809.08242}{{\tt
  arXiv:1809.08242}}].

\bibitem{Gould:2019qek}
O.~Gould, J.~Kozaczuk, L.~Niemi, M.~J. Ramsey-Musolf, T.~V. Tenkanen, and D.~J.
  Weir, {\it {Nonperturbative analysis of the gravitational waves from a
  first-order electroweak phase transition}},  {\em Phys. Rev. D} {\bf 100}
  (2019), no.~11 115024, [\href{http://arxiv.org/abs/1903.11604}{{\tt
  arXiv:1903.11604}}].

\bibitem{Alves:2019igs}
A.~Alves, D.~Gonçalves, T.~Ghosh, H.-K. Guo, and K.~Sinha, {\it {Di-Higgs
  Production in the $4b$ Channel and Gravitational Wave Complementarity}},
  {\em JHEP} {\bf 03} (2020) 053, [\href{http://arxiv.org/abs/1909.05268}{{\tt
  arXiv:1909.05268}}].

\bibitem{Alves:2020bpi}
A.~Alves, D.~Gon\c{c}alves, T.~Ghosh, H.-K. Guo, and K.~Sinha, {\it {Di-Higgs
  Blind Spots in Gravitational Wave Signals}},
  \href{http://arxiv.org/abs/2007.15654}{{\tt arXiv:2007.15654}}.

\bibitem{Alves:2018jsw}
A.~Alves, T.~Ghosh, H.-K. Guo, K.~Sinha, and D.~Vagie, {\it {Collider and
  Gravitational Wave Complementarity in Exploring the Singlet Extension of the
  Standard Model}},  {\em JHEP} {\bf 04} (2019) 052,
  [\href{http://arxiv.org/abs/1812.09333}{{\tt arXiv:1812.09333}}].

\bibitem{Carena:2019une}
M.~Carena, Z.~Liu, and Y.~Wang, {\it {Electroweak phase transition with
  spontaneous Z$_{2}$-breaking}},  {\em JHEP} {\bf 08} (2020) 107,
  [\href{http://arxiv.org/abs/1911.10206}{{\tt arXiv:1911.10206}}].

\bibitem{Chiang:2020yym}
C.-W. Chiang, D.~Huang, and B.-Q. Lu, {\it {Electroweak phase transition
  confronted with dark matter detection constraints}},
  \href{http://arxiv.org/abs/2009.08635}{{\tt arXiv:2009.08635}}.

\bibitem{Chao:2017vrq}
W.~Chao, H.-K. Guo, and J.~Shu, {\it {Gravitational Wave Signals of Electroweak
  Phase Transition Triggered by Dark Matter}},  {\em JCAP} {\bf 1709} (2017),
  no.~09 009, [\href{http://arxiv.org/abs/1702.02698}{{\tt arXiv:1702.02698}}].

\bibitem{Gonderinger:2012rd}
M.~Gonderinger, H.~Lim, and M.~J. Ramsey-Musolf, {\it {Complex Scalar Singlet
  Dark Matter: Vacuum Stability and Phenomenology}},  {\em Phys. Rev. D} {\bf
  86} (2012) 043511, [\href{http://arxiv.org/abs/1202.1316}{{\tt
  arXiv:1202.1316}}].

\bibitem{Chiang:2017nmu}
C.-W. Chiang, M.~J. Ramsey-Musolf, and E.~Senaha, {\it {Standard Model with a
  Complex Scalar Singlet: Cosmological Implications and Theoretical
  Considerations}},  {\em Phys. Rev. D} {\bf 97} (2018), no.~1 015005,
  [\href{http://arxiv.org/abs/1707.09960}{{\tt arXiv:1707.09960}}].

\bibitem{Cheng:2018ajh}
W.~Cheng and L.~Bian, {\it {From inflation to cosmological electroweak phase
  transition with a complex scalar singlet}},  {\em Phys. Rev.} {\bf D98}
  (2018), no.~2 023524, [\href{http://arxiv.org/abs/1801.00662}{{\tt
  arXiv:1801.00662}}].

\bibitem{Chiang:2019oms}
C.-W. Chiang and B.-Q. Lu, {\it {First-order electroweak phase transition in a
  complex singlet model with $\mathbb{Z}_3$ symmetry}},  {\em JHEP} {\bf 07}
  (2020) 082, [\href{http://arxiv.org/abs/1912.12634}{{\tt arXiv:1912.12634}}].

\bibitem{Bhoonah:2020oov}
A.~Bhoonah, J.~Bramante, S.~Nerval, and N.~Song, {\it {Gravitational Waves From
  Dark Sectors, Oscillating Inflatons, and Mass Boosted Dark Matter}},
  \href{http://arxiv.org/abs/2008.12306}{{\tt arXiv:2008.12306}}.

\bibitem{Jaeckel:2016jlh}
J.~Jaeckel, V.~V. Khoze, and M.~Spannowsky, {\it {Hearing the signal of dark
  sectors with gravitational wave detectors}},  {\em Phys. Rev. D} {\bf 94}
  (2016), no.~10 103519, [\href{http://arxiv.org/abs/1602.03901}{{\tt
  arXiv:1602.03901}}].

\bibitem{Addazi:2017gpt}
A.~Addazi and A.~Marciano, {\it {Gravitational waves from dark first order
  phase transitions and dark photons}},  {\em Chin. Phys. C} {\bf 42} (2018),
  no.~2 023107, [\href{http://arxiv.org/abs/1703.03248}{{\tt
  arXiv:1703.03248}}].

\bibitem{Chala:2016ykx}
M.~Chala, G.~Nardini, and I.~Sobolev, {\it {Unified explanation for dark matter
  and electroweak baryogenesis with direct detection and gravitational wave
  signatures}},  {\em Phys. Rev. D} {\bf 94} (2016), no.~5 055006,
  [\href{http://arxiv.org/abs/1605.08663}{{\tt arXiv:1605.08663}}].

\bibitem{Addazi:2020zcj}
A.~Addazi, Y.-F. Cai, Q.~Gan, A.~Marciano, and K.~Zeng, {\it {NANOGrav results
  and Dark First Order Phase Transitions}},
  \href{http://arxiv.org/abs/2009.10327}{{\tt arXiv:2009.10327}}.

\bibitem{Hambye:2008bq}
T.~Hambye, {\it {Hidden vector dark matter}},  {\em JHEP} {\bf 01} (2009) 028,
  [\href{http://arxiv.org/abs/0811.0172}{{\tt arXiv:0811.0172}}].

\bibitem{Boehm:2014bia}
C.~Boehm, M.~J. Dolan, and C.~McCabe, {\it {A weighty interpretation of the
  Galactic Centre excess}},  {\em Phys. Rev. D} {\bf 90} (2014), no.~2 023531,
  [\href{http://arxiv.org/abs/1404.4977}{{\tt arXiv:1404.4977}}].

\bibitem{Gross:2015cwa}
C.~Gross, O.~Lebedev, and Y.~Mambrini, {\it {Non-Abelian gauge fields as dark
  matter}},  {\em JHEP} {\bf 08} (2015) 158,
  [\href{http://arxiv.org/abs/1505.07480}{{\tt arXiv:1505.07480}}].

\bibitem{Baek:2013dwa}
S.~Baek, P.~Ko, and W.-I. Park, {\it {Hidden sector monopole, vector dark
  matter and dark radiation with Higgs portal}},  {\em JCAP} {\bf 10} (2014)
  067, [\href{http://arxiv.org/abs/1311.1035}{{\tt arXiv:1311.1035}}].

\bibitem{Khoze:2014woa}
V.~V. Khoze and G.~Ro, {\it {Dark matter monopoles, vectors and photons}},
  {\em JHEP} {\bf 10} (2014) 061, [\href{http://arxiv.org/abs/1406.2291}{{\tt
  arXiv:1406.2291}}].

\bibitem{Daido:2019tbm}
R.~Daido, S.-Y. Ho, and F.~Takahashi, {\it {Hidden monopole dark matter via
  axion portal and its implications for direct detection searches, beam-dump
  experiments, and the H$_{0}$ tension}},  {\em JHEP} {\bf 01} (2020) 185,
  [\href{http://arxiv.org/abs/1909.03627}{{\tt arXiv:1909.03627}}].

\bibitem{Ko:2020qlt}
P.~Ko, T.~Nomura, and H.~Okada, {\it {Dark matter physics in dark $SU(2)$ gauge
  symmetry with non-Abelian kinetic mixing}},
  \href{http://arxiv.org/abs/2007.08153}{{\tt arXiv:2007.08153}}.

\bibitem{Chen:2015nea}
C.-H. Chen and T.~Nomura, {\it {$SU(2)_X$ vector DM and Galactic Center
  gamma-ray excess}},  {\em Phys. Lett. B} {\bf 746} (2015) 351--358,
  [\href{http://arxiv.org/abs/1501.07413}{{\tt arXiv:1501.07413}}].

\bibitem{Chen:2015dea}
C.-H. Chen and T.~Nomura, {\it {Searching for vector dark matter via Higgs
  portal at the LHC}},  {\em Phys. Rev. D} {\bf 93} (2016), no.~7 074019,
  [\href{http://arxiv.org/abs/1507.00886}{{\tt arXiv:1507.00886}}].

\bibitem{Chiang:2013kqa}
C.-W. Chiang, T.~Nomura, and J.~Tandean, {\it {Nonabelian Dark Matter with
  Resonant Annihilation}},  {\em JHEP} {\bf 01} (2014) 183,
  [\href{http://arxiv.org/abs/1306.0882}{{\tt arXiv:1306.0882}}].

\bibitem{Hall:2019ank}
E.~Hall, T.~Konstandin, R.~McGehee, H.~Murayama, and G.~Servant, {\it
  {Baryogenesis From a Dark First-Order Phase Transition}},  {\em JHEP} {\bf
  04} (2020) 042, [\href{http://arxiv.org/abs/1910.08068}{{\tt
  arXiv:1910.08068}}].

\bibitem{Arcadi:2016kmk}
G.~Arcadi, C.~Gross, O.~Lebedev, Y.~Mambrini, S.~Pokorski, and T.~Toma, {\it
  {Multicomponent Dark Matter from Gauge Symmetry}},  {\em JHEP} {\bf 12}
  (2016) 081, [\href{http://arxiv.org/abs/1611.00365}{{\tt arXiv:1611.00365}}].

\bibitem{Tsumura:2017knk}
K.~Tsumura, M.~Yamada, and Y.~Yamaguchi, {\it {Gravitational wave from dark
  sector with dark pion}},  {\em JCAP} {\bf 07} (2017) 044,
  [\href{http://arxiv.org/abs/1704.00219}{{\tt arXiv:1704.00219}}].

\bibitem{Aoki:2017aws}
M.~Aoki, H.~Goto, and J.~Kubo, {\it {Gravitational Waves from Hidden QCD Phase
  Transition}},  {\em Phys. Rev.} {\bf D96} (2017), no.~7 075045,
  [\href{http://arxiv.org/abs/1709.07572}{{\tt arXiv:1709.07572}}].

\bibitem{Addazi:2016fbj}
A.~Addazi, {\it {Limiting First Order Phase Transitions in Dark Gauge Sectors
  from Gravitational Waves experiments}},  {\em Mod. Phys. Lett. A} {\bf 32}
  (2017), no.~08 1750049, [\href{http://arxiv.org/abs/1607.08057}{{\tt
  arXiv:1607.08057}}].

\bibitem{Archer-Smith:2019gzq}
P.~Archer-Smith, D.~Linthorne, and D.~Stolarski, {\it {Gravitational Wave
  Signals from Multiple Hidden Sectors}},  {\em Phys. Rev. D} {\bf 101} (2020),
  no.~9 095016, [\href{http://arxiv.org/abs/1910.02083}{{\tt
  arXiv:1910.02083}}].

\bibitem{Espinosa:2011ax}
J.~R. Espinosa, T.~Konstandin, and F.~Riva, {\it {Strong Electroweak Phase
  Transitions in the Standard Model with a Singlet}},  {\em Nucl. Phys. B} {\bf
  854} (2012) 592--630, [\href{http://arxiv.org/abs/1107.5441}{{\tt
  arXiv:1107.5441}}].

\bibitem{Saikawa:2017hiv}
K.~Saikawa, {\it {A review of gravitational waves from cosmic domain walls}},
  {\em Universe} {\bf 3} (2017), no.~2 40,
  [\href{http://arxiv.org/abs/1703.02576}{{\tt arXiv:1703.02576}}].

\bibitem{Arhrib:2011uy}
A.~Arhrib, R.~Benbrik, M.~Chabab, G.~Moultaka, M.~Peyranere, L.~Rahili, and
  J.~Ramadan, {\it {The Higgs Potential in the Type II Seesaw Model}},  {\em
  Phys. Rev. D} {\bf 84} (2011) 095005,
  [\href{http://arxiv.org/abs/1105.1925}{{\tt arXiv:1105.1925}}].

\bibitem{Poulin:2018kap}
A.~Poulin and S.~Godfrey, {\it {Multicomponent dark matter from a hidden gauged
  SU(3)}},  {\em Phys. Rev. D} {\bf 99} (2019), no.~7 076008,
  [\href{http://arxiv.org/abs/1808.04901}{{\tt arXiv:1808.04901}}].

\bibitem{Lopez-Val:2014jva}
D.~López-Val and T.~Robens, {\it {$\Delta$r and the W-boson mass in the
  singlet extension of the standard model}},  {\em Phys. Rev. D} {\bf 90}
  (2014) 114018, [\href{http://arxiv.org/abs/1406.1043}{{\tt
  arXiv:1406.1043}}].

\bibitem{Bian:2014cja}
L.~Bian, T.~Li, J.~Shu, and X.-C. Wang, {\it {Two component dark matter with
  multi-Higgs portals}},  {\em JHEP} {\bf 03} (2015) 126,
  [\href{http://arxiv.org/abs/1412.5443}{{\tt arXiv:1412.5443}}].

\bibitem{Robens:2015gla}
T.~Robens and T.~Stefaniak, {\it {Status of the Higgs Singlet Extension of the
  Standard Model after LHC Run 1}},  {\em Eur. Phys. J. C} {\bf 75} (2015) 104,
  [\href{http://arxiv.org/abs/1501.02234}{{\tt arXiv:1501.02234}}].

\bibitem{Aad:2015pla}
{\bf ATLAS} Collaboration, G.~Aad et~al., {\it {Constraints on new phenomena
  via Higgs boson couplings and invisible decays with the ATLAS detector}},
  {\em JHEP} {\bf 11} (2015) 206, [\href{http://arxiv.org/abs/1509.00672}{{\tt
  arXiv:1509.00672}}].

\bibitem{Falkowski:2015iwa}
A.~Falkowski, C.~Gross, and O.~Lebedev, {\it {A second Higgs from the Higgs
  portal}},  {\em JHEP} {\bf 05} (2015) 057,
  [\href{http://arxiv.org/abs/1502.01361}{{\tt arXiv:1502.01361}}].

\bibitem{Aaboud:2018bun}
{\bf ATLAS} Collaboration, M.~Aaboud et~al., {\it {Combination of searches for
  heavy resonances decaying into bosonic and leptonic final states using 36
  fb$^{-1}$ of proton-proton collision data at $\sqrt{s} = 13$ TeV with the
  ATLAS detector}},  {\em Phys. Rev. D} {\bf 98} (2018), no.~5 052008,
  [\href{http://arxiv.org/abs/1808.02380}{{\tt arXiv:1808.02380}}].

\bibitem{Huang:2017jws}
T.~Huang, J.~M. No, L.~Pernié, M.~Ramsey-Musolf, A.~Safonov, M.~Spannowsky,
  and P.~Winslow, {\it {Resonant di-Higgs boson production in the $b{\bar b}WW$
  channel: Probing the electroweak phase transition at the LHC}},  {\em Phys.
  Rev.} {\bf D96} (2017), no.~3 035007,
  [\href{http://arxiv.org/abs/1701.04442}{{\tt arXiv:1701.04442}}].

\bibitem{Adhikari:2020vqo}
S.~Adhikari, I.~M. Lewis, and M.~Sullivan, {\it {Beyond the Standard Model
  Effective Field Theory: The Singlet Extended Standard Model}},
  \href{http://arxiv.org/abs/2003.10449}{{\tt arXiv:2003.10449}}.

\bibitem{Baek:2011aa}
S.~Baek, P.~Ko, and W.-I. Park, {\it {Search for the Higgs portal to a singlet
  fermionic dark matter at the LHC}},  {\em JHEP} {\bf 02} (2012) 047,
  [\href{http://arxiv.org/abs/1112.1847}{{\tt arXiv:1112.1847}}].

\bibitem{ATLAS:2017muo}
{\bf ATLAS} Collaboration, {\it {Study of the double Higgs production channel
  $H(\rightarrow b\bar{b})H(\rightarrow \gamma\gamma)$ with the ATLAS
  experiment at the HL-LHC}}, .

\bibitem{CMS:2017cwx}
{\bf CMS} Collaboration, {\it {Projected performance of Higgs analyses at the
  HL-LHC for ECFA 2016}}, .

\bibitem{Tian:2013yda}
{\bf ILD} Collaboration, J.~Tian and K.~Fujii, {\it {Measurement of Higgs
  couplings and self-coupling at the ILC}},  {\em PoS} {\bf EPS-HEP2013} (2013)
  316, [\href{http://arxiv.org/abs/1311.6528}{{\tt arXiv:1311.6528}}].

\bibitem{Roloff:2019crr}
{\bf CLICdp} Collaboration, P.~Roloff, U.~Schnoor, R.~Simoniello, and B.~Xu,
  {\it {Double Higgs boson production and Higgs self-coupling extraction at
  CLIC}},  {\em Eur. Phys. J. C} {\bf 80} (2020), no.~11 1010,
  [\href{http://arxiv.org/abs/1901.05897}{{\tt arXiv:1901.05897}}].

\bibitem{Benedikt:2018csr}
{\bf FCC} Collaboration, A.~Abada et~al., {\it {FCC-hh: The Hadron Collider}:
  {Future Circular Collider Conceptual Design Report Volume 3}},  {\em Eur.
  Phys. J. ST} {\bf 228} (2019), no.~4 755--1107.

\bibitem{Han:2020pif}
T.~Han, D.~Liu, I.~Low, and X.~Wang, {\it {Electroweak couplings of the Higgs
  boson at a multi-TeV muon collider}},  {\em Phys. Rev. D} {\bf 103} (2021),
  no.~1 013002, [\href{http://arxiv.org/abs/2008.12204}{{\tt
  arXiv:2008.12204}}].

\bibitem{Feng:2008mu}
J.~L. Feng, H.~Tu, and H.-B. Yu, {\it {Thermal Relics in Hidden Sectors}},
  {\em JCAP} {\bf 10} (2008) 043, [\href{http://arxiv.org/abs/0808.2318}{{\tt
  arXiv:0808.2318}}].

\bibitem{Akrami:2018vks}
{\bf Planck} Collaboration, Y.~Akrami et~al., {\it {Planck 2018 results. I.
  Overview and the cosmological legacy of Planck}},
  \href{http://arxiv.org/abs/1807.06205}{{\tt arXiv:1807.06205}}.

\bibitem{Aghanim:2018eyx}
{\bf Planck} Collaboration, N.~Aghanim et~al., {\it {Planck 2018 results. VI.
  Cosmological parameters}},  \href{http://arxiv.org/abs/1807.06209}{{\tt
  arXiv:1807.06209}}.

\bibitem{Abazajian:2019eic}
K.~Abazajian et~al., {\it {CMB-S4 Science Case, Reference Design, and Project
  Plan}},  \href{http://arxiv.org/abs/1907.04473}{{\tt arXiv:1907.04473}}.

\bibitem{Ade:2015xua}
{\bf Planck} Collaboration, P.~Ade et~al., {\it {Planck 2015 results. XIII.
  Cosmological parameters}},  {\em Astron. Astrophys.} {\bf 594} (2016) A13,
  [\href{http://arxiv.org/abs/1502.01589}{{\tt arXiv:1502.01589}}].

\bibitem{Ackerman:mha}
L.~Ackerman, M.~R. Buckley, S.~M. Carroll, and M.~Kamionkowski, {\it {Dark
  Matter and Dark Radiation}},  \href{http://arxiv.org/abs/0810.5126}{{\tt
  arXiv:0810.5126}}.

\bibitem{Srednicki:1988ce}
M.~Srednicki, R.~Watkins, and K.~A. Olive, {\it {Calculations of Relic
  Densities in the Early Universe}},  {\em Nucl. Phys. B} {\bf 310} (1988) 693.

\bibitem{Cirelli:2007xd}
M.~Cirelli, A.~Strumia, and M.~Tamburini, {\it {Cosmology and Astrophysics of
  Minimal Dark Matter}},  {\em Nucl. Phys. B} {\bf 787} (2007) 152--175,
  [\href{http://arxiv.org/abs/0706.4071}{{\tt arXiv:0706.4071}}].

\bibitem{Belanger:2013oya}
G.~Belanger, F.~Boudjema, A.~Pukhov, and A.~Semenov, {\it {micrOMEGAs\_3: A
  program for calculating dark matter observables}},  {\em Comput. Phys.
  Commun.} {\bf 185} (2014) 960--985,
  [\href{http://arxiv.org/abs/1305.0237}{{\tt arXiv:1305.0237}}].

\bibitem{Hisano:2010yh}
J.~Hisano, K.~Ishiwata, N.~Nagata, and M.~Yamanaka, {\it {Direct Detection of
  Vector Dark Matter}},  {\em Prog. Theor. Phys.} {\bf 126} (2011) 435--456,
  [\href{http://arxiv.org/abs/1012.5455}{{\tt arXiv:1012.5455}}].

\bibitem{Aprile:2018dbl}
{\bf XENON} Collaboration, E.~Aprile et~al., {\it {Dark Matter Search Results
  from a One Ton-Year Exposure of XENON1T}},  {\em Phys. Rev. Lett.}
  [\href{http://arxiv.org/abs/1805.12562}{{\tt arXiv:1805.12562}}].

\bibitem{Springel:2005nw}
V.~Springel et~al., {\it {Simulating the joint evolution of quasars, galaxies
  and their large-scale distribution}},  {\em Nature} {\bf 435} (2005)
  629--636, [\href{http://arxiv.org/abs/astro-ph/0504097}{{\tt
  astro-ph/0504097}}].

\bibitem{Bullock:2017xww}
J.~S. Bullock and M.~Boylan-Kolchin, {\it {Small-Scale Challenges to the
  $\Lambda$CDM Paradigm}},  {\em Ann. Rev. Astron. Astrophys.} {\bf 55} (2017)
  343--387, [\href{http://arxiv.org/abs/1707.04256}{{\tt arXiv:1707.04256}}].

\bibitem{Feng:2009mn}
J.~L. Feng, M.~Kaplinghat, H.~Tu, and H.-B. Yu, {\it {Hidden Charged Dark
  Matter}},  {\em JCAP} {\bf 07} (2009) 004,
  [\href{http://arxiv.org/abs/0905.3039}{{\tt arXiv:0905.3039}}].

\bibitem{Agrawal:2016quu}
P.~Agrawal, F.-Y. Cyr-Racine, L.~Randall, and J.~Scholtz, {\it {Make Dark
  Matter Charged Again}},  {\em JCAP} {\bf 05} (2017) 022,
  [\href{http://arxiv.org/abs/1610.04611}{{\tt arXiv:1610.04611}}].

\bibitem{Zavala:2012us}
J.~Zavala, M.~Vogelsberger, and M.~G. Walker, {\it {Constraining
  Self-Interacting Dark Matter with the Milky Way's dwarf spheroidals}},  {\em
  Mon. Not. Roy. Astron. Soc.} {\bf 431} (2013) L20--L24,
  [\href{http://arxiv.org/abs/1211.6426}{{\tt arXiv:1211.6426}}].

\bibitem{Hindmarsh:2020hop}
M.~B. Hindmarsh, M.~L\"uben, J.~Lumma, and M.~Pauly, {\it {Phase transitions in
  the early universe}},  \href{http://arxiv.org/abs/2008.09136}{{\tt
  arXiv:2008.09136}}.

\bibitem{Farakos:1994xh}
K.~Farakos, K.~Kajantie, K.~Rummukainen, and M.~E. Shaposhnikov, {\it {3-d
  physics and the electroweak phase transition: A Framework for lattice Monte
  Carlo analysis}},  {\em Nucl. Phys. B} {\bf 442} (1995) 317--363,
  [\href{http://arxiv.org/abs/hep-lat/9412091}{{\tt hep-lat/9412091}}].

\bibitem{Kajantie:1995kf}
K.~Kajantie, M.~Laine, K.~Rummukainen, and M.~E. Shaposhnikov, {\it {The
  Electroweak phase transition: A Nonperturbative analysis}},  {\em Nucl. Phys.
  B} {\bf 466} (1996) 189--258,
  [\href{http://arxiv.org/abs/hep-lat/9510020}{{\tt hep-lat/9510020}}].

\bibitem{Moore:2000jw}
G.~D. Moore and K.~Rummukainen, {\it {Electroweak bubble nucleation,
  nonperturbatively}},  {\em Phys. Rev. D} {\bf 63} (2001) 045002,
  [\href{http://arxiv.org/abs/hep-ph/0009132}{{\tt hep-ph/0009132}}].

\bibitem{Linde:1980ts}
A.~D. Linde, {\it {Infrared Problem in Thermodynamics of the Yang-Mills Gas}},
  {\em Phys. Lett. B} {\bf 96} (1980) 289--292.

\bibitem{Brauner:2016fla}
T.~Brauner, T.~V.~I. Tenkanen, A.~Tranberg, A.~Vuorinen, and D.~J. Weir, {\it
  {Dimensional reduction of the Standard Model coupled to a new singlet scalar
  field}},  {\em JHEP} {\bf 03} (2017) 007,
  [\href{http://arxiv.org/abs/1609.06230}{{\tt arXiv:1609.06230}}].

\bibitem{Andersen:2017ika}
J.~O. Andersen, T.~Gorda, A.~Helset, L.~Niemi, T.~V.~I. Tenkanen, A.~Tranberg,
  A.~Vuorinen, and D.~J. Weir, {\it {Nonperturbative Analysis of the
  Electroweak Phase Transition in the Two Higgs Doublet Model}},  {\em Phys.
  Rev. Lett.} {\bf 121} (2018), no.~19 191802,
  [\href{http://arxiv.org/abs/1711.09849}{{\tt arXiv:1711.09849}}].

\bibitem{Niemi:2020hto}
L.~Niemi, M.~Ramsey-Musolf, T.~V. Tenkanen, and D.~J. Weir, {\it
  {Thermodynamics of a two-step electroweak phase transition}},
  \href{http://arxiv.org/abs/2005.11332}{{\tt arXiv:2005.11332}}.

\bibitem{Croon:2020cgk}
D.~Croon, O.~Gould, P.~Schicho, T.~V. Tenkanen, and G.~White, {\it {Theoretical
  uncertainties for cosmological first-order phase transitions}},
  \href{http://arxiv.org/abs/2009.10080}{{\tt arXiv:2009.10080}}.

\bibitem{Papaefstathiou:2020iag}
A.~Papaefstathiou and G.~White, {\it {The Electro-Weak Phase Transition at
  Colliders: Confronting Theoretical Uncertainties and Complementary
  Channels}},  \href{http://arxiv.org/abs/2010.00597}{{\tt arXiv:2010.00597}}.

\bibitem{Patel:2011th}
H.~H. Patel and M.~J. Ramsey-Musolf, {\it {Baryon Washout, Electroweak Phase
  Transition, and Perturbation Theory}},  {\em JHEP} {\bf 07} (2011) 029,
  [\href{http://arxiv.org/abs/1101.4665}{{\tt arXiv:1101.4665}}].

\bibitem{Quiros:1999jp}
M.~Quiros, {\it {Finite temperature field theory and phase transitions}},  in
  {\em {ICTP Summer School in High-Energy Physics and Cosmology}},
  pp.~187--259, 1, 1999.
\newblock \href{http://arxiv.org/abs/hep-ph/9901312}{{\tt hep-ph/9901312}}.

\bibitem{Vieu:2018nfq}
T.~Vieu, A.~P. Morais, and R.~Pasechnik, {\it {Electroweak phase transitions in
  multi-Higgs models: the case of Trinification-inspired THDSM}},  {\em JCAP}
  {\bf 07} (2018) 014, [\href{http://arxiv.org/abs/1801.02670}{{\tt
  arXiv:1801.02670}}].

\bibitem{Caprini:2015zlo}
C.~Caprini et~al., {\it {Science with the space-based interferometer eLISA. II:
  Gravitational waves from cosmological phase transitions}},  {\em JCAP} {\bf
  04} (2016) 001, [\href{http://arxiv.org/abs/1512.06239}{{\tt
  arXiv:1512.06239}}].

\bibitem{Turner:1992tz}
M.~S. Turner, E.~J. Weinberg, and L.~M. Widrow, {\it {Bubble nucleation in
  first order inflation and other cosmological phase transitions}},  {\em Phys.
  Rev.} {\bf D46} (1992) 2384--2403.

\bibitem{Wainwright:2011kj}
C.~L. Wainwright, {\it {CosmoTransitions: Computing Cosmological Phase
  Transition Temperatures and Bubble Profiles with Multiple Fields}},  {\em
  Comput. Phys. Commun.} {\bf 183} (2012) 2006--2013,
  [\href{http://arxiv.org/abs/1109.4189}{{\tt arXiv:1109.4189}}].

\bibitem{Guo:2020grp}
H.-K. Guo, K.~Sinha, D.~Vagie, and G.~White, {\it {Phase Transitions in an
  Expanding Universe: Stochastic Gravitational Waves in Standard and
  Non-Standard Histories}},  \href{http://arxiv.org/abs/2007.08537}{{\tt
  arXiv:2007.08537}}.

\bibitem{Apreda:2001us}
R.~Apreda, M.~Maggiore, A.~Nicolis, and A.~Riotto, {\it {Gravitational waves
  from electroweak phase transitions}},  {\em Nucl. Phys.} {\bf B631} (2002)
  342--368, [\href{http://arxiv.org/abs/gr-qc/0107033}{{\tt gr-qc/0107033}}].

\bibitem{Hindmarsh:2019phv}
M.~Hindmarsh and M.~Hijazi, {\it {Gravitational waves from first order
  cosmological phase transitions in the Sound Shell Model}},  {\em JCAP} {\bf
  1912} (2019), no.~12 062, [\href{http://arxiv.org/abs/1909.10040}{{\tt
  arXiv:1909.10040}}].

\bibitem{Bodeker:2017cim}
D.~Bodeker and G.~D. Moore, {\it {Electroweak Bubble Wall Speed Limit}},  {\em
  JCAP} {\bf 1705} (2017), no.~05 025,
  [\href{http://arxiv.org/abs/1703.08215}{{\tt arXiv:1703.08215}}].

\bibitem{Cai:2017cbj}
R.-G. Cai, Z.~Cao, Z.-K. Guo, S.-J. Wang, and T.~Yang, {\it {The
  Gravitational-Wave Physics}},  {\em Natl. Sci. Rev.} {\bf 4} (2017) 687--706,
  [\href{http://arxiv.org/abs/1703.00187}{{\tt arXiv:1703.00187}}].

\bibitem{Caprini:2018mtu}
C.~Caprini and D.~G. Figueroa, {\it {Cosmological Backgrounds of Gravitational
  Waves}},  {\em Class. Quant. Grav.} {\bf 35} (2018), no.~16 163001,
  [\href{http://arxiv.org/abs/1801.04268}{{\tt arXiv:1801.04268}}].

\bibitem{Romano:2016dpx}
J.~D. Romano and N.~J. Cornish, {\it {Detection methods for stochastic
  gravitational-wave backgrounds: a unified treatment}},  {\em Living Rev.
  Rel.} {\bf 20} (2017) 2, [\href{http://arxiv.org/abs/1608.06889}{{\tt
  arXiv:1608.06889}}].

\bibitem{Christensen:2018iqi}
N.~Christensen, {\it {Stochastic Gravitational Wave Backgrounds}},  {\em Rept.
  Prog. Phys.} {\bf 82} (2019), no.~1 016903,
  [\href{http://arxiv.org/abs/1811.08797}{{\tt arXiv:1811.08797}}].

\bibitem{Jinno:2016vai}
R.~Jinno and M.~Takimoto, {\it {Gravitational waves from bubble collisions: An
  analytic derivation}},  {\em Phys. Rev.} {\bf D95} (2017), no.~2 024009,
  [\href{http://arxiv.org/abs/1605.01403}{{\tt arXiv:1605.01403}}].

\bibitem{Huber:2008hg}
S.~J. Huber and T.~Konstandin, {\it {Gravitational Wave Production by
  Collisions: More Bubbles}},  {\em JCAP} {\bf 0809} (2008) 022,
  [\href{http://arxiv.org/abs/0806.1828}{{\tt arXiv:0806.1828}}].

\bibitem{Kosowsky:1992rz}
A.~Kosowsky, M.~S. Turner, and R.~Watkins, {\it {Gravitational waves from first
  order cosmological phase transitions}},  {\em Phys. Rev. Lett.} {\bf 69}
  (1992) 2026--2029.

\bibitem{Kosowsky:1991ua}
A.~Kosowsky, M.~S. Turner, and R.~Watkins, {\it {Gravitational radiation from
  colliding vacuum bubbles}},  {\em Phys. Rev.} {\bf D45} (1992) 4514--4535.

\bibitem{Kosowsky:1992vn}
A.~Kosowsky and M.~S. Turner, {\it {Gravitational radiation from colliding
  vacuum bubbles: envelope approximation to many bubble collisions}},  {\em
  Phys. Rev.} {\bf D47} (1993) 4372--4391,
  [\href{http://arxiv.org/abs/astro-ph/9211004}{{\tt astro-ph/9211004}}].

\bibitem{Jinno:2017fby}
R.~Jinno and M.~Takimoto, {\it {Gravitational waves from bubble dynamics:
  Beyond the Envelope}},  \href{http://arxiv.org/abs/1707.03111}{{\tt
  arXiv:1707.03111}}.

\bibitem{Child:2012qg}
H.~L. Child and J.~Giblin, John~T., {\it {Gravitational Radiation from
  First-Order Phase Transitions}},  {\em JCAP} {\bf 10} (2012) 001,
  [\href{http://arxiv.org/abs/1207.6408}{{\tt arXiv:1207.6408}}].

\bibitem{Cutting:2018tjt}
D.~Cutting, M.~Hindmarsh, and D.~J. Weir, {\it {Gravitational waves from vacuum
  first-order phase transitions: from the envelope to the lattice}},  {\em
  Phys. Rev. D} {\bf 97} (2018), no.~12 123513,
  [\href{http://arxiv.org/abs/1802.05712}{{\tt arXiv:1802.05712}}].

\bibitem{Hindmarsh:2013xza}
M.~Hindmarsh, S.~J. Huber, K.~Rummukainen, and D.~J. Weir, {\it {Gravitational
  waves from the sound of a first order phase transition}},  {\em Phys. Rev.
  Lett.} {\bf 112} (2014) 041301, [\href{http://arxiv.org/abs/1304.2433}{{\tt
  arXiv:1304.2433}}].

\bibitem{Hindmarsh:2015qta}
M.~Hindmarsh, S.~J. Huber, K.~Rummukainen, and D.~J. Weir, {\it {Numerical
  simulations of acoustically generated gravitational waves at a first order
  phase transition}},  {\em Phys. Rev.} {\bf D92} (2015), no.~12 123009,
  [\href{http://arxiv.org/abs/1504.03291}{{\tt arXiv:1504.03291}}].

\bibitem{Hindmarsh:2017gnf}
M.~Hindmarsh, S.~J. Huber, K.~Rummukainen, and D.~J. Weir, {\it {Shape of the
  acoustic gravitational wave power spectrum from a first order phase
  transition}},  {\em Phys. Rev.} {\bf D96} (2017), no.~10 103520,
  [\href{http://arxiv.org/abs/1704.05871}{{\tt arXiv:1704.05871}}].

\bibitem{Cutting:2019zws}
D.~Cutting, M.~Hindmarsh, and D.~J. Weir, {\it {Vorticity, kinetic energy, and
  suppressed gravitational wave production in strong first order phase
  transitions}},  \href{http://arxiv.org/abs/1906.00480}{{\tt
  arXiv:1906.00480}}.

\bibitem{Hindmarsh:2016lnk}
M.~Hindmarsh, {\it {Sound shell model for acoustic gravitational wave
  production at a first-order phase transition in the early Universe}},  {\em
  Phys. Rev. Lett.} {\bf 120} (2018), no.~7 071301,
  [\href{http://arxiv.org/abs/1608.04735}{{\tt arXiv:1608.04735}}].

\bibitem{Weir:2017wfa}
D.~J. Weir, {\it {Gravitational waves from a first order electroweak phase
  transition: a brief review}},  {\em Phil. Trans. Roy. Soc. Lond.} {\bf A376}
  (2018), no.~2114 20170126, [\href{http://arxiv.org/abs/1705.01783}{{\tt
  arXiv:1705.01783}}].

\bibitem{Espinosa:2010hh}
J.~R. Espinosa, T.~Konstandin, J.~M. No, and G.~Servant, {\it {Energy Budget of
  Cosmological First-order Phase Transitions}},  {\em JCAP} {\bf 1006} (2010)
  028, [\href{http://arxiv.org/abs/1004.4187}{{\tt arXiv:1004.4187}}].

\bibitem{Ellis:2019oqb}
J.~Ellis, M.~Lewicki, J.~M. No, and V.~Vaskonen, {\it {Gravitational wave
  energy budget in strongly supercooled phase transitions}},  {\em JCAP} {\bf
  06} (2019) 024, [\href{http://arxiv.org/abs/1903.09642}{{\tt
  arXiv:1903.09642}}].

\bibitem{Ellis:2020awk}
J.~Ellis, M.~Lewicki, and J.~M. No, {\it {Gravitational waves from first-order
  cosmological phase transitions: lifetime of the sound wave source}},
  \href{http://arxiv.org/abs/2003.07360}{{\tt arXiv:2003.07360}}.

\bibitem{Caprini:2019egz}
C.~Caprini et~al., {\it {Detecting gravitational waves from cosmological phase
  transitions with LISA: an update}},  {\em JCAP} {\bf 03} (2020) 024,
  [\href{http://arxiv.org/abs/1910.13125}{{\tt arXiv:1910.13125}}].

\bibitem{Pen:2015qta}
U.-L. Pen and N.~Turok, {\it {Shocks in the Early Universe}},  {\em Phys. Rev.
  Lett.} {\bf 117} (2016), no.~13 131301,
  [\href{http://arxiv.org/abs/1510.02985}{{\tt arXiv:1510.02985}}].

\bibitem{Giese:2020rtr}
F.~Giese, T.~Konstandin, and J.~van~de Vis, {\it {Model-independent energy
  budget of cosmological first-order phase transitions}},
  \href{http://arxiv.org/abs/2004.06995}{{\tt arXiv:2004.06995}}.

\bibitem{Pol:2019yex}
A.~Roper~Pol, S.~Mandal, A.~Brandenburg, T.~Kahniashvili, and A.~Kosowsky, {\it
  {Numerical Simulations of Gravitational Waves from Early-Universe
  Turbulence}},  {\em Phys. Rev. D} {\bf 102} (2020) 083512,
  [\href{http://arxiv.org/abs/1903.08585}{{\tt arXiv:1903.08585}}].

\bibitem{no:2011fi}
J.~M. No, {\it {Large Gravitational Wave Background Signals in Electroweak
  Baryogenesis Scenarios}},  {\em Phys. Rev.} {\bf D84} (2011) 124025,
  [\href{http://arxiv.org/abs/1103.2159}{{\tt arXiv:1103.2159}}].

\bibitem{Alves:2018oct}
A.~Alves, T.~Ghosh, H.-K. Guo, and K.~Sinha, {\it {Resonant Di-Higgs Production
  at Gravitational Wave Benchmarks: A Collider Study using Machine Learning}},
  \href{http://arxiv.org/abs/1808.08974}{{\tt arXiv:1808.08974}}.

\bibitem{Audley:2017drz}
{\bf LISA} Collaboration, P.~Amaro-Seoane et~al., {\it {Laser Interferometer
  Space Antenna}},  \href{http://arxiv.org/abs/1702.00786}{{\tt
  arXiv:1702.00786}}.

\bibitem{Gong:2014mca}
X.~Gong et~al., {\it {Descope of the ALIA mission}},  {\em J. Phys. Conf. Ser.}
  {\bf 610} (2015), no.~1 012011, [\href{http://arxiv.org/abs/1410.7296}{{\tt
  arXiv:1410.7296}}].

\bibitem{Luo:2015ght}
{\bf TianQin} Collaboration, J.~Luo et~al., {\it {TianQin: a space-borne
  gravitational wave detector}},  {\em Class. Quant. Grav.} {\bf 33} (2016),
  no.~3 035010, [\href{http://arxiv.org/abs/1512.02076}{{\tt
  arXiv:1512.02076}}].

\bibitem{Kudoh:2005as}
H.~Kudoh, A.~Taruya, T.~Hiramatsu, and Y.~Himemoto, {\it {Detecting a
  gravitational-wave background with next-generation space interferometers}},
  {\em Phys. Rev. D} {\bf 73} (2006) 064006,
  [\href{http://arxiv.org/abs/gr-qc/0511145}{{\tt gr-qc/0511145}}].

\bibitem{Chala:2019rfk}
M.~Chala, V.~V. Khoze, M.~Spannowsky, and P.~Waite, {\it {Mapping the shape of
  the scalar potential with gravitational waves}},  {\em Int. J. Mod. Phys. A}
  {\bf 34} (2019), no.~33 1950223, [\href{http://arxiv.org/abs/1905.00911}{{\tt
  arXiv:1905.00911}}].

\end{thebibliography}\endgroup

\end{document}